\documentclass[onecolumn]{pasj02} 
\usepackage[switch,mathlines]{lineno} 
\usepackage{url}
\usepackage{enumitem}
\usepackage{graphicx}
\usepackage{overpic}
\newlist{inparaenum}{enumerate}{2}
\setlist[inparaenum]{nosep}
\setlist[inparaenum,1]{label=\bfseries\arabic*.}
\setlist[inparaenum,2]{label=\arabic{inparaenumi}\emph{\alph*})}

\jyear{2025}
\Received{}
\Accepted{}

\begin{document} 
\title{Cosmic-ray and Interstellar Gas Properties in the Solar Neighborhood Revealed by Diffuse Gamma Rays}

\author{
 Tsunefumi \textsc{Mizuno},\altaffilmark{1}\altemailmark\orcid{0000-0001-7263-0296} \email{mizuno@astro.hiroshima-u.ac.jp} 
 Katsuhiro \textsc{Hayashi},\altaffilmark{2}\orcid{0000-0001-6922-6583}
 Hinako \textsc{Ochi},\altaffilmark{3}
 Igor V. \textsc{Moskalenko},\altaffilmark{4}\orcid{0000-0001-6141-458X}
 Elena \textsc{Orlando},\altaffilmark{4,5,6}\orcid{0000-0001-6406-9910}
and
 Andrew W. \textsc{Strong},\altaffilmark{7}\orcid{0000-0003-3799-5489}
}
\altaffiltext{1}{Hiroshima Astrophysical Science Center, Hiroshima University, 1-3-1 Kagamiyama, Higashi-Hiroshima, Hiroshima 739-8526, Japan}
\altaffiltext{2}{Institute of Space and Astronautical Science, Japan Aerospace Exploration Agency, Kanagawa 252-5210, Japan}
\altaffiltext{3}{Department of Physics, Hiroshima University, Hiroshima 739-8526, Japan}
\altaffiltext{4}{W. W. Hansen Experimental Physics Laboratory, Kavli Institute for Particle Astrophysics and Cosmology, Stanford University, CA 94305, USA}
\altaffiltext{5}{Department of Physics, University of Trieste and INFN, I-34127 Trieste, Italy}
\altaffiltext{6}{Eureka Scientific, Oakland, CA 94602-3017, USA}
\altaffiltext{7}{Max-Planck Institut f\"ur extraterrestrische Physik, D-85748 Garching, Germany}




\newcommand{\HI}{H\,\emissiontype{I}}
\newcommand{\Htwo}{H$_{2}$}

\newcommand{\WHI}{W_\mathrm{HI}}
\newcommand{\NHI}{N_\mathrm{HI}}
\newcommand{\VHI}{V_\mathrm{HI}}
\newcommand{\NH}{N_\mathrm{H}}
\newcommand{\WCO}{W_\mathrm{CO}}
\newcommand{\CHI}{C_{\scriptsize {H}{i}}}
\newcommand{\CHIi}{C_{\mathrm{HI}, i}}
\newcommand{\NHIi}{N_{\mathrm{HI}, i}}
\newcommand{\CHIzro}{C_\mathrm{HI,0}}
\newcommand{\CHIone}{C_\mathrm{HI,1}}
\newcommand{\CHItwo}{C_\mathrm{HI,2}}
\newcommand{\CHIthr}{C_\mathrm{HI,3}}
\newcommand{\CHIonetwo}{C_\mathrm{HI,1+2}}
\newcommand{\CHItwothr}{C_\mathrm{HI,2+3}}
\newcommand{\NHtwo}{N_\mathrm{H_{2}}}
\newcommand{\MHtwo}{M_\mathrm{H_{2}}}
\newcommand{\MHtwoCO}{M_\mathrm{H_{2}}^\mathrm{CO}}

\newcommand\mycolor[2]{\textcolor{#2}{#1}}

\maketitle

\begin{abstract}
To investigate the interstellar medium (ISM) and Galactic cosmic rays (CRs) in the solar neighborhood,
we analyzed ${\gamma}$-ray data by Fermi Large Area Telescope (LAT) for five nearby molecular cloud regions.
Our data includes the MBM/Pegasus region (MBM~53, 54, 55 clouds and Pegasus loop),
R CrA region (R Coronae Australis clouds),
Chamaeleon region (Chamaeleon clouds),
Cep/Pol region (Cepheus and Polaris flare), and 
Orion region (Orion clouds).
The ISM templates are constructed by a
component decomposition of the 21~cm {\HI} line, the Planck dust emission model, and the
carbon monoxide (CO) 2.6~mm line.
Through $\gamma$-ray data analysis the ISM gas is successfully decomposed into 
non-local {\HI}, narrow-line and optically thick {\HI}, broad-line and optically thin {\HI}, 
CO-bright {\Htwo}, and CO-dark {\Htwo} for all five regions.
CR intensities evaluated by the ${\gamma}$-ray emissivity of broad {\HI} agree well with a model based on directly-measured CR spectra at the Earth,
with a gradient giving a higher CR intensity toward the inner Galaxy at the 10\% level in ${\sim}$ 500~pc.
The ratio of CO-dark {\Htwo} to CO-bright {\Htwo} anti-correlates with the {\Htwo} mass traced by the CO 2.6~mm line,
and reaches 5--10 for small systems of ${\sim}$1000 solar mass.
\end{abstract}


\clearpage

\section{Introduction}
Interstellar space in the Milky Way is permeated with ordinary matter (gas or dust) known as the interstellar medium (ISM). 
It also contains high-energy charged particles known as cosmic rays (CRs), an interstellar radiation field (ISRF), and a magnetic field. 
These components mutually interact and play important roles in physical and chemical processes (e.g., star formation and cycle of matter). 
Hence, they have been studied in various wavebands from radio to $\gamma$-rays (e.g., \cite{Ferriere2001}).

Cosmic ${\gamma}$-ray emission (with photon energies $E \ge 100~\mathrm{MeV}$) is a powerful probe for studying the ISM and Galactic CRs. 
High-energy CR protons and electrons interact with the ISM gas or the ISRF and produce $\gamma$-rays through nucleon--nucleon interactions, 
electron bremsstrahlung, and inverse-Compton (IC) scattering. 
Because the $\gamma$-ray production cross section is independent of the chemical or thermodynamic state of the ISM gas, 
and the interstellar space is transparent to those ${\gamma}$-rays (e.g., \cite{Moskalenko2006}),
cosmic $\gamma$-rays have been recognized as a unique tracer of the total column density of the gas, 
regardless of its atomic or molecular state. 
If observations in other wavebands provide an accurate estimate of the gas column density, 
we can also examine the CR spectrum and intensity distribution.

Usually, the distribution of neutral atomic hydrogen ({\HI}) is measured directly via 21~cm line surveys
(e.g., \cite{Dickey1990}; \cite{Kalberla2009}), assuming the optically thin approximation, 
and the distribution of molecular hydrogen ({\Htwo}) is estimated indirectly from carbon monoxide (CO) 
line-emission surveys (e.g., \cite{Dame2001}), assuming a linear conversion factor. 
However, these line surveys likely miss some fraction of the ISM gas in optically thick {\HI} or CO-dark {\Htwo} phases.
Such "dark gas" can be studied using total gas tracers such as dust extinction, reddening, 
and emission (e.g., \cite{Reach1994}) and $\gamma$-rays (e.g., \cite{Grenier2005}). 
The pioneering work by \citet{Grenier2005} based on CGRO-EGRET data has been confirmed and improved by subsequent studies with the Fermi Large Area Telescope 
(LAT; \cite{Atwood2009}). In addition, the Planck mission has provided an all-sky model of thermal emission from dust 
(\cite{Planck2011}; \cite{Planck2014}). It is useful for studying the ISM gas distribution because of its sensitivity and high angular resolution.

Nevertheless, uncertainties in the ISM gas column density and CR intensity are still uncomfortably large, 
by as much as ${\sim}$50\% even in the local environment (e.g., \cite{Grenier2015}). 
This is mainly due to the uncertainty in the spin-temperature ($T_\mathrm{s}$) of {\HI} gas, 
which affects the conversion from the 21~cm line intensity to the {\HI} gas column density. 
Also, the composition of dark gas 
(i.e., the fractions of optically thick {\HI} and CO-dark {\Htwo}) is quite uncertain and controversial. 
For example, while \citet{Fukui2015} proposed that optically thick {\HI} dominates dark gas, 
\citet{Murray2018} claimed that dark gas is mainly molecular. 
Again, this is because the value of $T_\mathrm{s}$ is usually unknown, 
and neither dust nor $\gamma$-rays can distinguish between atomic and molecular gas phases.

One may solve these issues using {\HI} linewidth.
{\HI} absorption features (e.g., the optical depth profile) are often well represented by Gaussians, 
supporting the idea that gas motions within {\HI} clouds have a random velocity distribution. 
{\HI} emission profiles can also be decomposed into Gaussian lines, 
and components with narrow or broad line widths could be assumed to arise from a cold neutral medium (CNM) or a warm neutral medium (WNM), 
respectively (e.g., \cite{Kalberla2020}). Recently, \citet{Kalberla2018} analyzed the all-sky HI4PI survey data 
\citep{HI4PI} and decomposed the {\HI} 21~cm line emission into Gaussians by taking spatial coherence into account. 
Although their analysis uses emission spectra only and hence suffers from systematic uncertainties,
it allows them to separate the CNM and WNM over the entire sky.
Subsequently, \citet{Kalberla2020} found that narrow-line {\HI} gas 
(hereafter called "narrow {\HI}")
is associated with the dark gas estimated from infrared dust-emission maps 
by \citet{Schlegel1998}. 
Specifically, {\HI} lines with a Doppler temperature $T_\mathrm{D} \le 1000~\mathrm{K}$ 
are associated with gas for which the estimated column density is significantly larger than the optically thin case.
Here, $T_\mathrm{D}$ is defined as $22 \times \delta v^{2}$ where $\delta v$ is the observed Gaussian linewidth
in full width at the half maximum (FWHM) in $\mathrm{km~s^{-1}}$.
On the other hand, broad-line {\HI} (hereafter called "broad {\HI}") is associated with translucent gas.
Their work allows for identifying optically thin {\HI} and dark gas using {\HI} line profile information.
Following their work,
\citet{Mizuno2022} employed an {\HI}-line-profile-based analysis in analyzing Fermi-LAT ${\gamma}$-ray data toward
the MBM 53, 54, and 55 clouds and the Pegasus loop.
By also employing the dust emission model by the Planck mission, they succeeded in decomposing ISM gas phases
into optically thin {\HI} (traced by broad {\HI}), optically thick {\HI} (traced by narrow {\HI}), 
CO-bright {\Htwo} (traced by CO-line emission), and CO-dark {\Htwo} (traced by dust emission not correlated
with {\HI}- nor CO-lines).
$\gamma$-ray yield of broad {\HI} agrees well with the expectation from diredtly-measured CR spectra,
supporting it to be optically thin.
We aim to extend their work to other nearby, high-latitude molecular clouds, to study CRs and ISM gas properties
in the solar neighborhood.

This paper is organized as follows: In Section~2, we describe the preparation of ISM gas templates,
along with the $\gamma$-ray observations, data selection, and modeling.
In Section~3, we present the results of the data analysis which confirm that narrow {\HI} traces optically thick {\HI}.
In Section~4, we discuss CR and ISM properties. 
Finally, in Section~5, we present a summary of the study and prospects.

\clearpage
\section{Gamma-Ray Data and Modeling}
\subsection{Gamma-Ray Observation and Data Selection}
The LAT onboard the Fermi Gamma-ray Space Telescope (launched in June 2008)
is a pair-tracking $\gamma$-ray telescope that detects photons from {$\sim$}20~MeV to more than
300~GeV. Details of the LAT instrument and the pre-launch performance expectations
can be found in \citet{Atwood2009}, and the on-orbit calibration is described in \citet{Abdo2009}.
Thanks to its wide field of view (${\sim}$2.4~sr), Fermi-LAT is an ideal telescope for studying
Galactic diffuse $\gamma$-rays.
Although the angular resolution is worse than those of gas tracers and energy-dependent
(the 68\% containment radiuses are ${\sim}\timeform{5D}$ and \timeform{0.8D} at 100~MeV and 1~GeV, respectively),
it will be taken into account in the $\gamma$-ray data analysis described in Section~3.
Examples of relevant studies on Galactic diffuse emission include \citet{FermiPaper2} and \citet{FermiHI2}.

Routine science operations with the LAT started on 
August 4, 2008.
We have accumulated events 
from August 4, 2008 to August 2, 2023
(i.e., 15 years) to study diffuse 
$\gamma$-rays in our targets. Specifically, we defined five regions of interest (ROIs) according to Galactic longitude ($l$)
and latitude ($b$);
$60^{\circ} \le l \le 120^{\circ}$ and $-60^{\circ} \le b \le -28^{\circ}$ for the MBM/Pegasus region,
$-30^{\circ} \le l \le 30^{\circ}$ and $-40^{\circ} \le b \le -15^{\circ}$ for the R CrA region,
$280^{\circ} \le l \le 320^{\circ}$ and $-35^{\circ} \le b \le -12^{\circ}$ for the Camaeleon region,
$100^{\circ} \le l \le 120^{\circ}$ and $15^{\circ} \le b \le 30^{\circ}$ for the Cep/Pol region,
and
$195^{\circ} \le l \le 225^{\circ}$ and $-35^{\circ} \le b \le -10^{\circ}$ for the Orion region.
See also Appendix~1.
As described in Section~3, they are all nearby molecular cloud regions (within 500~pc) and cover a wide range of molecular mass
(${\sim}10^{3}\mbox{--}10^{5}~M_{\odot}$ where $M_{\odot}$ stands for the solar mass).
During most of these observations,
the LAT was operated in sky-survey mode, obtaining complete sky coverage every
two orbits (which corresponds to ${\sim}$3~h), with relatively uniform exposure over time.
We used the latest release of the Pass~8 \citep{Atwood2013,Bruel2018}
data, namely P8R3.
We used the standard LAT analysis software, Fermitools\footnote{
\url{https://fermi.gsfc.nasa.gov/ssc/data/analysis/software/}
}
version 2.2.0,
to select events satisfying the ULTRACLEAN class
to obtain low contamination from background events.
We also required that the reconstructed zenith angles of the
arrival directions of the photons be less than $100^{\circ}$ and $90^{\circ}$ for energies
above and below 400~MeV, respectively,
to reduce contamination by photons from Earth's atmosphere.
We used events and the responses of point-spread-function (PSF) event types 2 and 3 below 400~MeV
to accommodate the relatively poor angular resolution at low energy.
We did not apply selections based on PSF event types to maximize the photon statistics above 400~MeV.
We used the {\tt gtselect} command to apply the selections described above.

Among five targets, data for the MBM/Pegasus region is contaminated by a bright active galactic nucleus 3C~454.3.
Accordingly, 
we referred to the Monitored Source List
light curves\footnote{
\url{https://fermi.gsfc.nasa.gov/ssc/data/access/lat/msl_lc/}
},
and we excluded the periods ({$\sim$}1870~days in total) during which the LAT detected flares from the source
by using the {\tt gtmktime} command.
This cut\footnote{
The lists of the mission elapsed time (the number of seconds since January 1, 2001)
that passed the criteria are 2.45--2.72, 3.23--4.17, 5.05--5.45, 5.47--6.83, and larger than 6.99
in $10^{8}$.}
is applied only to data of the MBM/Pegasus region.
We also excluded the periods during which the LAT
detected bright $\gamma$-ray bursts or solar flares. 
The integrated time excluded in this procedure is negligible
compared to the total exposure.
Then we prepared a livetime cube by using the {\tt gtltcube} command.
We used the latest response functions that match
our dataset and event selection, P8R3\_ULTRACLEAN\_V3, in the following analysis. 

\clearpage
\subsection{Model Preparation}
Since the ISM is optically thin to $\gamma$-rays in the energy range considered in this paper and the CR spectrum is not expected to vary
significantly within each of the small regions, the $\gamma$-ray intensity from CR protons and electrons interacting with the 
ISM gas and the ISRF can be modeled as a sum of emission from separate gas phases and other emission components. 
The key to the analysis is to prepare good templates,
which we will give details below. All gas maps are stored in
HEALPix \citep{Gorski2005} equal-area sky map of order 9\footnote{
This corresponds to the total number of pixels of $12 \times (2^{9})^{2} = 3145728.$ (9 comes from the resolution index.)
}
with a mean distance of adjacent pixels of \timeform{6.9'} ($0.114~\mathrm{deg}$) and a pixel size of $0.0131~\mathrm{deg^{2}}$.
\subsubsection{{\HI} and CO Maps}
We downloaded {\HI} line profile data files by \citet{Kalberla2020}\footnote{
\url{https://www.astro.uni-bonn.de/hisurvey/AllSky_gauss/}
}
for our ROIs and peripheral regions (${\le}5\arcdeg$ from the boundaries). 
They modeled
the {\HI} 21-cm emission 
of the HI4PI survey data 
with an angular resolution of \timeform{16.2'} in FWHM
in each sky direction using several Gaussians.
They also gave
the normalization, center, and width of each line.
As described in \citet{Kalberla2018}, they required the residuals to be consistent with the noise level,
and also required the number of the used Gaussians to be as low as possible by considering the information from the 
neighboring pixels.
Negative normalizations or widths are given to suspicious lines (weak lines likely being artifacts due to the noise), 
and we discarded them when preparing the map.
We then separated the {\HI} column densities into three components along the line of sight; one corresponds to the non-local region,
and the others correspond to narrow {\HI} and broad {\HI} in the local region.
Doppler temperature threshold of 1000~K was adopted to separate narrow {\HI} and broad {\HI} (see Appendix~2).
Following the previous works, we determined the velocity boundaries to separate local and non-local clouds as a function of ($l$, $b$). 
Specifically, the local velocity range is 
$-30 \le v_\mathrm{LSR} \le 20~\mathrm{km~s^{-1}}$ for the MBM/Pegasus region (same as \cite{Mizuno2022}),
where $v_\mathrm{LSR}$ is the line-of-sight velocity relative to the local standard of rest.
Other ranges are $-15 \le v_\mathrm{LSR} \le 15~\mathrm{km~s^{-1}}$ for the R CrA region,
$-15 \le v_\mathrm{LSR} \le 15~\mathrm{km~s^{-1}}$ for the Chamaeleon region,
$-8~\mathrm{km~s^{-1}} \le v_\mathrm{LSR}$ for the Cep/Pol region (the same as those by \citet{Ackermann2012}),
and $-10 \le v_\mathrm{LSR} \le 25~\mathrm{km~s^{-1}}$ for the Orion region.
For R CrA and Chamaeleon regions, $| v_\mathrm{LSR} |$ is increased below $-10^{\circ}$ to $80~\mathrm{km~s^{-1}}$ at 
$b=\timeform{20.5D}$.
Similarly, $v_\mathrm{LSR}$ is decreased above $15^{\circ}$ to $-100~\mathrm{K~km~s^{-1}}$ at $b=24^{\circ}$ for Cep/Pol region.
For Orion, $| v_\mathrm{LSR} |$ is increased by $30~\mathrm{K~km~s^{-1}}$ from $b= -10^{\circ}$ to $-25^{\circ}$. Those
velocity boundaries are determined as in \citet{Ackermann2012} with slight simplification for R CrA, Chamaeleon, and Cep/Pol regions, 
and by inspecting the velocity-latitude profile for the Orion region.
We converted the $\WHI$ (integrated 21~cm line intensity) map into the $\NHI$ ({\HI} column density) map 
for the optically thin case 
[$\NHI \mathrm{(cm^{-2})} = 1.82 \times 10^{18}\ \WHI \mathrm{(K~km~s^{-1})}$]
and used them in the $\gamma$-ray data analysis (Section~2.2.4).
$\NHI$ maps of narrow {\HI} and broad {\HI} for each of our ROIs are shown in Figures~1--5,
and those of non-local {\HI} are summarized in Appendix~3.

As was done in \citet{Mizuno2022}, we used a $\WCO$ map 
[map of the integrated $^{12}$CO (J=1--0) 2.6-mm line intensity]
internally available to the LAT team.
It combines 
the work by \citet{Dame2001} and new data at high Galactic latitudes.
Those data were taken by two 1.2~m telescopes (one in the northern hemisphere and the other in the southern hemisphere)
with a spacing of \timeform{0.25D} or better for most areas.
Then the data were
smoothed to give an angular resolution of $18^{'}$ (FWHM) and sampled in \timeform{0.25D} intervals.
Some CO clouds in the Chamaeleon region are not covered, and we referred to NANTEN data used by \citet{Hayashi2019} to complement this.
These maps are also shown in Figures~1-5. 

\begin{figure}[htbp]
\begin{tabular}{cc}
\begin{minipage}{0.5\textwidth}
\centering
\begin{overpic}[width=\textwidth]{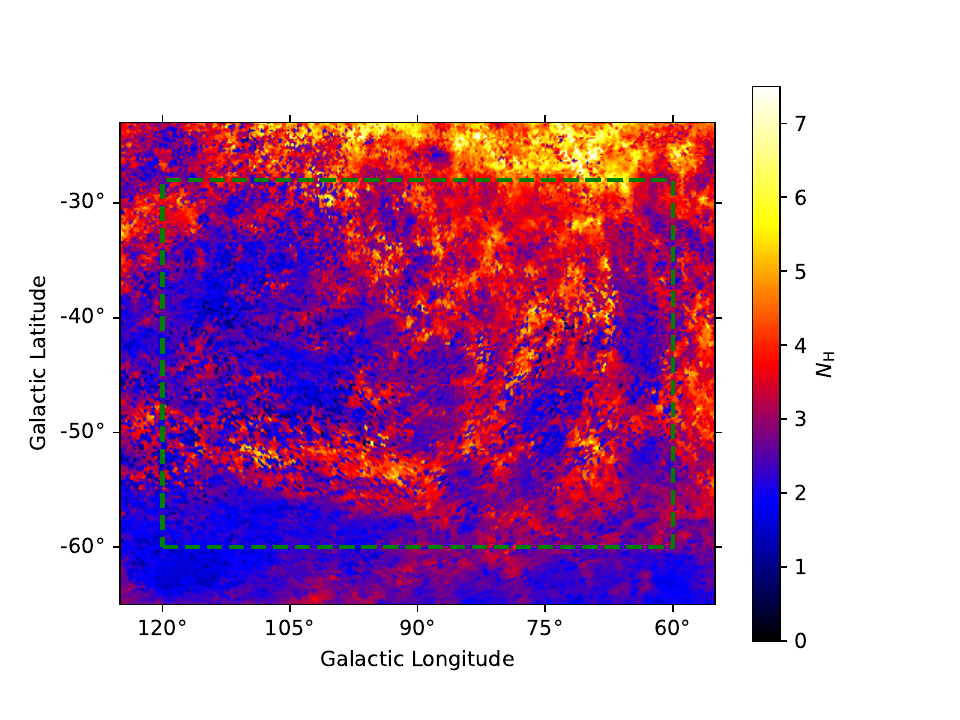}
\put(15,65){(a)}
\end{overpic}
\end{minipage}
\begin{minipage}{0.5\textwidth}
\centering
\begin{overpic}[width=\textwidth]{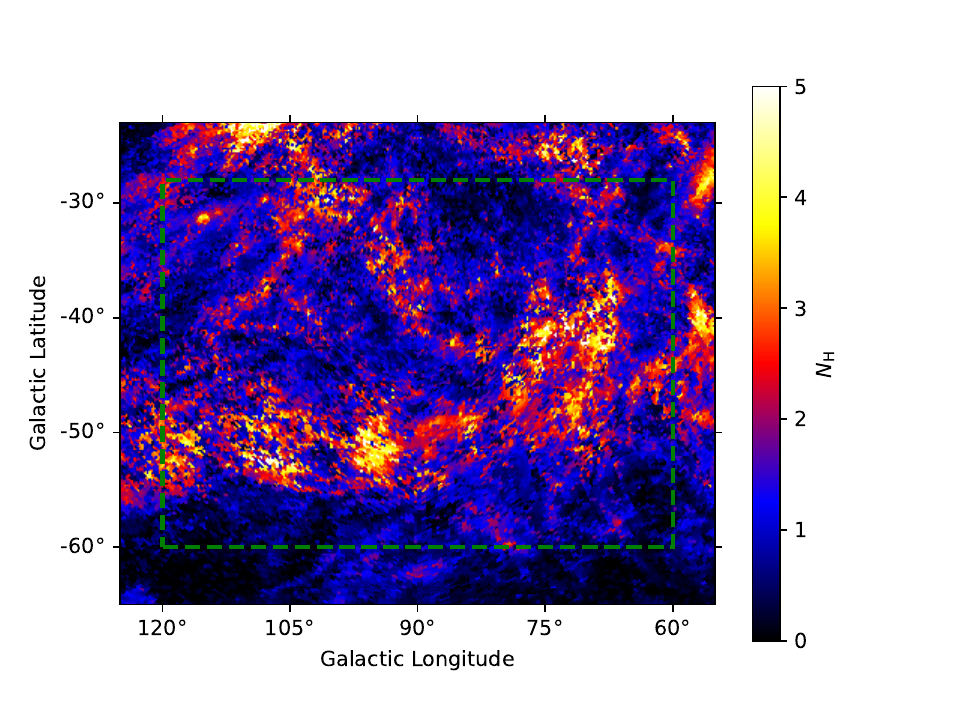}
\put(15,65){(b)}
\end{overpic}
\end{minipage} \\
\\
\begin{minipage}{0.5\textwidth}
\centering
\begin{overpic}[width=\textwidth]{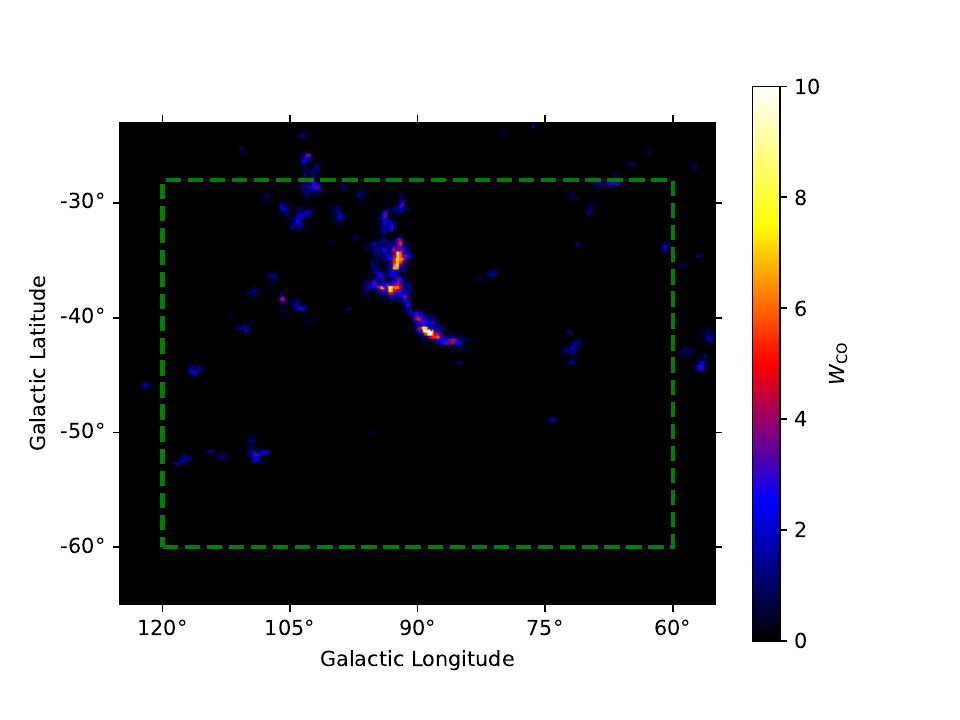}
\put(15,65){(c)}
\end{overpic}
\end{minipage}
\begin{minipage}{0.5\textwidth}
\centering
\begin{overpic}[width=\textwidth]{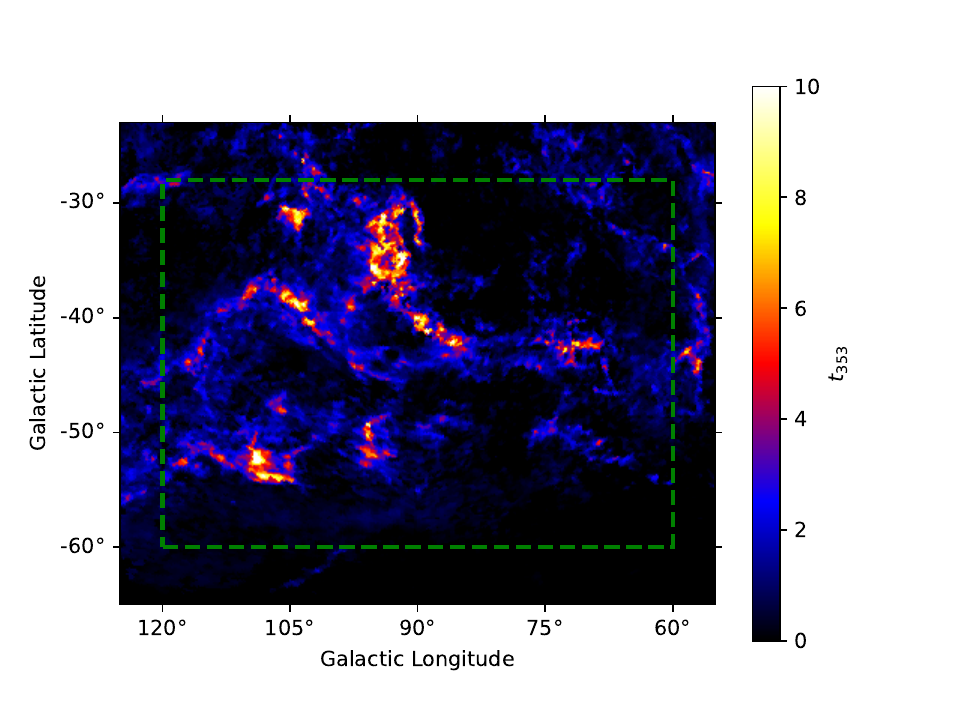}
\put(15,65){(d)}
\end{overpic}
\end{minipage}
\end{tabular}
\caption{
Template maps for the MBM/Pegasus region. (a) $\NHI$ map of broad {\HI}, 
(b) $\NHI$ map of narrow {\HI}, (c) $\WCO$ map, 
and (d) $D_\mathrm{em, res}$ map.
Two $\NHI$ maps are converted from $\WHI$ assuming an optically thin case and in $10^{20}~\mathrm{cm^{-2}}$, 
$\WCO$ map is in $\mathrm{K~km~s^{-1}}$, and $D_\mathrm{em, res}$ map is constructed from $\tau_{353}$ and in $10^{-6}$.
{Alt text: Four template maps for the MBM/Pegasus region.}}\label{..}
\end{figure}

\begin{figure}[htbp]
\begin{tabular}{cc}
\begin{minipage}{0.5\textwidth}
\centering
\begin{overpic}[width=\textwidth]{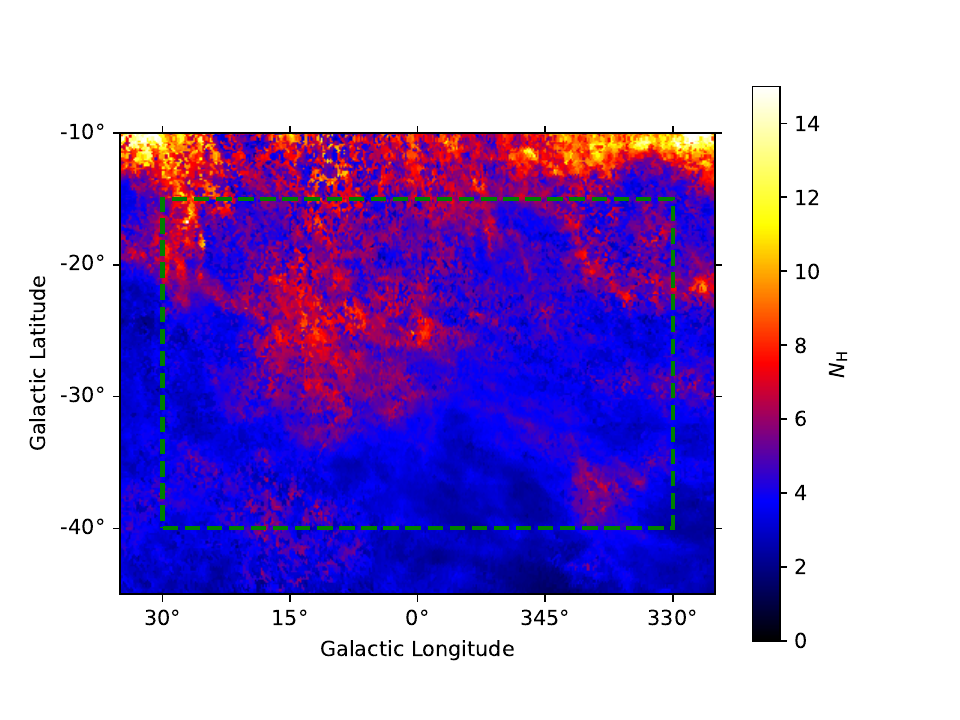}
\put(15,65){(a)}
\end{overpic}
\end{minipage}
\begin{minipage}{0.5\textwidth}
\centering
\begin{overpic}[width=\textwidth]{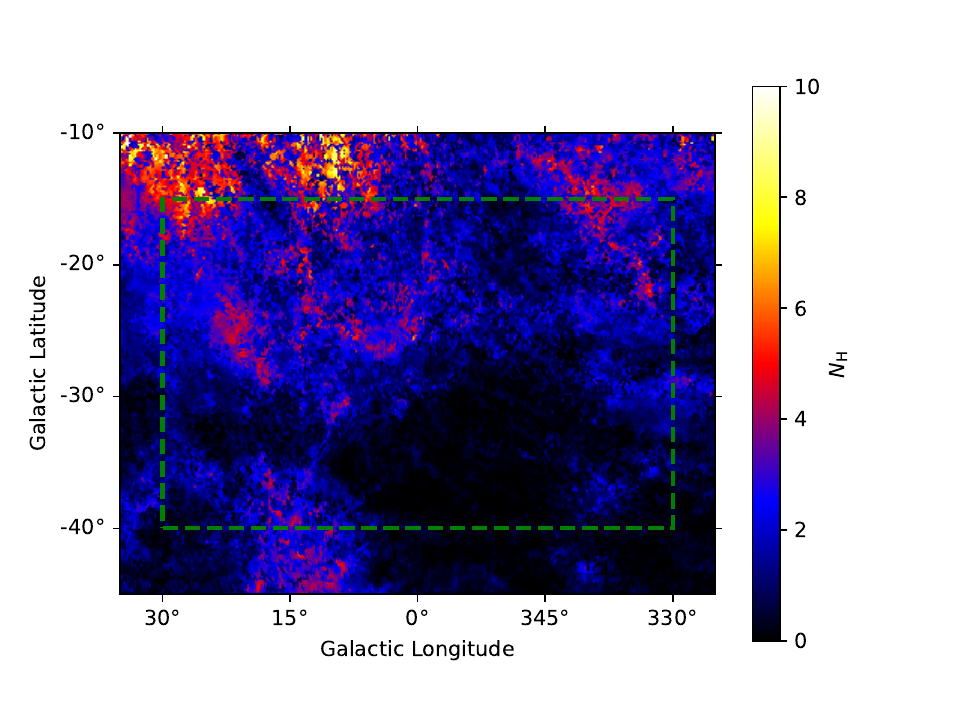}
\put(15,65){(b)}
\end{overpic}
\end{minipage} \\
\\
\begin{minipage}{0.5\textwidth}
\centering
\begin{overpic}[width=\textwidth]{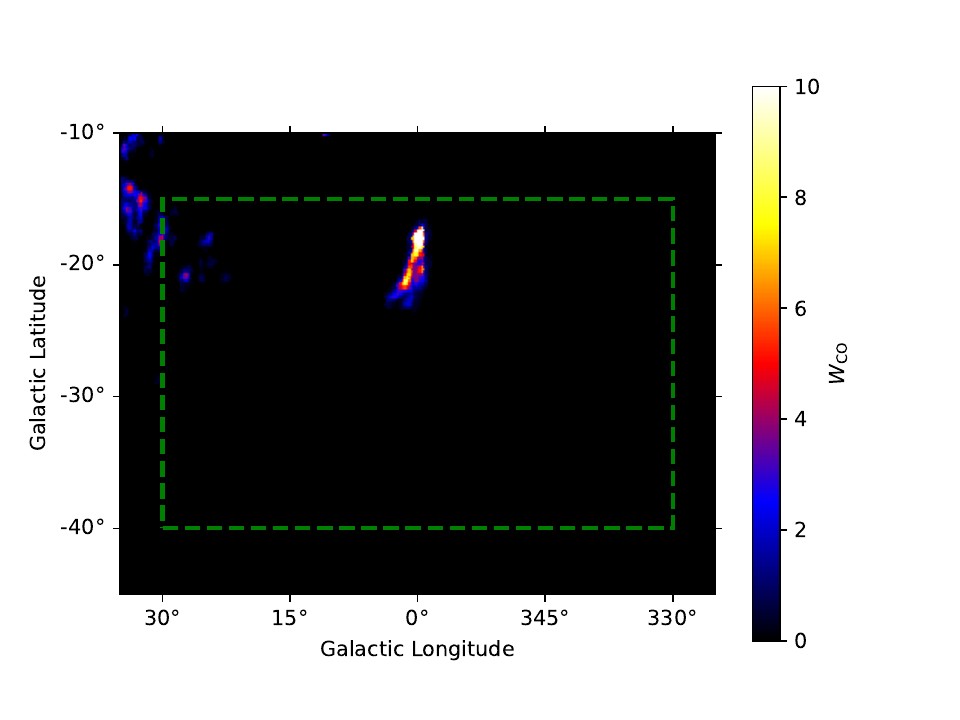}
\put(15,65){(c)}
\end{overpic}
\end{minipage}
\begin{minipage}{0.5\textwidth}
\centering
\begin{overpic}[width=\textwidth]{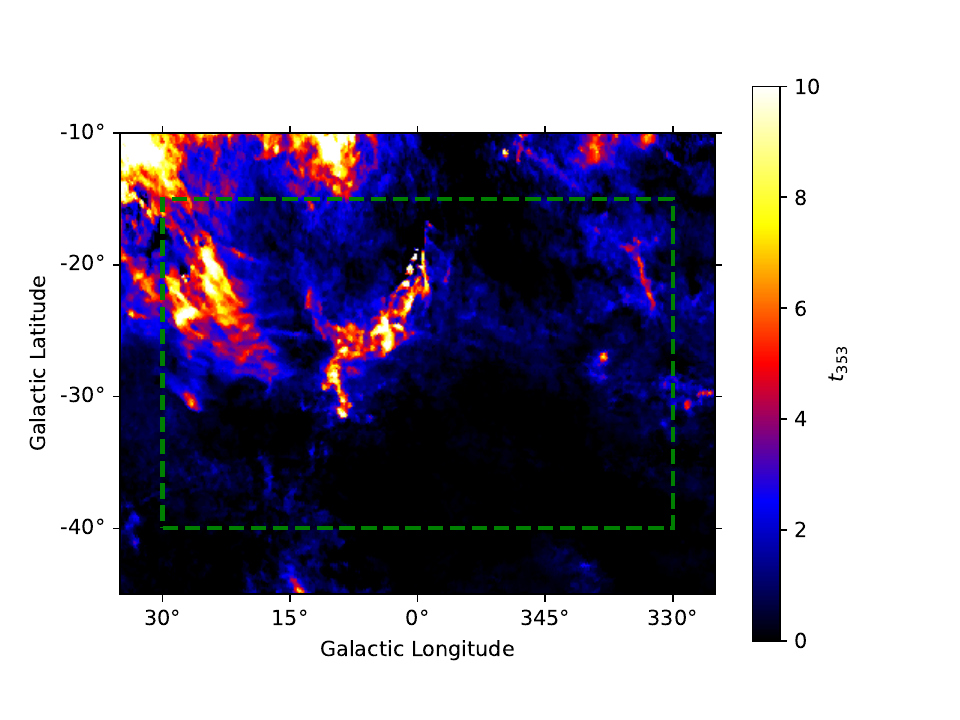}
\put(15,65){(d)}
\end{overpic}
\end{minipage}
\end{tabular}
\caption{Same as Figure~1 but for the R CrA region instead of the MBM/Pegasus region.
{Alt text: Four template maps for the R CrA region.}}\label{..}
\end{figure}

\begin{figure}[htbp]
\begin{tabular}{cc}
\begin{minipage}{0.5\textwidth}
\centering
\begin{overpic}[width=\textwidth]{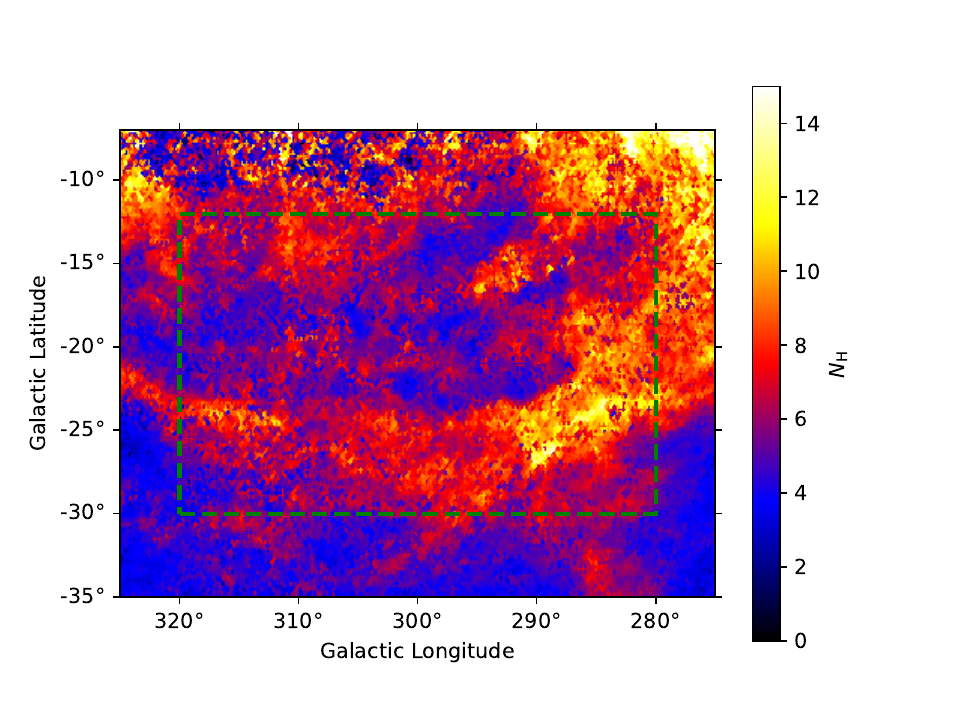}
\put(15,65){(a)}
\end{overpic}
\end{minipage}
\begin{minipage}{0.5\textwidth}
\centering
\begin{overpic}[width=\textwidth]{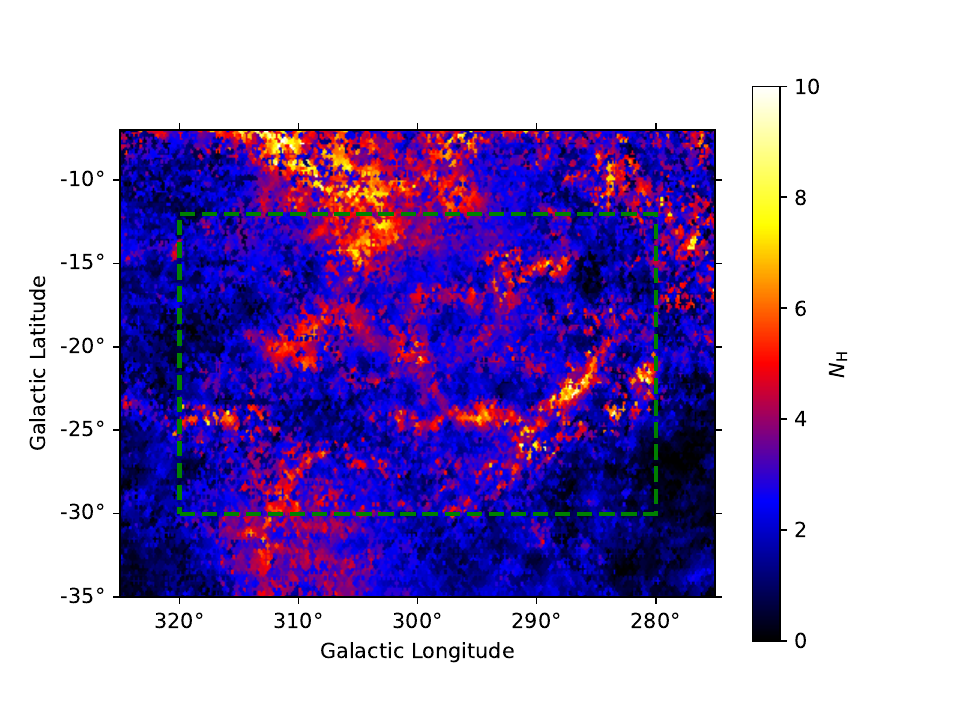}
\put(15,65){(b)}
\end{overpic}
\end{minipage} \\
\\
\begin{minipage}{0.5\textwidth}
\centering
\begin{overpic}[width=\textwidth]{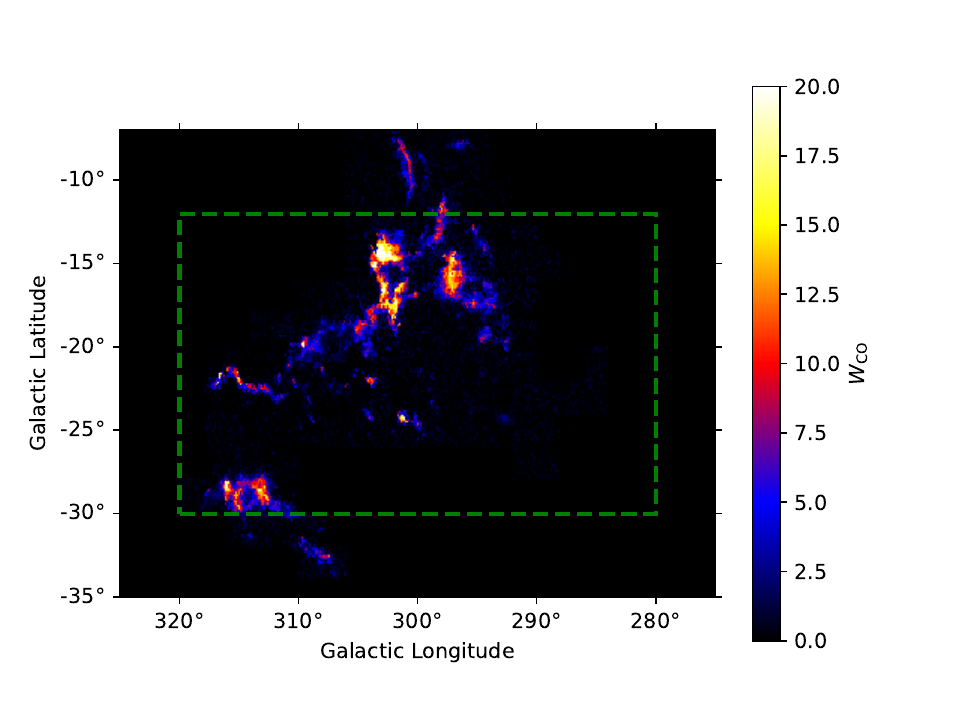}
\put(15,65){(c)}
\end{overpic}
\end{minipage}
\begin{minipage}{0.5\textwidth}
\centering
\begin{overpic}[width=\textwidth]{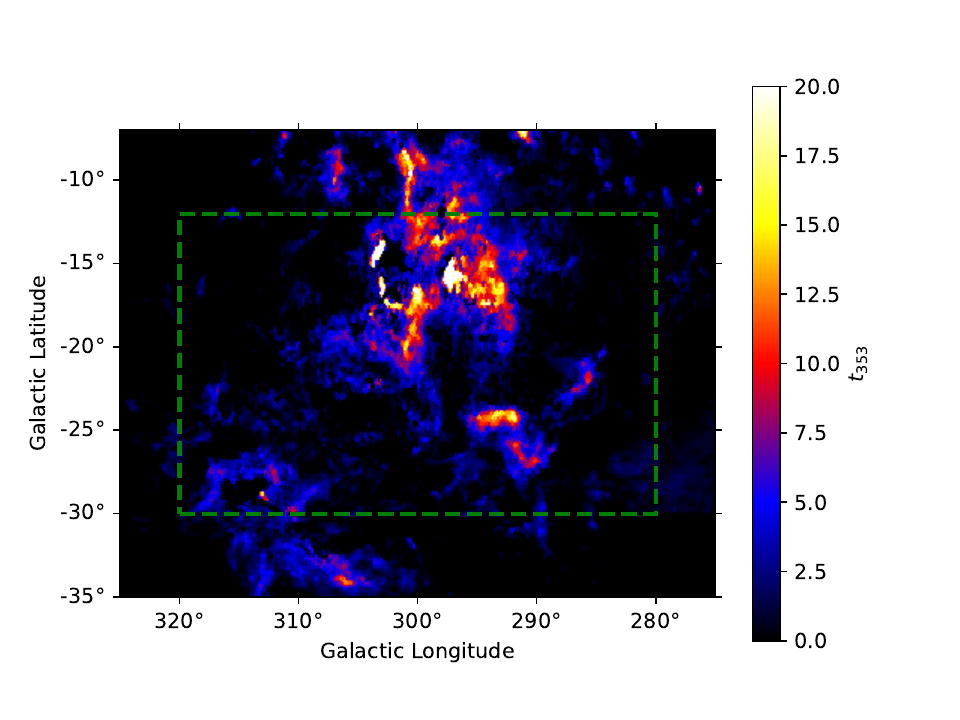}
\put(15,65){(d)}
\end{overpic}
\end{minipage}
\end{tabular}
\caption{Same as Figure~1 but for the Chamaeleon region instead of the MBM/Pegasus region.
{Alt text: Four template maps for the Chamaeleon region.}}\label{..}
\end{figure}

\begin{figure}[htbp]
\begin{tabular}{cc}
\begin{minipage}{0.5\textwidth}
\centering
\begin{overpic}[width=\textwidth]{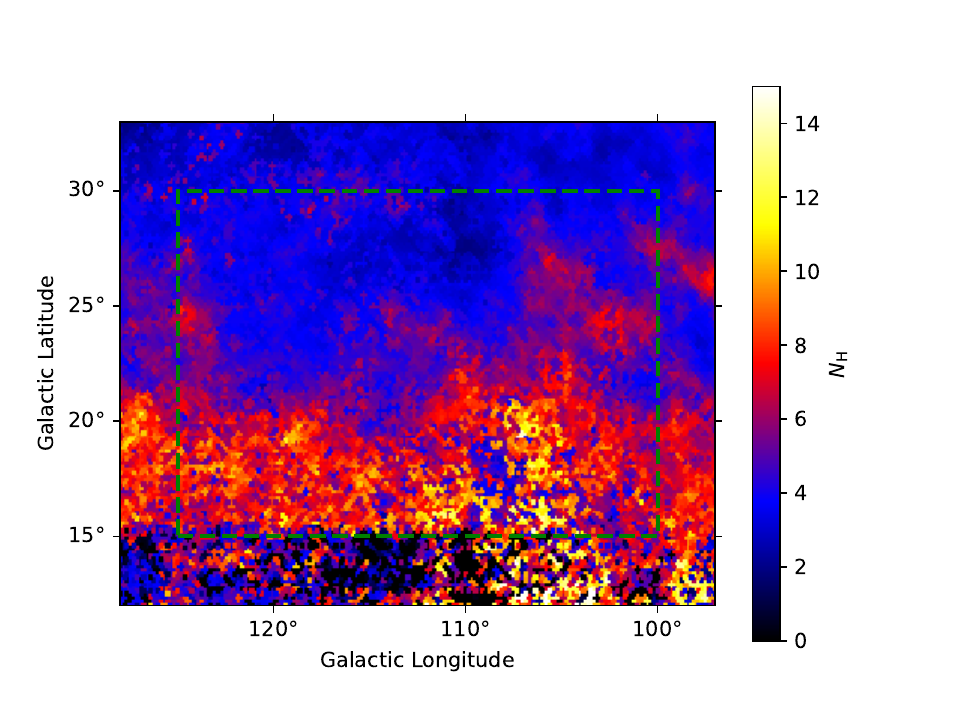}
\put(15,65){(a)}
\end{overpic}
\end{minipage}
\begin{minipage}{0.5\textwidth}
\centering
\begin{overpic}[width=\textwidth]{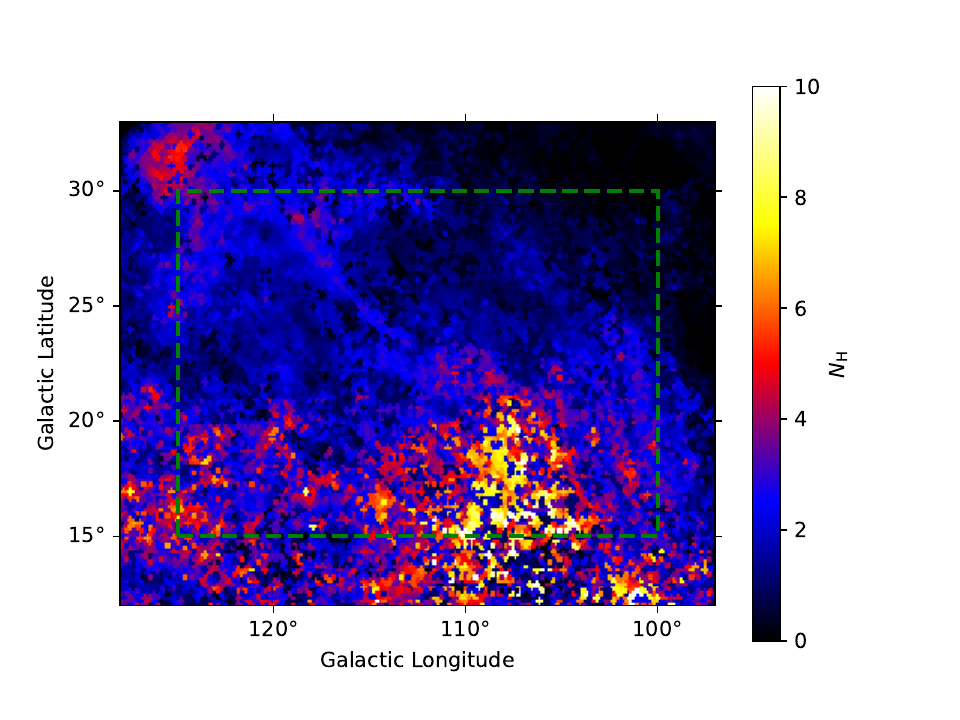}
\put(15,65){(b)}
\end{overpic}
\end{minipage} \\
\\
\begin{minipage}{0.5\textwidth}
\centering
\begin{overpic}[width=\textwidth]{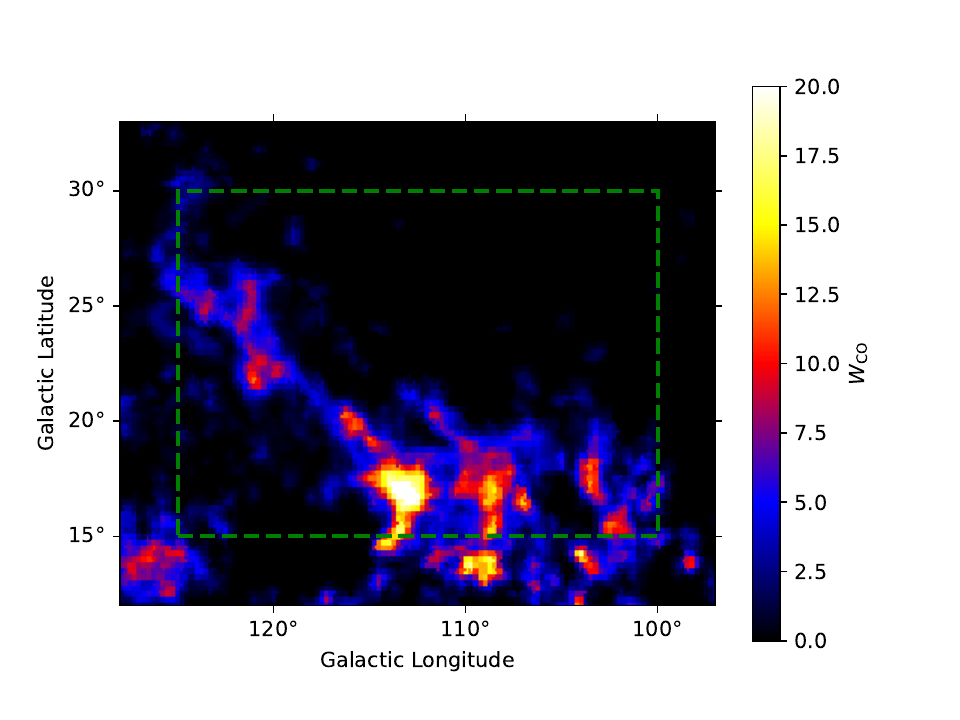}
\put(15,65){(c)}
\end{overpic}
\end{minipage}
\begin{minipage}{0.5\textwidth}
\centering
\begin{overpic}[width=\textwidth]{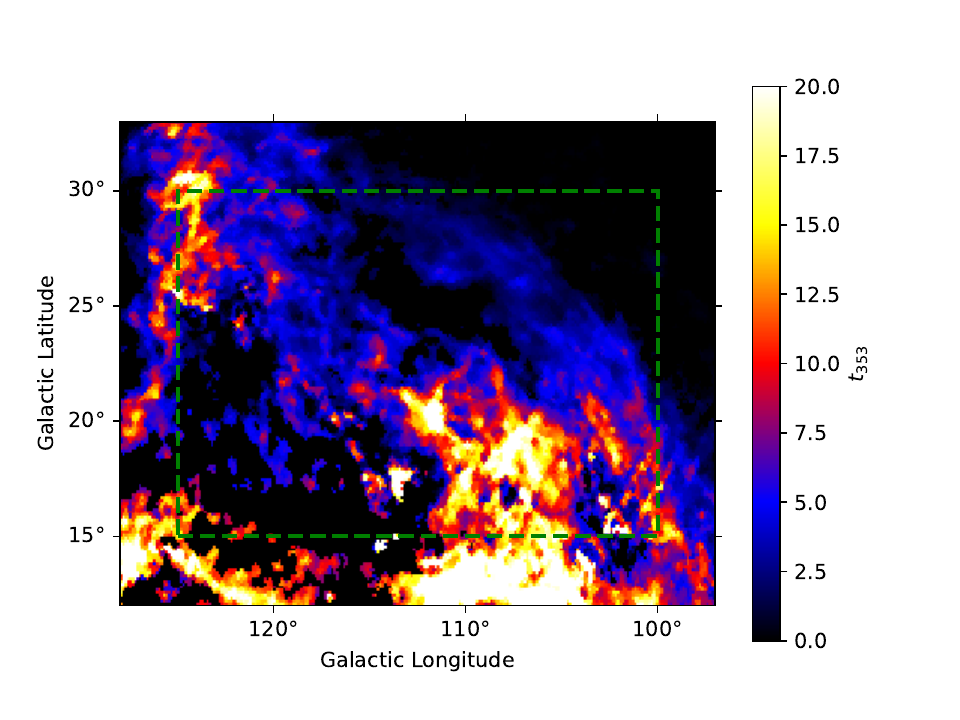}
\put(15,65){(d)}
\end{overpic}
\end{minipage}
\end{tabular}
\caption{Same as Figure~1 but for the Cep/Pol region instead of the MBM/Pegasus region.
{Alt text: Four template maps for the Cep/Pol region.}}\label{..}
\end{figure}

\begin{figure}[htbp]
\begin{tabular}{cc}
\begin{minipage}{0.5\textwidth}
\centering
\begin{overpic}[width=\textwidth]{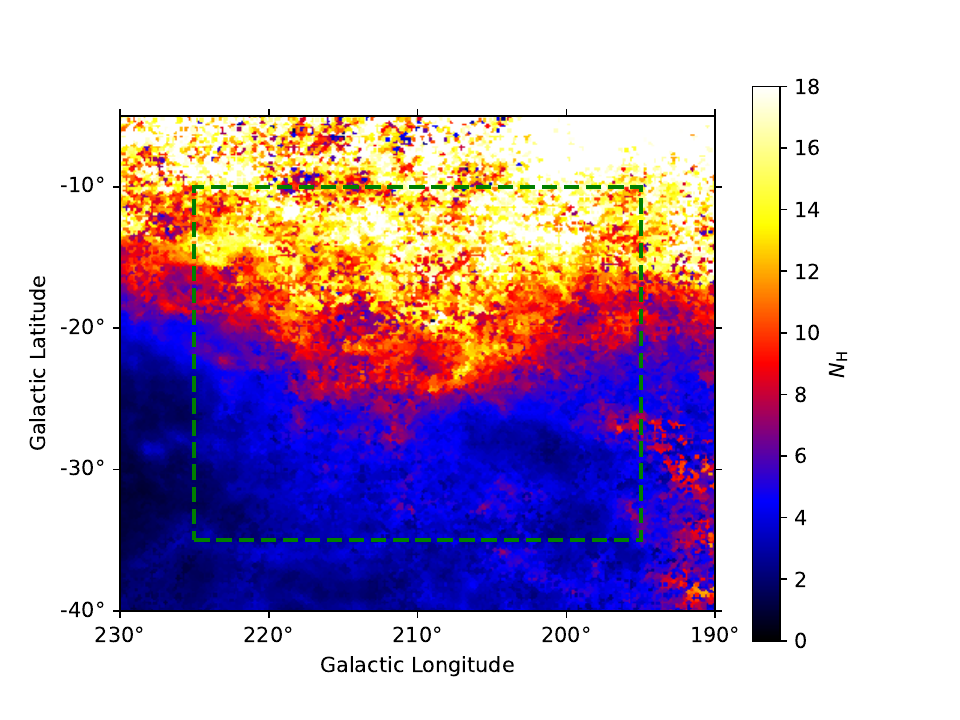}
\put(15,65){(a)}
\end{overpic}
\end{minipage}
\begin{minipage}{0.5\textwidth}
\centering
\begin{overpic}[width=\textwidth]{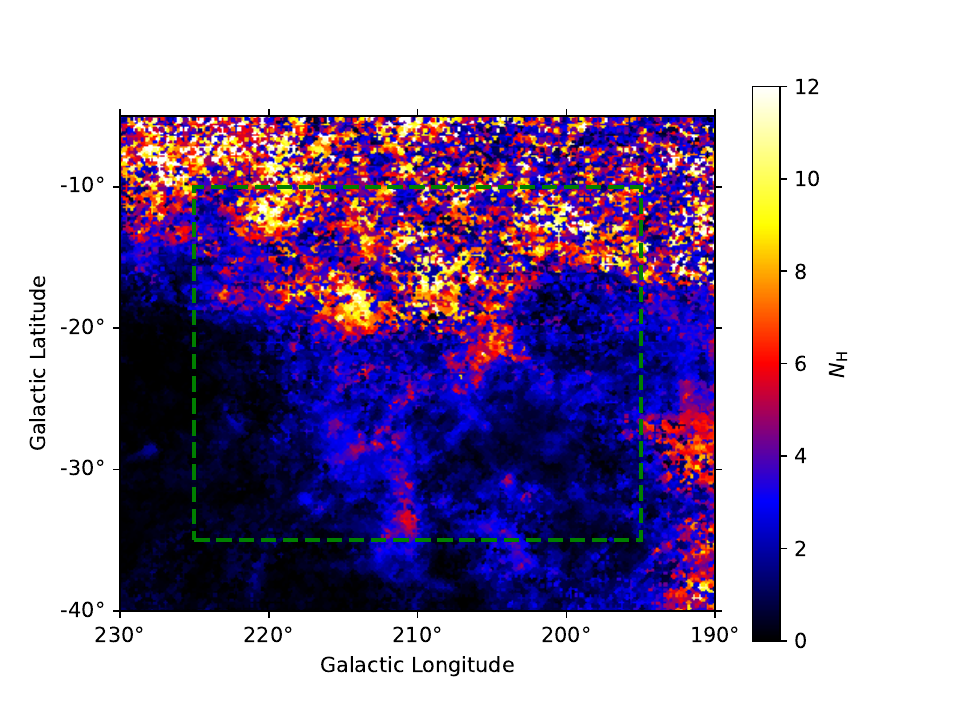}
\put(15,65){(b)}
\end{overpic}
\end{minipage} \\
\\
\begin{minipage}{0.5\textwidth}
\centering
\begin{overpic}[width=\textwidth]{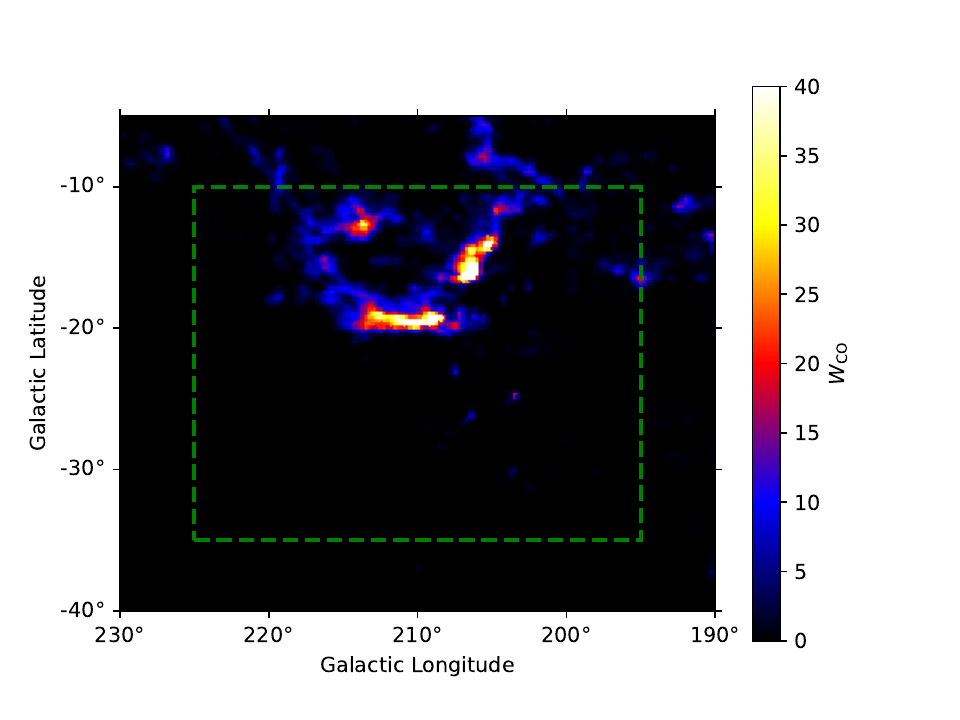}
\put(15,65){(c)}
\end{overpic}
\end{minipage}
\begin{minipage}{0.5\textwidth}
\centering
\begin{overpic}[width=\textwidth]{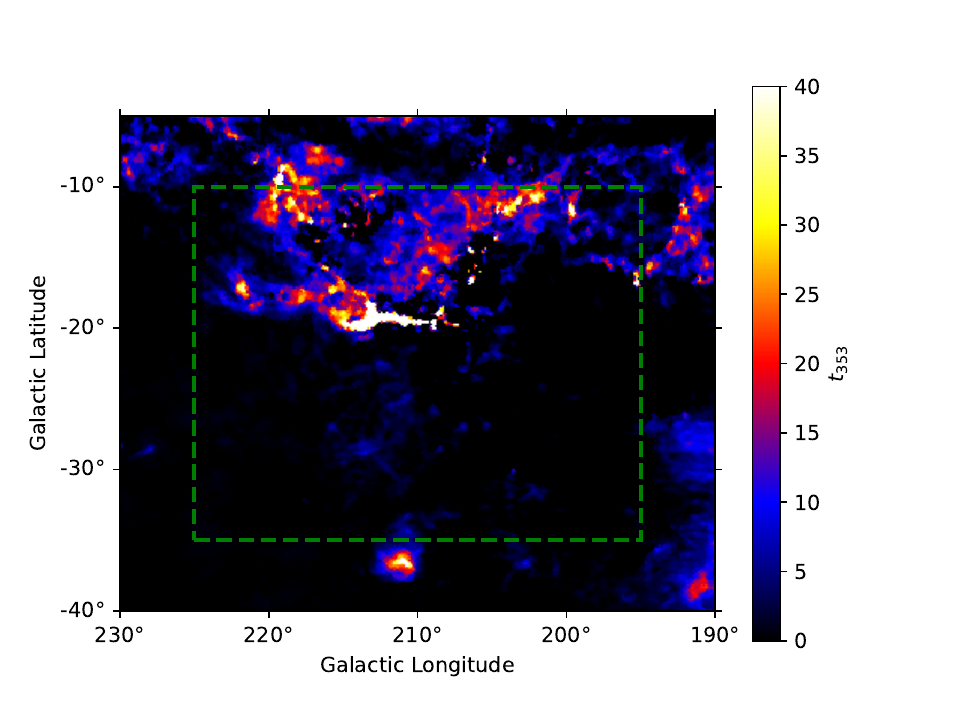}
\put(15,65){(d)}
\end{overpic}
\end{minipage}
\end{tabular}
\caption{Same as Figure~1 but for the Orion region instead of the MBM/Pegasus region.
{Alt text: Four template maps for the Orion region.}}\label{..}
\end{figure}

\subsubsection{Residual Gas Template}
Even though narrow and broad {\HI} are separated in templates, one 
cannot reproduce $\gamma$-ray data due to
gas not being traced by {\HI} 21~cm line survey.
Accordingly we prepared a "residual gas" template from Planck dust model maps
under the assumption that dust emission ($D_\mathrm{em}$) is represented by a linear combination of
narrow {\HI} ($W_\mathrm{HI, narrow}$),
broad {\HI} ($W_\mathrm{HI, broad}$),
offset, 
$\WCO$ and residual gas. 
Based on previous work by \citet{Mizuno2022}, we
developed an iterative procedure as described below. Since we observed possible contamination from infrared sources 
in the Orion region,
we removed them as described in Appendix~4 in advance.
\begin{inparaenum}
  \item Determine coefficients for narrow {\HI} and broad {\HI} and offset to represent atomic hydrogen distribution traced by $D_\mathrm{em}$.
To avoid contamination from CO-bright {\Htwo} and residual gas, we select areas of low $\WCO$ 
($\WCO \le 0.1~\mathrm{K~km~s^{-1}}$),
high dust temperature with $T_\mathrm{d} \ge T_\mathrm{d, th}$, and rich in broad {\HI}
with a fraction of broad {\HI} in $\WHI$ $f_\mathrm{broad\mathchar`-HI} \ge f_\mathrm{broad\mathchar`-HI, th}$. 
Throughout this step~1, we apply this selection iteratively.
  \begin{inparaenum}
    \item Fit $D_\mathrm{em}$ with $W_\mathrm{HI, narrow}$, $W_\mathrm{HI, broad}$, and offset. Narrow {\HI} is included in $W_\mathrm{HI, narrow}$.
If coefficients change less than 1\% from the previous fit, quit the loop and move to step~2. Otherwise move to step 1b.
    \item (Update of the fit region) Use three parameters (two coefficients and an offset) of the previous fit, 
calculate residuals of all pixels that pass the selection described above and select pixels within 
the peak of the residual distribution $\mathrm{\pm 3 \times rms}$ (root mean square).
The residual distribution is mostly symmetric with some tail and this step is to remove those 
outliers, presumably contamination from the residual gas.
  \end{inparaenum}
  \item Use obtained coefficients and offset, and calculate residuals of $D_\mathrm{em}$ in high $\WCO$ areas
($\WCO \ge 1~\mathrm{K~km~s^{-1}}$).
Fit the residual with a linear equation of $\WCO$. We allow an offset free to vary to approximate $D_\mathrm{em}$ from residual gas on average.
  \item Use three coefficients (steps~1 and 2) and an offset (step~1) and construct 
the residual $D_\mathrm{em}$ map (denoted as $D_\mathrm{em, res}$)
in the same unit of $D_\mathrm{em}$ as a residual gas template.
\end{inparaenum}

The obtained template is noisy and fluctuates considerably around 0. We applied the median-filter with $\sigma = 10^{'}$ 
and used the filtered template
in $\gamma$-ray data analysis.
We expect the $D_\mathrm{em, res}$ map 
to trace CO-dark {\Htwo}, which is quite uncertain. To cope with this uncertainty we prepared 
three templates.
Specifically, we constructed templates
using $\tau_{353}$ and radiance maps of Planck Public Data Release 2 (version R2.00)\footnote{
\url{https://irsa.ipac.caltech.edu/data/Planck/release_2/all-sky-maps/}
}
and a revised $\tau_{353}$ map by \citet{Casandjian2022}. 
They carefully examined the zero-level of dust emission and found that revised $\tau_{353}$ is proportional to
$\WHI$ in low-density areas in a broader range of $\WHI$.
We tested these three templates against the $\gamma$-ray data
(see Section~3.1).

We remind that we assumed the $D_\mathrm{em, res}$ map 
is proportional to the residual gas distribution, which is not granted.
In particular, $\tau_{353}$ to gas column density ratio is known to rise steeply with increasing gas density
in the molecular gas phase because of the dust grain evolution (e.g., \citet{Remy2017}).
In order to be very precise one need to prepare an additional template to compensate the non-linearity,
but we defer such a sophisticated analysis to future studies.

We also note that CO observations of the R CrA region were done with \timeform{0.25D} sampling \citep{Dame2001} which is not fine enough
to map the dense core of CO at $(l, b) \sim (\timeform{0D}, \timeform{-18D})$.
This produces the artificial structure in the residual gas template 
around the CO core, and makes the $\gamma$-ray fitting unstable.
To cope with this issue, we masked a circle of a radius of \timeform{0.8D}
centered at $(l, b) = (\timeform{0D}, \timeform{-18D})$ in the fitting,
and filled this circle with 0 in the $D_\mathrm{em, res}$ map. 
Since the angular resolution of Fermi-LAT is at the \timeform{0.1D} level even at 10~GeV and the ISM gas is expected to be dominated
by {\Htwo} at the CO core,
the $D_\mathrm{em, res}$ map 
is good enough for $\gamma$-ray data analysis.

\subsubsection{Other Templates and Point Sources}
In addition to the gas-related components described above,
we also need to consider $\gamma$-ray from IC scattering, the contribution of point sources,
extragalactic diffuse emission, and instrumental residual background.
To model the IC emission, we used GALPROP\footnote{
\url{http://galprop.stanford.edu}}
(e.g., \cite{Galprop1,Galprop2}).
GALPROP is a numerical code that solves the CR transport equation within the Milky Way and predicts
the $\gamma$-ray emission produced via the interactions of CRs with the ISM gas
and the ISRF (IC scattering). 
It calculates the IC emission from the distribution
of propagated electrons and the modeled ISRF.
Specifically, we utilized the work by \citet{Porter2017}
to construct the IC model template. 
They employed 3D spatial models for the CR source distribution and the ISRF.
They considered different spatial distributions for the CR sources
(differentiated by the ratio of the smooth-disk component to the spiral-arm component) and three ISRF models.
The three CR source distributions are labeled SA0, SA50, and SA100;
SA0 corresponds to a 100\% (2D) disk, and SA100 corresponds to a 100\% spiral-arm contribution.
For the ISRF, they used a standard 2D model and two 3D models.
\citet{Mizuno2022} tested all nine IC models
against the $\gamma$-ray data for the MBM/Pegasus region and found that
the difference among the three ISRF models in log-likelihood in $\gamma$-ray data analysis is minor.
We therefore will use a standard ISRF (labeled Std) with SA0, SA50, and SA100 to model IC emission
and test them against $\gamma$-ray data.

To model the individual $\gamma$-ray point sources, we referred to the fourth Fermi-LAT catalog (4FGL-DR3)
described by \citet{Fermi4FGLDR3}, 
which is based on the first 12 yr of the mission's science phase and includes more than
6000 sources detected at a significance of ${\ge4}~\sigma$. For our analysis, we included all sources detected at a significance level 
${\ge}5~\sigma$ in our ROIs with spectral parameters free to vary (see Section~2.2.4 and 2.3 for details).
We also included bright sources (${\ge}20\sigma$) just outside it (within $5^{\circ}$),
with the parameters fixed at those in the 4FGL, to consider their possible contamination. We added an isotropic component
representing extragalactic diffuse emission and the residual instrumental background from misclassified CR interaction
in the LAT detector. We adopted the isotropic template provided by the Fermi Science Support Center\footnote{
\url{https://fermi.gsfc.nasa.gov/ssc/data/access/lat/BackgroundModels.html}
} for the ULTRACLEAN class.

\subsubsection{Model to Reproduce Gamma-Ray Emission}
Then, the $\gamma$-ray intensities $I_{\gamma}(l, b, E)~{\rm(ph~s^{-1}~cm^{-2}~sr^{-1}~MeV^{-1})}$
can be modeled as
\begin{equation}
\begin{split}
I_{\gamma}(l, b, E) & = 
\left[\sum_{i} \CHIi(E) \cdot \NHIi(l, b)  + 
C_\mathrm{CO}(E) \cdot 2X_\mathrm{CO}^{0} \cdot \WCO(l, b) +
C_\mathrm{dust}(E) \cdot X_\mathrm{dust}^{0} \cdot D_\mathrm{em, res}(l, b) \right] \cdot q_{\gamma}(E) \\
& + C_\mathrm{IC}(E) \cdot I_\mathrm{IC}(l, b, E) +
C_\mathrm{iso}(E) \cdot I_\mathrm{iso}(E) + \sum_{j} C_{\mathrm{PS}, j}(E) \cdot {\rm PS}_{j}(l, b, E)~~,
\end{split}
\end{equation}
where the $\NHIi$ is the atomic gas column density (${\rm cm^{-2}}$) model maps,
$\WCO$ is the integrated $^{12}$CO (J=1--0) intensity map ($\mathrm{K~km~s^{-1}}$), 
$D_\mathrm{em, res}$ is the dust emission residual map,
and  $q_{\gamma}$ (${\rm ph~s^{-1}~sr^{-1}~MeV^{-1}}$) is the model of the 
differential $\gamma$-ray yield 
($\gamma$-ray emissivity) per H atom.
The quantities $I_{\rm IC}$ and $I_{\rm iso}$ are the IC model and the isotropic template intensities
(${\rm ph~s^{-1}~cm^{-2}~sr^{-1}~MeV^{-1}}$), respectively, and
${\rm PS}_{j}$ represents point-source models in the 4FGL.
The subscript $i$ allows using $\NHI$ maps from separate {\HI} line profiles.
Specifically, $N_\mathrm{HI,0}$, $N_\mathrm{HI,1}$, and $N_\mathrm{HI,2}$ correspond to non-local {\HI}, narrow {\HI}, and broad {\HI}, respectively.
We adopted the $\gamma$-ray emissivity model by \citet{Mizuno2022}
that is based on directly-measured CR spectra \citep{Maurin2014}, 
hadronic interaction models (AAfrag; \cite{AAfrag}), and an electron/positron bremsstrahlung model \citep{Orlando2018}.

To investigate the energy spectrum of each component of the model,
we devided data into eight energy bands (Section~2.3).
To accommodate the uncertainties, in either the emissivity model or the gas templates,
we included normalization factors
[$\CHIi$, $C_\mathrm{CO}$, and $C_\mathrm{dust}$ in Equation~(1)] as free parameters in each energy band.
The quantities $X_\mathrm{CO}^{0}$ and $X_\mathrm{dust}^{0}$ are constant scale factors that make the fitting coefficients
($C_\mathrm{CO}$ and $C_\mathrm{dust}$) close to 1. Specifically, we used
$1 \times 10^{20}~\mathrm{cm^{-2}~(K~km~s^{-1})^{-1}}$ and
$1.82 \times 10^{26}~\mathrm{cm^{-2}}$ for
$X_\mathrm{CO}^{0}$ and $X_\mathrm{dust}^{0}$ (for $\tau_\mathrm{353}$), respectively.
While $\CHIi$ will be 1 if $\NHIi$ represents the true gas column density
and the adopted $\gamma$-ray emissivity model agrees with the true $\gamma$-ray yield,
$C_\mathrm{CO}$ and $C_\mathrm{dust}$ provide the
CO-to-{\Htwo} conversion factor ($X_\mathrm{CO} \equiv \NHtwo / \WCO$, where $\NHtwo$ is the
molecular gas column density)
and the dust-to-gas conversion factor ($X_\mathrm{dust} \equiv \NH / D_\mathrm{em, res}$, where $\NH$ is the gas column density), 
respectively.
$X_\mathrm{CO}$ will be $X_\mathrm{CO}^{0}$ if $\CHIi$ for optically thin {\HI} 
(i.e., $\CHItwo$)
and $C_\mathrm{CO}$ are equal 
[$X_\mathrm{CO} = X_\mathrm{CO}^{0} \times \left( C_\mathrm{CO}/\CHItwo \right)$].
$X_\mathrm{dust}$ will be $X_\mathrm{dust}^{0}$ if 
$\CHItwo$
and $C_\mathrm{dust}$ are equal
[$X_\mathrm{dust} = X_\mathrm{dust}^{0} \times \left( C_\mathrm{dust}/\CHItwo \right)$].
The IC emission and isotropic models are also uncertain,
and we have included other
normalization factors [$C_\mathrm{IC}$ and $C_\mathrm{iso}$ in Equation~(1)] as free parameters.
For the point sources, 
we also have included normalization factors ($C_{\mathrm{PS}, j}$) to capture a possible change of the spectrum from 4FGL.

\clearpage
\subsection{Model-fitting Procedure}
We divided the $\gamma$-ray data into eight energy bands, extending from 0.1 to 25.6~GeV and stored them in HEALPix maps
of order 8 (by using the {\tt gtbin} command).
We used only the HEALPix pixels inside each ROI.
We used a pixel size two times larger than those of the gas maps 
(mean distance of adjacent pixels is \timeform{13.7'} instead of \timeform{6.9'}) to 
perform the $\gamma$-ray fitting reasonably fast,
while keeping the $\gamma$-ray map fine enough to evaluate the ISM gas distributions.
We used energy bins equally spaced logarithmically
(e.g., 0.1--0.2, 0.2--0.4, and 0.4-0.8~GeV).
To accurately evaluate the model spectral shape,
the data were subdivided into three grids within each energy band.
Then, in each energy band, 
we prepared exposure and source maps 
with finer grids taken into account (by using the {\tt gtexpcube2} and {\tt gtsrcmaps} commands),
and fitted Equation~(1) to the $\gamma$-ray data
using the binned maximum-likelihood method with Poisson statistics taken into account
(by importing the {\tt BinnedAnalysis} module in python).
The angular resolution of $\gamma$-ray data is taken into account in this step.
In each energy band, we modeled $\CHIi$, $C_\mathrm{CO}$, $C_\mathrm{dust}$,
$C_\mathrm{IC}$, $C_\mathrm{iso}$, and $C_{\mathrm{PS}, j}$ 
 as energy-independent normalization factors.

When modeling the point sources, we sorted them by significance in the 4FGL
and divided them into several groups (each group has 10 sources at the maximum).
We then iteratively fitted them in order of decreasing significance.
First, we fitted the normalizations of the 10 most significant sources;
then, we fitted the normalizations of the second group with parameters of the first group
fixed at the values already determined.
In this way, we walked down to the sources detected at more than 5$\sigma$ in the 4FGL.
For each step, the parameters of the diffuse-emission model
were always left free to vary.
After reaching the least significant sources, we returned back
to the brightest ones
and let them and the diffuse-emission models be free to vary.
In contrast, the parameters of the other sources were fixed
at the values already determined.
We repeated this process until the increments of the log-likelihoods, $\ln{L}$\footnote{
$L$ is conventionally calculated as $\ln{L}=\sum_{i}n_{i} \ln(\theta_{i})-\sum_{i}\theta_{i}$,
where $n_{i}$ and $\theta_{i}$ are the data and the model-predicted counts in each pixel (for each energy grid)
denoted by the subscript, respectively (see, e.g., \cite{Mattox1996}).
}, 
were less than 0.1 over one loop in each energy band.

\clearpage
\section{Data Analysis}
\subsection{MBM 53-55 clouds and Pegasus loop}
MBM~53, 54, 55 clouds and Pegasus loop
are nearby (100--150~pc) high-latitude clouds \citep{Welty1989, Yamamoto2006},
having ${\sim}1000~M_{\odot}$ molecular gas traced by $\WCO$ \citep{Dame1987}.
In \citet{Mizuno2022}, we studied the $\gamma$-ray data in the region of
$60\arcdeg \le l \le 120\arcdeg$ and
$-60\arcdeg \le b \le -28\arcdeg$, which encompasses those ISM structures.
Here we re-analyzed $\gamma$-ray data (of 15 yr instead of 12 yr) to compare CR and ISM properties among five regions
in a consistent way.

Dust is a known tracer of the total gas column density, and we employed dust maps of the \textit{Planck} public Data Release 2
(version R2.00)\footnote{\url{https://irsa.ipac.caltech.edu/data/Planck/release_2/all-sky-maps/}}, which is 
less affected by infrared sources
than Data Release 1. Following \citet{Mizuno2022}, we masked a variable star 
RAFGL 3068 in dust maps (temperature, $\tau_{353}$, and radiance). Then we constructed
a residual gas template map as described in Section~2.2.2. We used the original $\tau_{353}$ and radiance maps, 
and also used the revised $\tau_{353}$ map
by \citet{Casandjian2022}. 
We adopted 
$T_\mathrm{d, th}=19~\mathrm{K}$ and
$f_\mathrm{broad\mathchar`-HI, th}=0.8$
to construct a residual gas template from
original dust maps.
We adopted $T_\mathrm{d, th}=19.5~\mathrm{K}$ for the revised $\tau_{353}$ map since \citet{Casandjian2022}
reported higher dust temperature by ${\sim}$0.5~K. The same values of $T_\mathrm{d, th}$ and $f_\mathrm{broad\mathchar`-HI, th}$
are used for other regions unless otherwise stated.
To examine the effect of these selection criteria, we also constructed templates with
$T_\mathrm{d, th}$ and $f_\mathrm{broad\mathchar`-HI, th}$
changed by ${\pm}$0.5~K and ${\pm}$0.1, respectively, and tested them against $\gamma$-ray data.
IC model of Std-SA0 was used in this test.
We confirmed that the change of the emissivity of broad {\HI} is ${\sim}$3\% for the same dust map and will use the 
selection criteria described above hereafter if otherwise stated.
We found that the $\tau_{353}$-based template gives a better fit
than the radiance-based one ($\Delta \ln{L} \sim 40$ in ${\ge}400~\mathrm{MeV}$),
and a template based on the revised $\tau_{353}$ map is slightly better
than the original $\tau_\mathrm{353}$ map does ($\Delta \ln{L} \sim 5$).
Therefore we decided to use the revised $\tau_\mathrm{353}$ to construct the residual gas template (Figure~1).
We tested different IC templates (labeled as Std-SA50 and Std0-SA100) and confirmed that Std-SA0 
best fits $\gamma$-ray data.

The obtained
{\HI} emissivity spectra are summarized in Figure~6 (left). We found that narrow {\HI}
gives higher emissivity than broad {\HI} by ${\sim}$30\%. 
This higher emissivity can be naturally explained by that narrow {\HI} is optically thick.
We also found that low energy spectra are not stable due to the large PSF of $\gamma$-ray data;
broad {\HI} gives larger emissivity
than narrow {\HI} does at the lowest energy band. 
To mitigate this, we constructed a single $\NHI$ map for local {\HI}.
We evaluated a correction factor for narrow {\HI} by
looking at the fit-coefficient ratio about broad {\HI} in ${\ge}$400~MeV; 
the average of $\CHIone$ and $\CHItwo$ are $1.392 \pm 0.031$ and $1.092 \pm 0.033$, respectively, giving the correction factor of 1.27. 
Assuming that CR intensity is uniform over the scale studied here and broad {\HI} is optically thin,
we multiplied this factor to the $\NHI$ map of the narrow {\HI} and added it to the $\NHI$ map of broad {\HI}.
The final fit results, by using the single local $\NHI$ map instead of narrow {\HI} and broad {\HI} maps, 
are summarized in Table~1, Figure~6 (right panel), and Figure~7.
The local {\HI} emissivity spectrum agrees well with the adopted model based on directly-measured CR spectra.
The {\HI} emissivity above 100~MeV and 400~MeV are $(1.499 \pm 0.028) \times 10^{-26}$ and $(0.511 \pm 0.008) \times 10^{-26}$ in units of 
${\rm ph~s^{-1}~sr^{-1}}$ per H atom, respectively.
$X_\mathrm{CO}$ is evaluated, using averages of the fit coefficients above 400~MeV for the $\WCO$ map and corrected local $\NHI$ map,
to be $(0.559 \pm 0.038) \times 10^{20}~\mathrm{cm^{-2}~(K~km~s^{-1})^{-1}}.$

The correction of narrow {\HI} using a single scaling factor described above is a simplified procedure, 
and one may apply a spin-temperature correction to take account of the brightness temperature of each {\HI} line along the line of sight.
Following the procedure by \citet{Mizuno2022}, we tested
a $T_\mathrm{s}$ correction on narrow {\HI}. We found that the impact is minor, at the 1\% level. 
Effects on other regions are also small (at the {$\le$}5~\% level); see Appendix~5 for details. 
To take advantage of simplicity, we will use single scaling factors for MBM/Pegasus and other regions.

\begin{figure}[htbp]
\begin{tabular}{cc}
\begin{minipage}{0.5\textwidth}
\centering
\begin{overpic}[width=\textwidth]{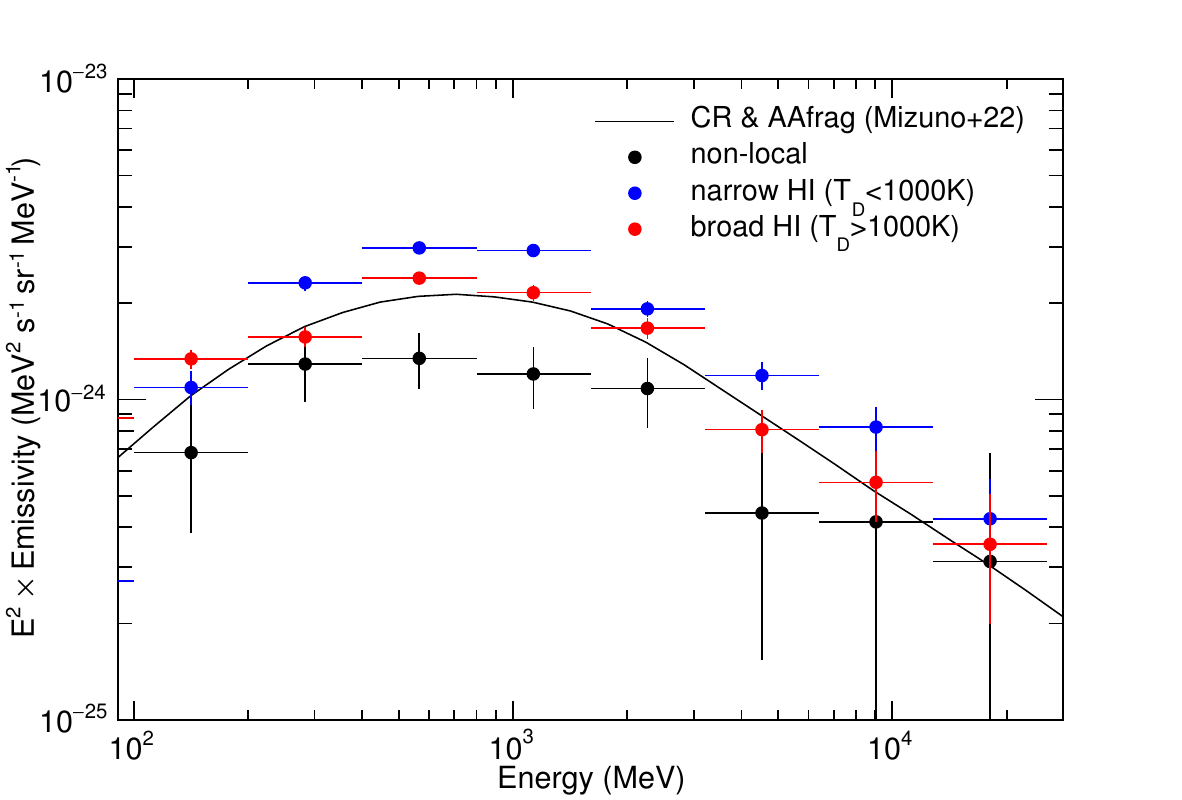}
\put(15,63){(a)}
\end{overpic}
\end{minipage}
\begin{minipage}{0.5\textwidth}
\centering
\begin{overpic}[width=\textwidth]{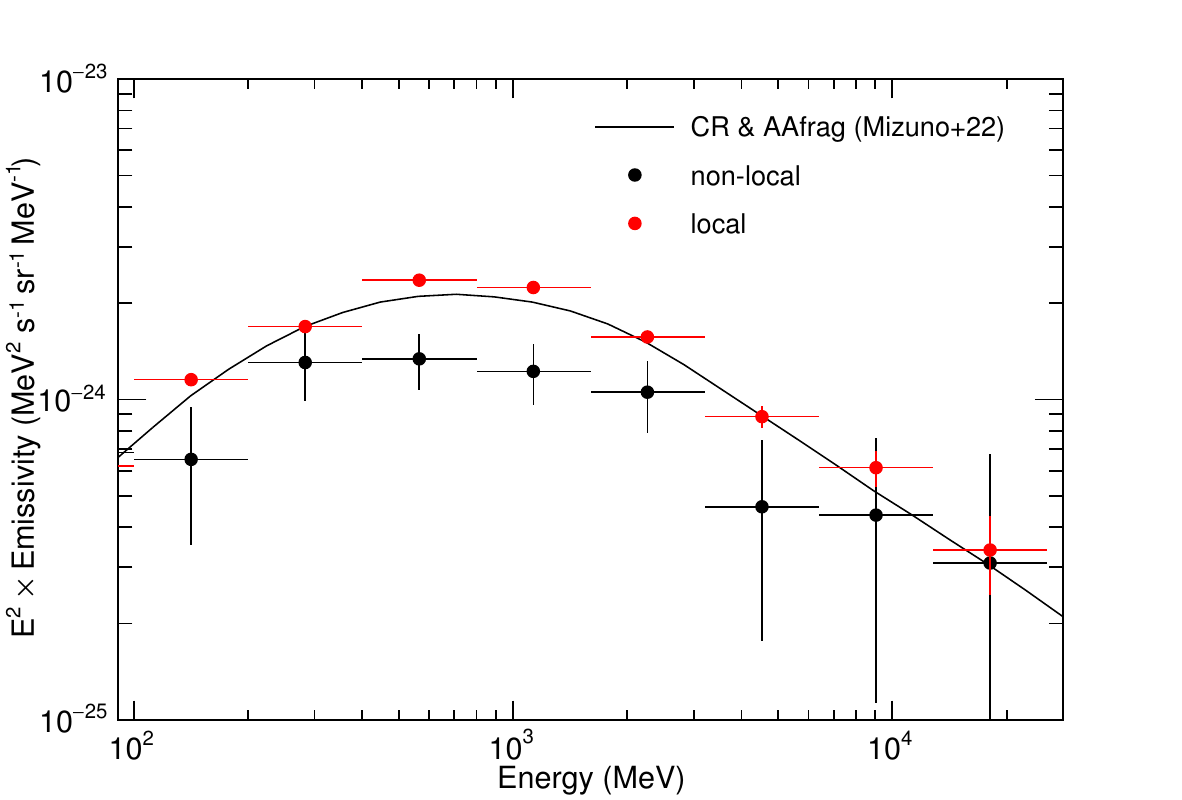}
\put(15,63){(b)}
\end{overpic}
\end{minipage}
\end{tabular}
\caption{(a) {\HI} emissivity spectra of three {\HI} phases, and (b) those of local and non-local {\HI} gas
for the MBM/Pegasus region.
{Alt text: Two emissivity spectra for the MBM/Pegasus region.}
}\label{..}
\end{figure}

\begin{figure}[htbp]
\begin{tabular}{cc}
\begin{minipage}{0.5\textwidth}
\centering
\begin{overpic}[width=\textwidth]{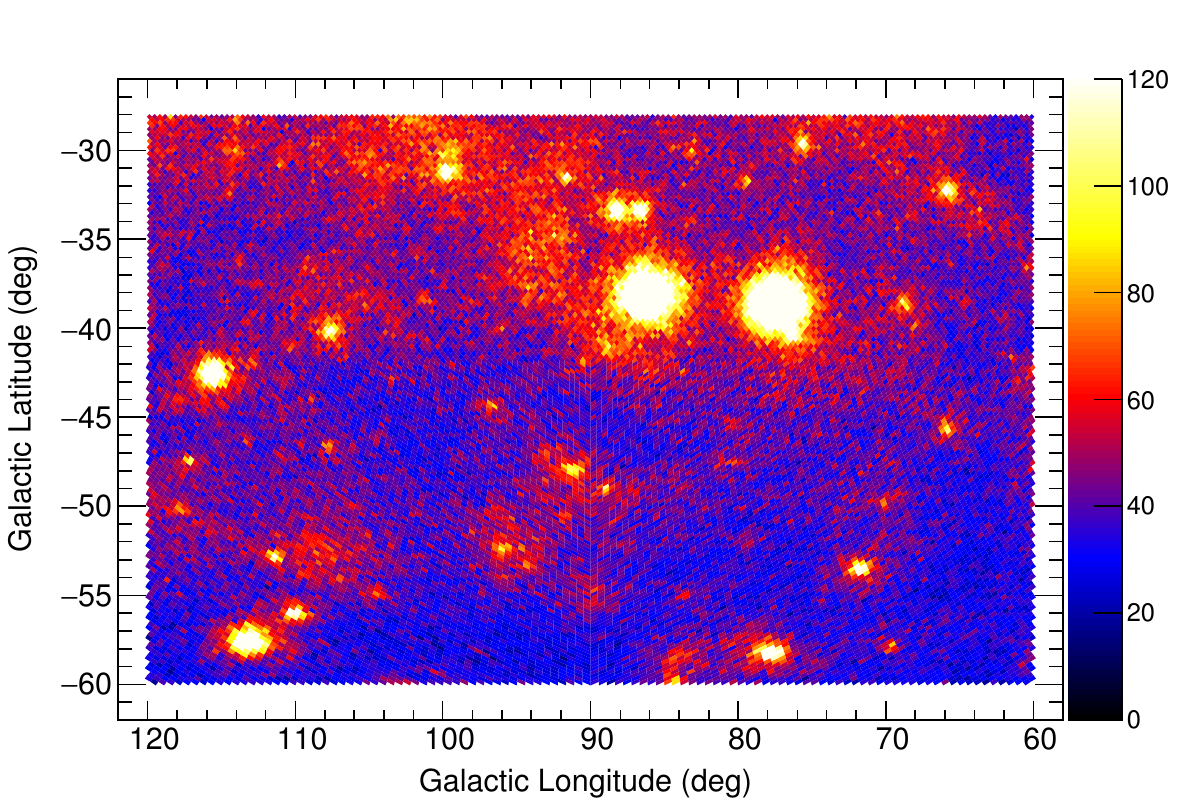}
\put(15,63){(a)}
\end{overpic}
\end{minipage}
\begin{minipage}{0.5\textwidth}
\centering
\begin{overpic}[width=\textwidth]{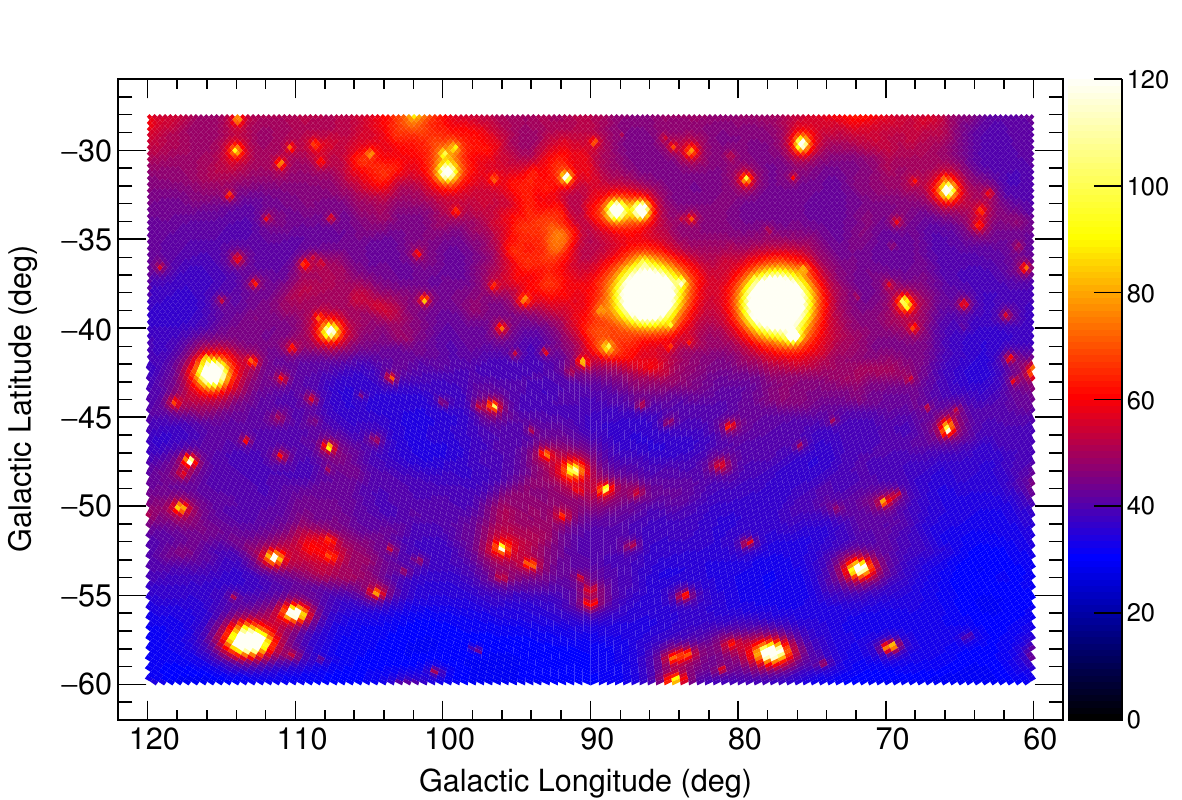}
\put(15,63){(b)}
\end{overpic}
\end{minipage}\\
\\
\begin{minipage}{0.5\textwidth}
\centering
\begin{overpic}[width=\textwidth]{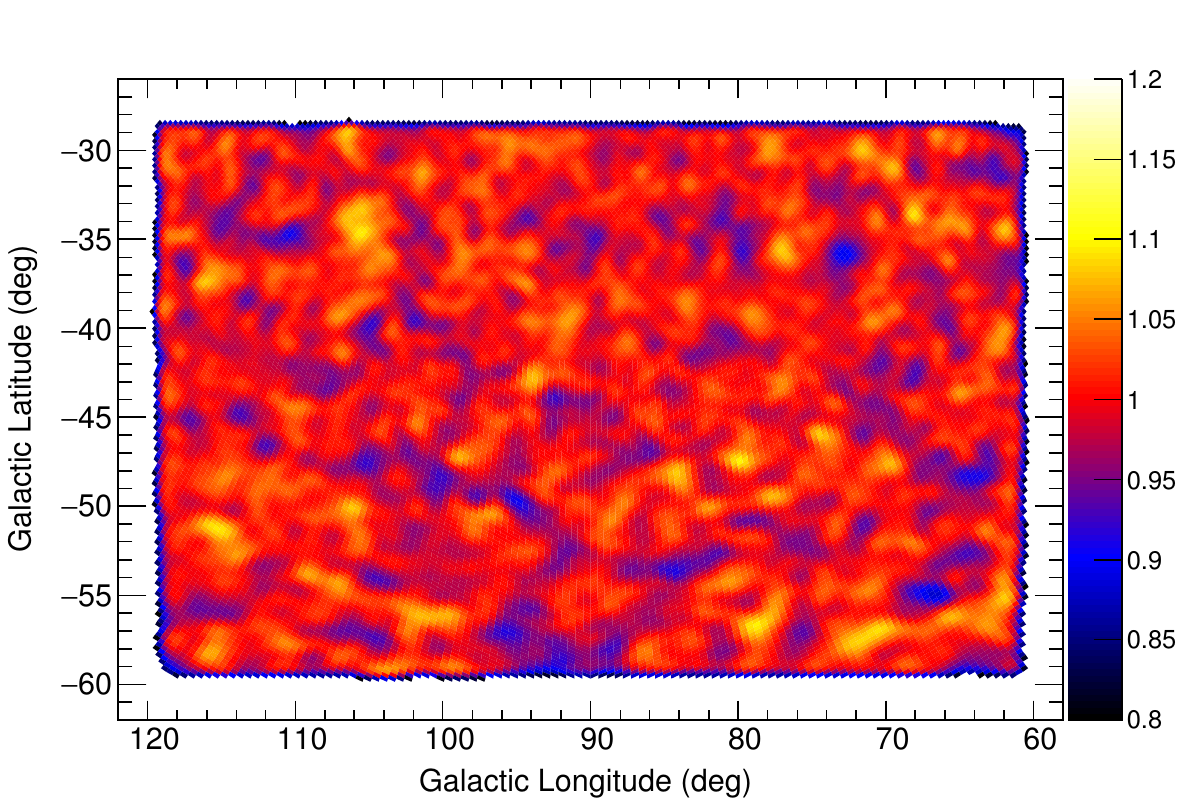}
\put(15,63){(c)}
\end{overpic}
\end{minipage}
\begin{minipage}{0.5\textwidth}
\centering
\begin{overpic}[width=\textwidth]{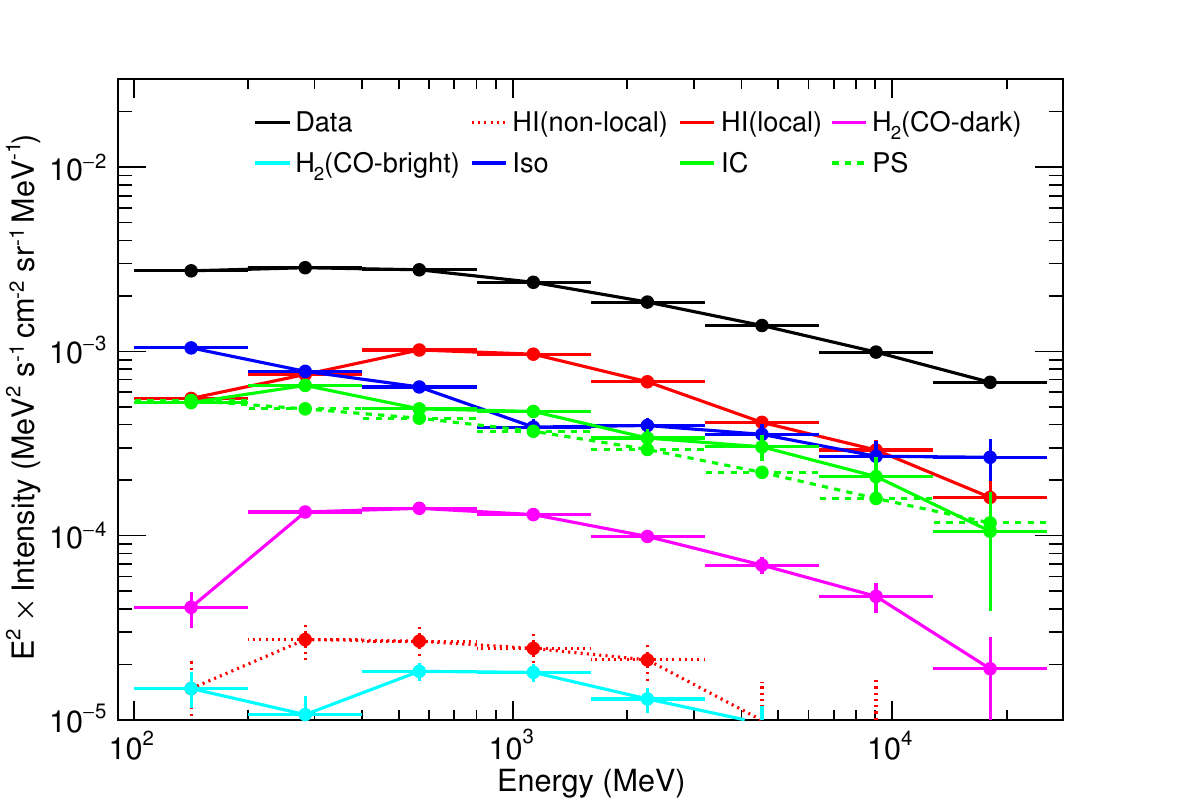}
\put(15,63){(d)}
\end{overpic}
\end{minipage}\\
\end{tabular}
\caption{(a) $\gamma$-ray data count map, (b) model count map, (c) data/model ratio,
and (d) spectrum of each component for the MBM/Pegasus region, given in HEALPix maps of order 8 
(mean area of 0.0524~$\mathrm{deg^{2}}$). They are obtained with the final modeling.
The data/model ratio is smoothed using a Gaussian kernel with $\sigma=60^{'}$ for display.
{Alt text: Three maps and one plot showing the spectrum of the MBM/Pegasus region.}
}\label{..}
\end{figure}

\begin{table}[htbp]
  \tbl{Best-fit parameters, with 1$\sigma$ statistical uncertainties, obtained by the final modeling for the MBM/Pegasus region}{%
  \begin{tabular}{ccccccc}
      \hline
      Energy & $\CHIzro$ & $\CHIonetwo$ & $C_\mathrm{CO}$ & $C_\mathrm{dust}$ & $C_\mathrm{IC}$ & $C_\mathrm{iso}$ \\ 
      (GeV)  & (non-local) & (local)    &                 &                   &                 &                  \\
      \hline
      0.10--0.20 & $0.632\pm0.290$ & $1.122\pm0.046$ & $0.894\pm0.193$ & $0.238\pm0.052$ & $1.048\pm0.070$ & $1.156\pm0.042$ \\
      0.20--0.40 & $0.769\pm0.182$ & $0.995\pm0.035$ & $0.417\pm0.107$ & $0.512\pm0.032$ & $1.460\pm0.087$ & $0.963\pm0.051$ \\
      0.40--0.80 & $0.634\pm0.124$ & $1.117\pm0.026$ & $0.602\pm0.071$ & $0.446\pm0.021$ & $1.295\pm0.086$ & $0.915\pm0.046$ \\
      0.80--1.60 & $0.607\pm0.129$ & $1.109\pm0.030$ & $0.618\pm0.068$ & $0.429\pm0.021$ & $1.573\pm0.119$ & $0.711\pm0.068$ \\
      1.60--3.20 & $0.696\pm0.176$ & $1.035\pm0.043$ & $0.580\pm0.088$ & $0.429\pm0.029$ & $1.460\pm0.172$ & $0.951\pm0.099$ \\
      3.20--6.40 & $0.515\pm0.319$ & $0.985\pm0.076$ & $0.674\pm0.160$ & $0.474\pm0.052$ & $1.703\pm0.271$ & $0.977\pm0.127$ \\
      6.40--12.8 & $0.836\pm0.619$ & $1.174\pm0.153$ & $0.789\pm0.294$ & $0.542\pm0.101$ & $1.586\pm0.438$ & $0.823\pm0.176$ \\
      12.8--25.6 & $1.005\pm1.192$ & $1.106\pm0.308$ & $0.579\pm0.532$ & $0.373\pm0.181$ & $1.098\pm0.692$ & $1.006\pm0.263$ \\
      \hline
    \end{tabular}}\label{tab:first}
\begin{tabnote}
\end{tabnote}
\end{table}

\clearpage

\subsection{R CrA Clouds}
The R Coronae Australis molecular cloud, or R CrA cloud, is one of the nearest star-forming regions at a distance of ${\sim}$150~pc \citep{Gali2020},
with approximately 50 young stars concentrated in an area of $\mathrm{0.5~pc^{2}}$.
The morphology of the molecular gas has a characteristic head-tail structure, with ${\sim}2000~M_{\odot}$ of molecular gas
concentrated in the head region \citep{Tachihara2024}.
Past studies in $\gamma$-ray can be found in \citet{Ackermann2012} and \citet{Yang2014}.

The cloud is located toward the Galactic center. There, the hard and very extended structure known as the "Fermi Bubble" (FB) is observed 
by \citet{Su2010}
and may affect the $\gamma$-ray analysis of diffuse emission.
To constrain the gas-related emission accurately, we adopted a larger ROI than \citet{Ackermann2012}. Specifically,
we extended the ROI by $10^{\circ}$ toward the Galactic plane and positive/negative Galactic longitude.
We also employed a template of FB developed by \citet{Ackermann2017}, with some modifications described in Appendix~6.
We approximated the FB spectrum by \citet{Ackermann2017} with double broken power-law; we assumed $E^{-1}$ below 0.5~GeV,
$E^{-1.9}$ in 0.5--10~GeV, and $E^{-2.1}$ above 10~GeV. We added the FB model to Equation~(1) 
with the normalization free to vary in each energy bin.

The value of $T_\mathrm{d}$ in the R CrA region is higher than that of the other clouds.
Most of the pixels pass the selection using $T_\mathrm{d, th}$ in constructing the residual gas template (step~1 in section~2.2.2) and
the dust fit by narrow and broad {\HI} may be biased by residual gas remaining.
We therefore tested $T_\mathrm{d, th}$ higher by 0.5, 1.0, and 1.5~K and found that the threshold higher by 1.0~K 
best fits $\gamma$-ray data.
We also confirmed that $\tau_{353}$-based templates give a better fit than radiance-based one, 
and adopted
the template based on the revised $\tau_{353}$ map with $T_\mathrm{d, th}=20.5~\mathrm{K}$. Like MBM/Pegasus region, IC of Std-SA0 gives the best fit.
The correction factor for narrow {\HI}, obtained as the average of fit coefficients for narrow {\HI} to broad {\HI} above 400~MeV, 
was found to be 1.12.
This scale factor was used in constructing a single, local $\NHI$ template.
The obtained results with the final model configuration are summarized in Figure~8 and Table~2.
The non-local $\NHI$ is almost 0 (see Appendix~2) and their coefficients are fixed to 1.
The {\HI} emissivity above 100~MeV and 400~MeV are $(1.495 \pm 0.023) \times 10^{-26}$ and $(0.510 \pm 0.007) \times 10^{-26}$ in units of 
${\rm ph~s^{-1}~sr^{-1}}$ per H atom, respectively.
$X_\mathrm{CO}$ is obtained to be $(1.668 \pm 0.034) \times 10^{20}~\mathrm{cm^{-2}~(K~km~s^{-1})^{-1}}.$

\begin{figure}[htbp]
\begin{tabular}{cc}
\begin{minipage}{0.5\textwidth}
\centering
\begin{overpic}[width=\textwidth]{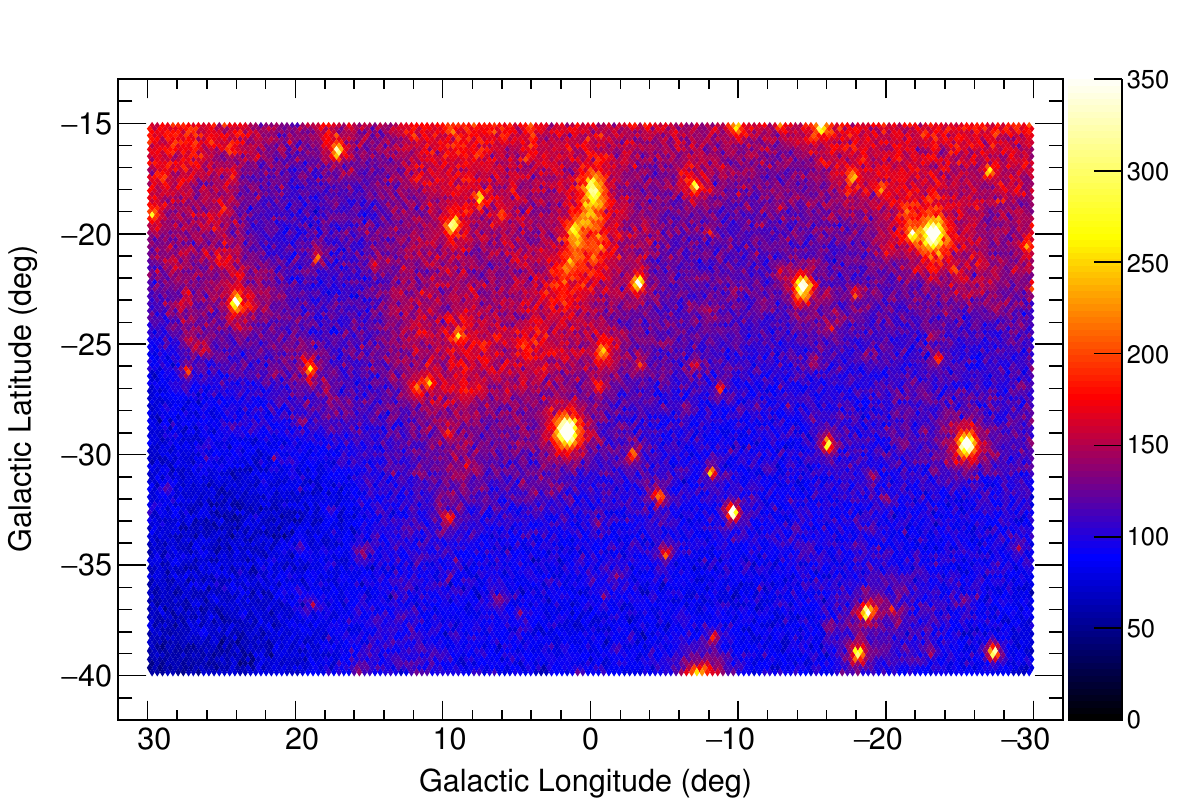}
\put(15,63){(a)}
\end{overpic}
\end{minipage}
\begin{minipage}{0.5\textwidth}
\centering
\begin{overpic}[width=\textwidth]{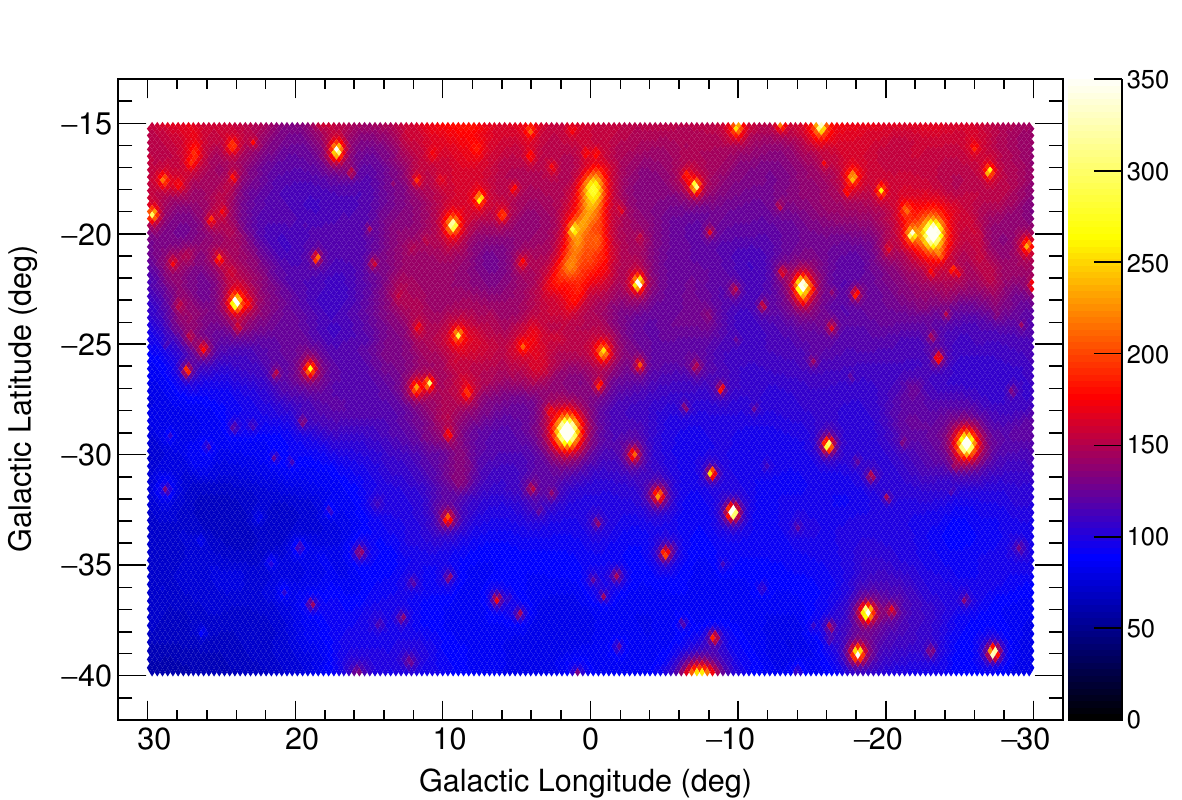}
\put(15,63){(b)}
\end{overpic}
\end{minipage}\\
\\
\begin{minipage}{0.5\textwidth}
\centering
\begin{overpic}[width=\textwidth]{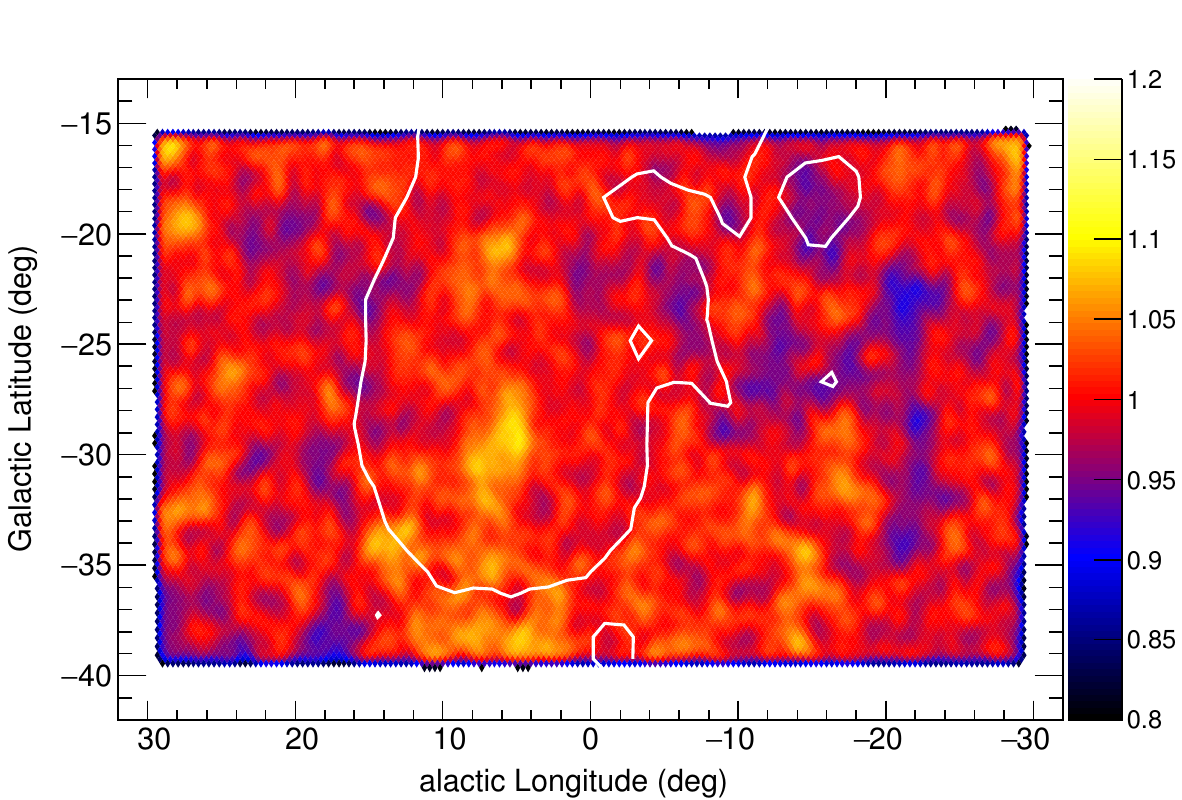}
\put(15,63){(c)}
\end{overpic}
\end{minipage}
\begin{minipage}{0.5\textwidth}
\centering
\begin{overpic}[width=\textwidth]{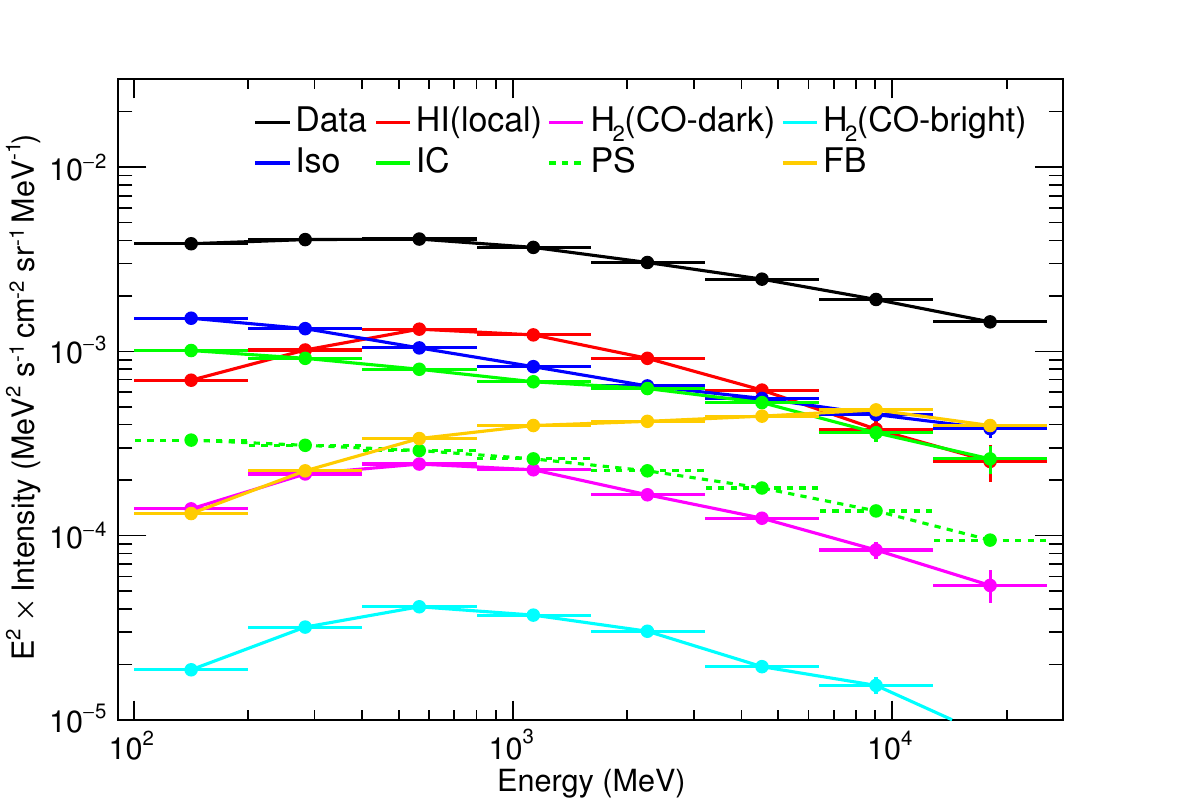}
\put(15,63){(d)}
\end{overpic}
\end{minipage}\\
\end{tabular}
\caption{The same as Figure~7, but for the R CrA region instead of the MBM/Pegasus region.
Contours of intensity = 0.8 (arbitrary unit) of the FB template are overlaid in data/model ratio map for reference.
{Alt text: Three maps and one plot showing the spectrum of the R CraA region.}}\label{..}
\end{figure}

\begin{table}[htbp]
  \tbl{Best-fit parameters, with 1$\sigma$ statistical uncertainties, obtained by the final modeling for the R CrA region}{%
  \begin{tabular}{ccccccc}
      \hline
      Energy & $\CHIonetwo$ & $C_\mathrm{CO}$ & $C_\mathrm{dust}$ & $C_\mathrm{IC}$ & $C_\mathrm{iso}$ & $C_\mathrm{FB}$ \\
      (GeV)  & (local)      &                 &                   &                 &                  &               \\
      \hline
      0.10--0.20 & $1.090\pm0.037$ & $1.469\pm0.108$ & $0.689\pm0.030$ & $0.855\pm0.012$ & $1.684\pm0.018$ & $2.254\pm0.117$ \\
      0.20--0.40 & $1.050\pm0.028$ & $1.763\pm0.064$ & $0.712\pm0.020$ & $0.914\pm0.017$ & $1.653\pm0.023$ & $2.029\pm0.064$ \\
      0.40--0.80 & $1.113\pm0.021$ & $1.868\pm0.046$ & $0.646\pm0.014$ & $0.978\pm0.020$ & $1.482\pm0.023$ & $1.903\pm0.035$ \\
      0.80--1.60 & $1.080\pm0.025$ & $1.769\pm0.049$ & $0.630\pm0.016$ & $1.092\pm0.030$ & $1.512\pm0.035$ & $1.995\pm0.035$ \\
      1.60--3.20 & $1.061\pm0.037$ & $1.906\pm0.071$ & $0.605\pm0.023$ & $1.344\pm0.048$ & $1.559\pm0.052$ & $1.967\pm0.038$ \\
      3.20--6.40 & $1.129\pm0.070$ & $1.941\pm0.131$ & $0.719\pm0.044$ & $1.524\pm0.083$ & $1.522\pm0.073$ & $1.913\pm0.044$ \\
      6.40--12.8 & $1.171\pm0.145$ & $2.605\pm0.279$ & $0.822\pm0.090$ & $1.443\pm0.146$ & $1.388\pm0.102$ & $1.918\pm0.050$ \\
      12.8--25.6 & $1.338\pm0.303$ & $2.332\pm0.549$ & $0.903\pm0.183$ & $1.485\pm0.258$ & $1.445\pm0.160$ & $1.620\pm0.064$ \\
      \hline
    \end{tabular}}\label{tab:first}
\begin{tabnote}
\end{tabnote}
\end{table}

\clearpage
\subsection{Chamaeleon Clouds}
Chamaeleon molecular cloud complex is a famous star-forming region located in the solar neighborhood with a
distance of ${\sim}$150~pc (e.g., \cite{Mizuno2001} and \cite{Luhman2008}).
Owing to the moderate molecular mass of the order of $10^{4}~M_{\odot}$ \citep{Mizuno2001} 
and the relatively uniform ISRF
suggested by the lack of OB stellar clusters, it is a valuable target for studying the ISM. Past studies of Chamaeleon cloud in $\gamma$-rays can be found in,
e.g., \citet{Ackermann2012}, \citet{Planck2015}, and \citet{Hayashi2019}.

In our ROI, the gas at $280^{\circ} \le l \le 290^{\circ}$ and $b \le -22^{\circ}$ has different velocity features than the local {\HI} emission.
Following \citet{Hayashi2019}, we masked this area in constructing the residual gas template and $\gamma$-ray data analysis.
Again, we found that $\tau_{353}$-based residual gas templates give a better fit than radiance-based one, and narrow {\HI} gives larger
emissivity than broad {\HI}. Based on $\ln{L}$,
we adopted the revised $\tau_{353}$ map to construct the residual gas template and Std-SA100 as 
an IC model.
The correction factor for narrow {\HI}, obtained from the fit coefficients above 400~MeV, was 1.21.
Obtained results with the final model configuration are summarized in Figure~9 and Table~3.
The {\HI} emissivity above 100~MeV and 400~MeV are $(1.571 \pm 0.021) \times 10^{-26}$ and $(0.518 \pm 0.007) \times 10^{-26}$ in units of 
${\rm ph~s^{-1}~sr^{-1}}$ per H atom, respectively.
$X_\mathrm{CO}$ is $(0.932 \pm 0.016) \times 10^{20}~\mathrm{cm^{-2}~(K~km~s^{-1})^{-1}}.$

\begin{figure}[htbp]
\begin{tabular}{cc}
\begin{minipage}{0.5\textwidth}
\centering
\begin{overpic}[width=\textwidth]{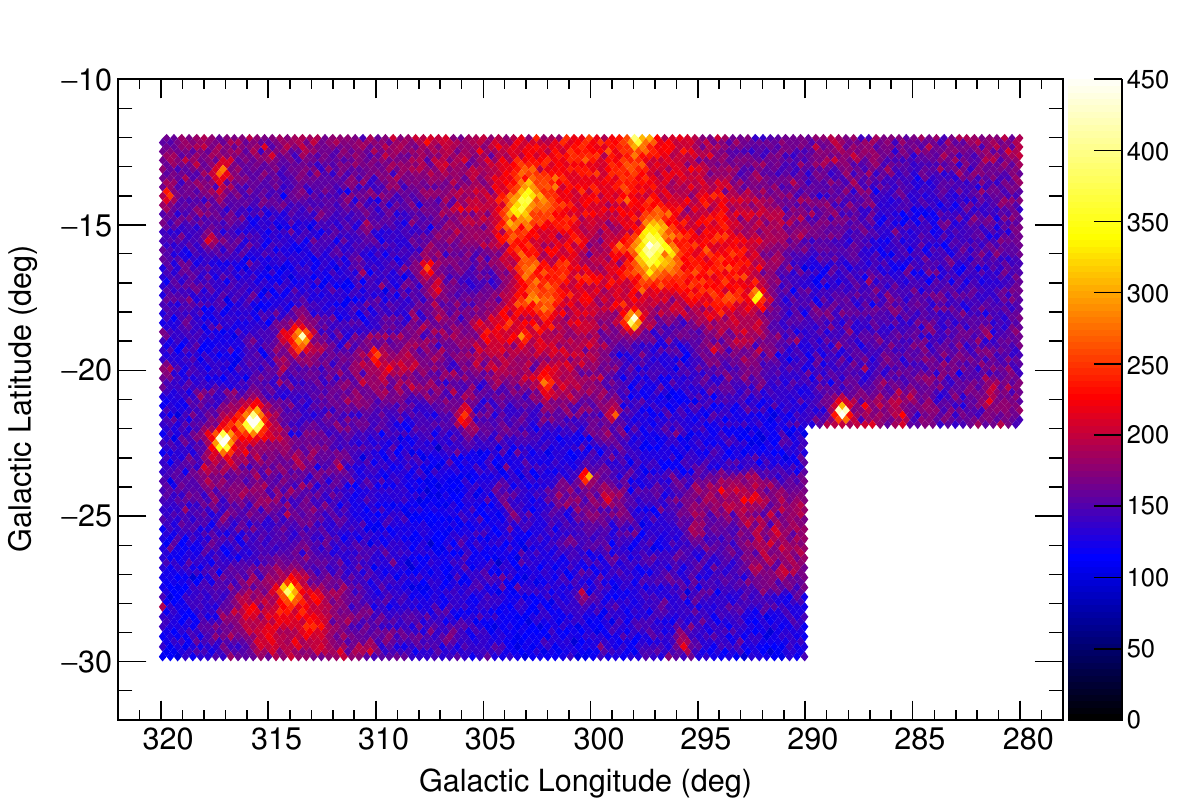}
\put(15,63){(a)}
\end{overpic}
\end{minipage}
\begin{minipage}{0.5\textwidth}
\centering
\begin{overpic}[width=\textwidth]{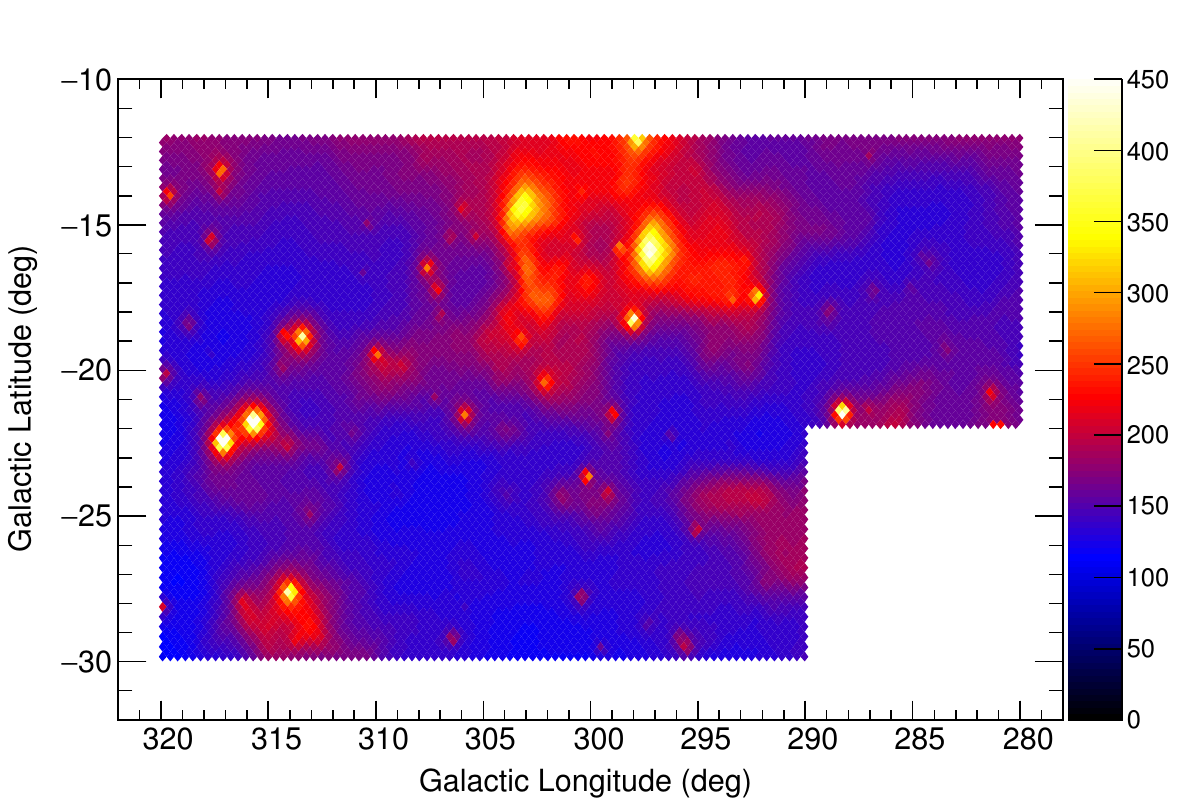}
\put(15,63){(b)}
\end{overpic}
\end{minipage}\\
\\
\begin{minipage}{0.5\textwidth}
\centering
\begin{overpic}[width=\textwidth]{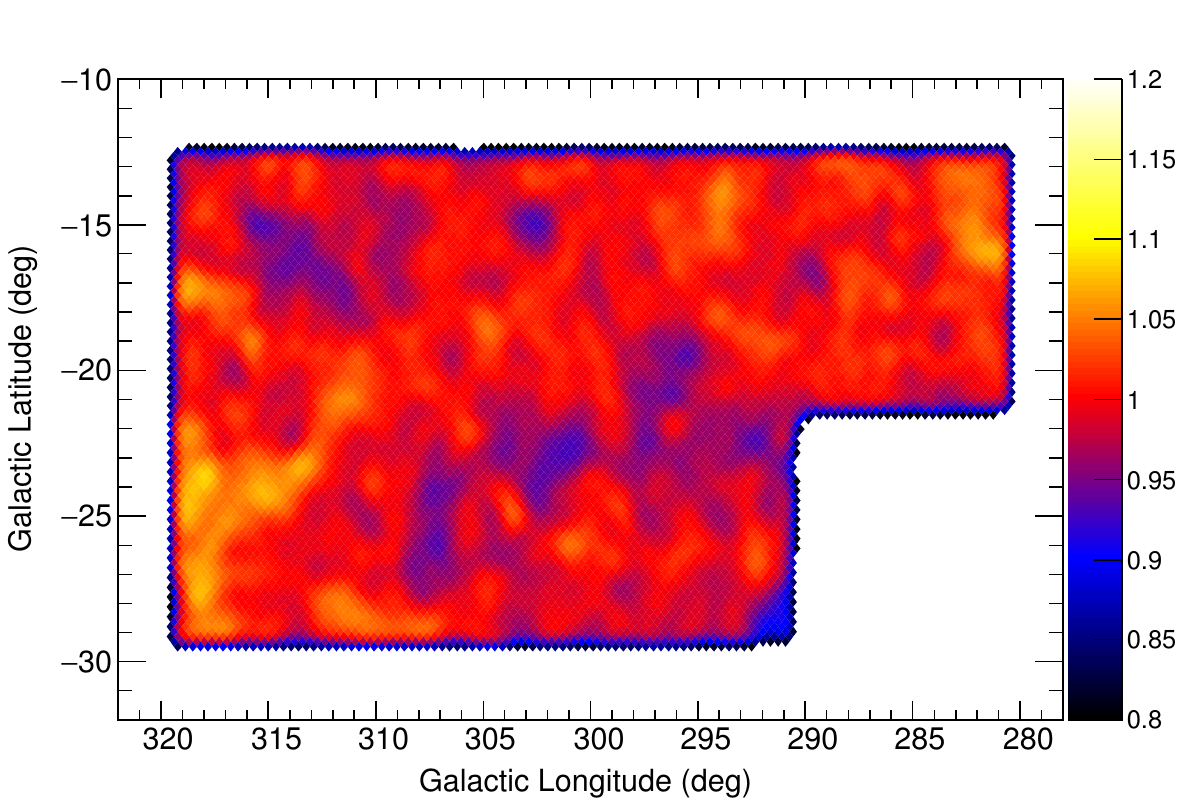}
\put(15,63){(c)}
\end{overpic}
\end{minipage}
\begin{minipage}{0.5\textwidth}
\centering
\begin{overpic}[width=\textwidth]{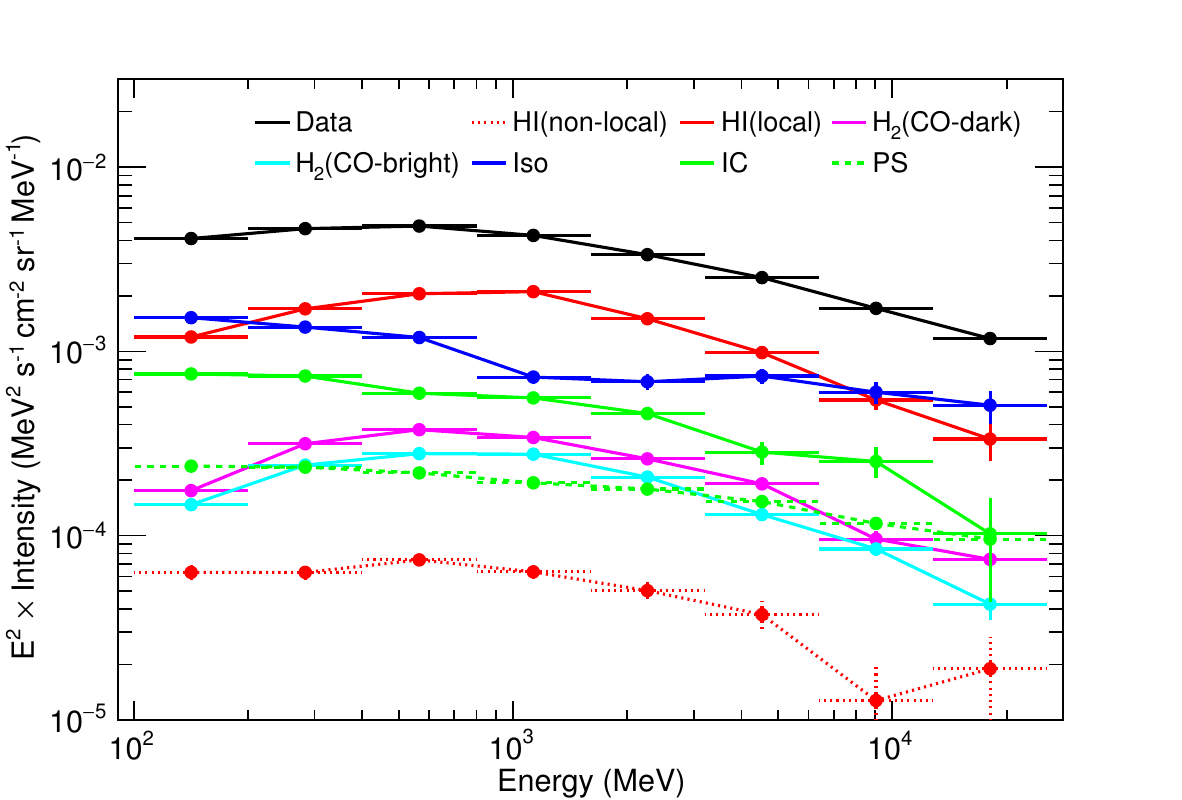}
\put(15,63){(d)}
\end{overpic}
\end{minipage}\\
\end{tabular}
\caption{The same as Figure~7, but for the Chamaeleon region instead of the MBM/Pegasus region.
{Alt text: Three maps and one plot showing the spectrum of the Chamaeleon region.}}\label{..}
\end{figure}

\begin{table}[htbp]
  \tbl{Best-fit parameters, with 1$\sigma$ statistical uncertainties, obtained by the final modeling for the Chamaeleon region}{%
  \begin{tabular}{ccccccc}
      \hline
      Energy & $\CHIzro$   & $\CHIonetwo$ & $C_\mathrm{CO}$ & $C_\mathrm{dust}$ & $C_\mathrm{IC}$ & $C_\mathrm{iso}$ \\ 
      (GeV)  & (non-local) & (local)      &                 &                   &                 &                  \\
      \hline
      0.10--0.20 & $2.851\pm0.256$ & $1.180\pm0.033$ & $1.053\pm0.055$ & $0.476\pm0.026$ & $0.598\pm0.021$ & $1.705\pm0.050$ \\
      0.20--0.40 & $2.537\pm0.225$ & $1.106\pm0.028$ & $1.098\pm0.030$ & $0.559\pm0.015$ & $0.638\pm0.024$ & $1.696\pm0.068$ \\
      0.40--0.80 & $2.564\pm0.174$ & $1.090\pm0.022$ & $1.033\pm0.021$ & $0.547\pm0.010$ & $0.589\pm0.025$ & $1.701\pm0.072$ \\
      0.80--1.60 & $2.654\pm0.223$ & $1.170\pm0.028$ & $1.061\pm0.021$ & $0.513\pm0.011$ & $0.683\pm0.035$ & $1.334\pm0.112$ \\
      1.60--3.20 & $2.892\pm0.336$ & $1.095\pm0.038$ & $1.035\pm0.028$ & $0.516\pm0.015$ & $0.713\pm0.051$ & $1.657\pm0.153$ \\
      3.20--6.40 & $3.487\pm0.643$ & $1.134\pm0.062$ & $1.025\pm0.049$ & $0.593\pm0.027$ & $0.564\pm0.080$ & $2.014\pm0.185$ \\
      6.40--12.8 & $2.041\pm1.223$ & $1.068\pm0.128$ & $1.134\pm0.093$ & $0.568\pm0.052$ & $0.608\pm0.126$ & $1.838\pm0.248$ \\
      12.8--25.6 & $5.234\pm2.542$ & $1.120\pm0.273$ & $0.972\pm0.171$ & $0.672\pm0.102$ & $0.363\pm0.207$ & $1.941\pm0.396$ \\
      \hline
    \end{tabular}}\label{tab:first}
\begin{tabnote}
\end{tabnote}
\end{table}

\clearpage

\subsection{Cepheus and Polaris Flare}
Cepheus and Polaris flare clouds are other nearby molecular clouds (${\sim}$450~pc; \cite{Dame1987}). They are characterized by prominent extended structures toward
the high latitude and have been studied extensively in various wavebands (e.g., \cite{Panopoulou2016}).
They are located in the direction almost opposite to the Chamaeleon region in the Gould Belt (e.g., \cite{Perrot2003}), allowing us to investigate
the CR and ISM properties over several hundred pc but still inside the coherent environment. Past studies in $\gamma$-rays can be found in, 
e.g., \citet{Ackermann2012}.

While the $\tau_{353}$-based residual gas template gives a better fit above 1.6~GeV ($\Delta \ln{L} \sim 50$), 
the radiance-based one gives a better fit
below that energy ($\Delta \ln{L} \sim 50$). 
We also found that the $\tau_{353}$-based template has many pixels with positive values close to 0.
We interpret this because $\tau_{353}$ and radiance are better gas tracers in high-density and low-density areas, respectively.
Accordingly we modified the template based on the revised $\tau_{353}$ map by filling pixels of no residual in 
the radiance-based map with 0. We confirmed that
this "merged" template gives a much-improved fit to $\gamma$-ray data ($\Delta \ln{L} \sim 100$).

Once again, we found that narrow {\HI} gives larger
emissivity than broad {\HI}, with a rather large correction factor of 1.61.
We adopted Std-SA100 which best fits $\gamma$-ray data among the three IC models.
The obtained results with the final model configuration are summarized in Figure~10 and Table~4.
The {\HI} emissivity above 100~MeV and 400~MeV are $(1.358 \pm 0.025) \times 10^{-26}$ and $(0.435 \pm 0.008) \times 10^{-26}$ in units of 
${\rm ph~s^{-1}~sr^{-1}}$ per H atom, respectively.
$X_\mathrm{CO}$ is $(0.912 \pm 0.017) \times 10^{20}~\mathrm{cm^{-2}~(K~km~s^{-1})^{-1}}.$

\begin{figure}[htbp]
\begin{tabular}{cc}
\begin{minipage}{0.5\textwidth}
\centering
\begin{overpic}[width=\textwidth]{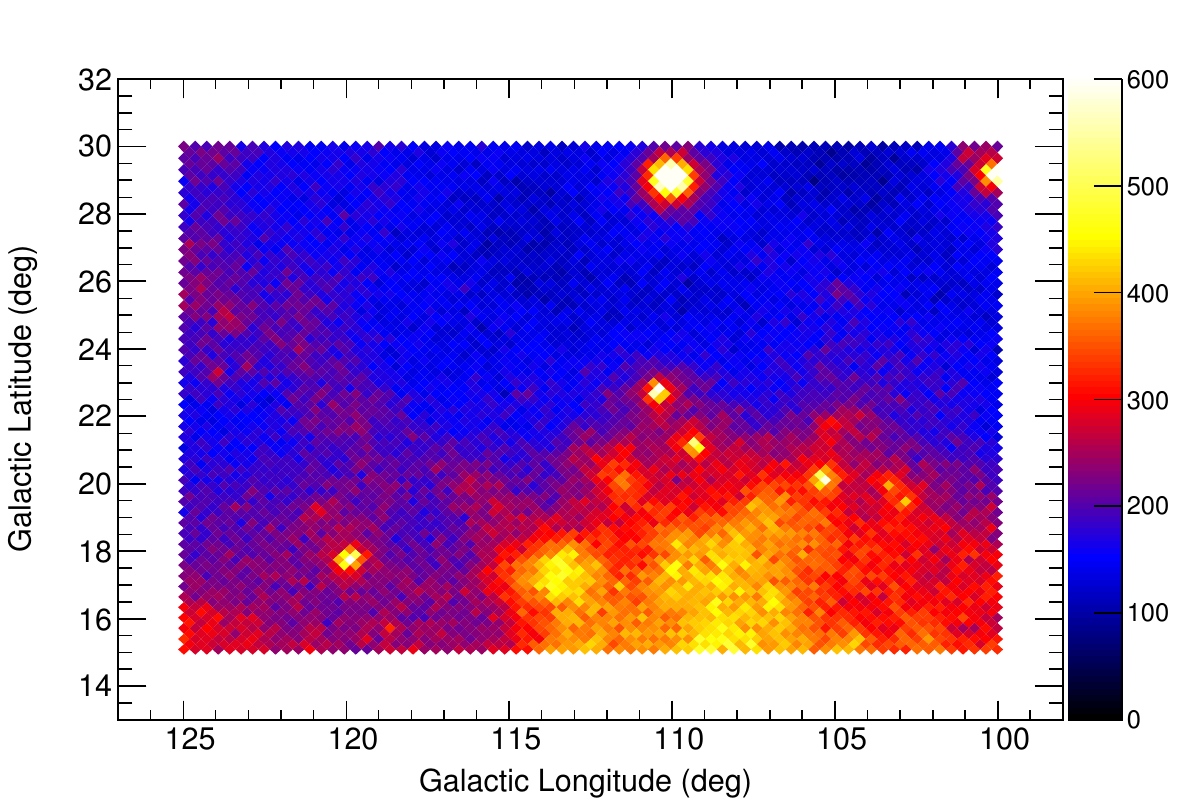}
\put(15,63){(a)}
\end{overpic}
\end{minipage}
\begin{minipage}{0.5\textwidth}
\centering
\begin{overpic}[width=\textwidth]{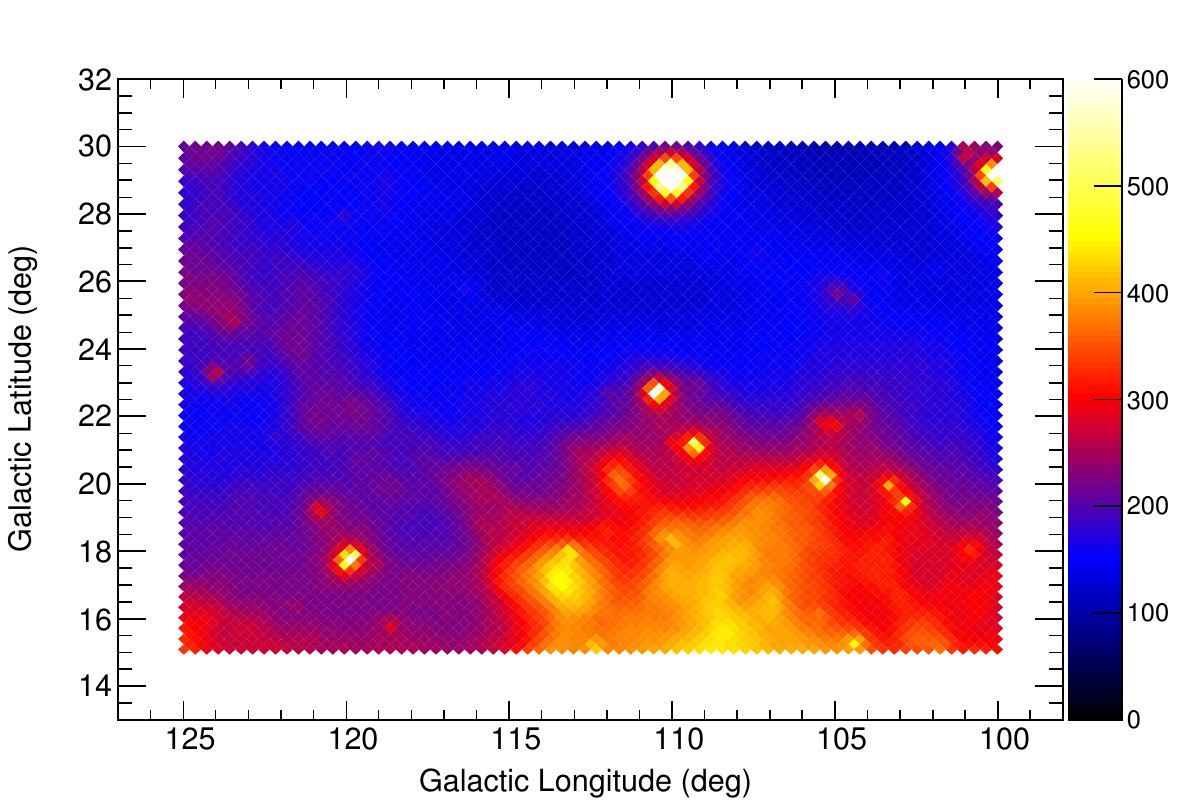}
\put(15,63){(b)}
\end{overpic}
\end{minipage}\\
\\
\begin{minipage}{0.5\textwidth}
\centering
\begin{overpic}[width=\textwidth]{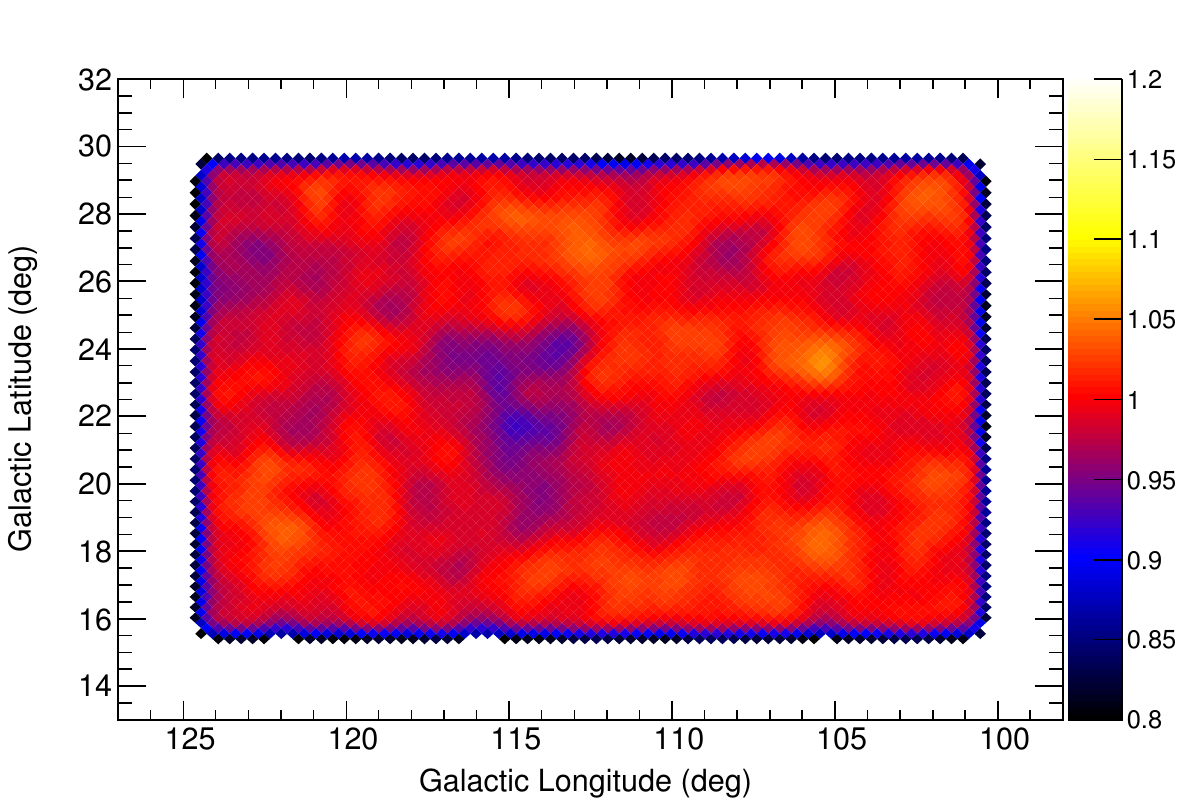}
\put(15,63){(c)}
\end{overpic}
\end{minipage}
\begin{minipage}{0.5\textwidth}
\centering
\begin{overpic}[width=\textwidth]{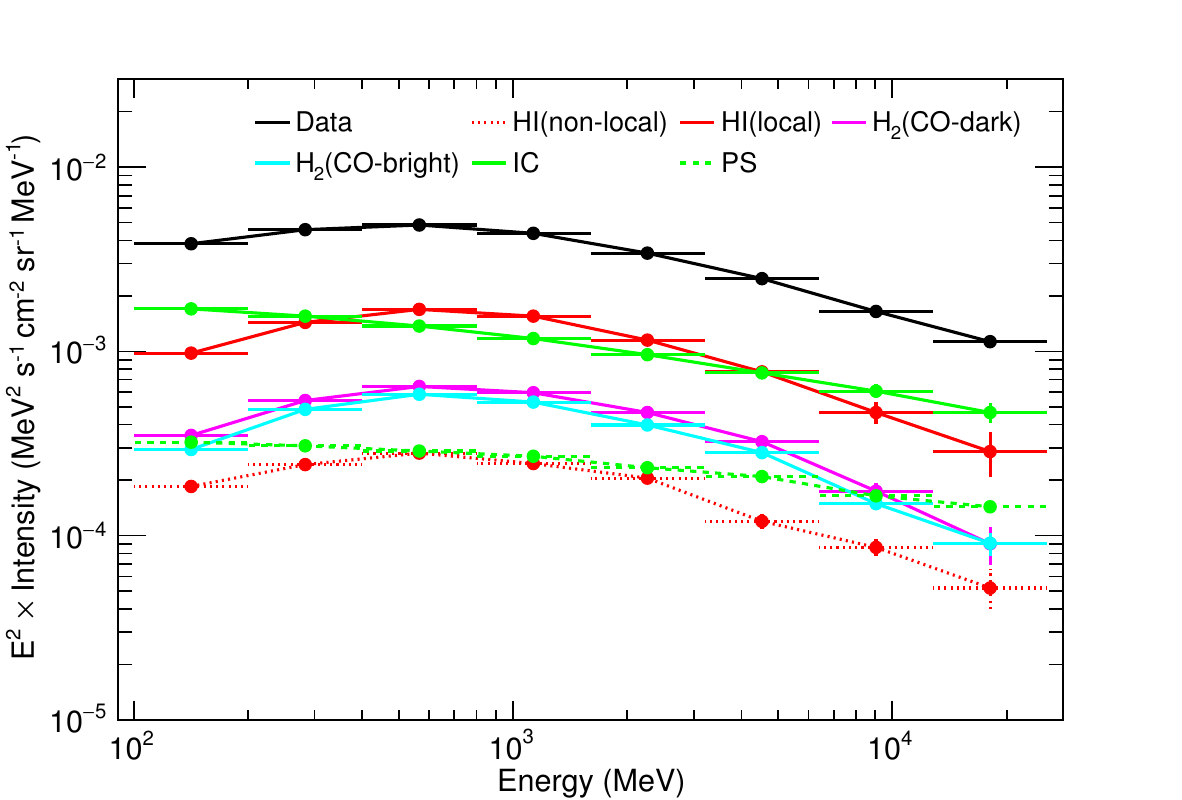}
\put(15,63){(d)}
\end{overpic}
\end{minipage}\\
\end{tabular}
\caption{The same as Figure~7, but for the Cep/Pol region instead of the MBM/Pegasus region.
{Alt text: Three maps and one plot showing the spectrum of the Cep/Pol region.}}\label{..}
\end{figure}

\begin{table}[htbp]
  \tbl{Best-fit parameters, with 1$\sigma$ statistical uncertainties, obtained by the final modeling for the Cep/Pol region.
Parameters for Isotropic are fixed to 0 since best-fit values are very small.}{%
  \begin{tabular}{ccccccc}
      \hline
      Energy & $\CHIzro$   & $\CHIonetwo$ & $C_\mathrm{CO}$ & $C_\mathrm{dust}$ & $C_\mathrm{IC}$ & $C_\mathrm{iso}$ \\ 
      (GeV)  & (non-local) & (local)      &                 &                   &                 &                  \\
      \hline
      0.10--0.20 & $1.132\pm0.042$ & $1.014\pm0.040$ & $0.855\pm0.028$ & $0.415\pm0.018$ & $3.416\pm0.052$ & $0$ \\
      0.20--0.40 & $1.102\pm0.035$ & $0.994\pm0.031$ & $0.898\pm0.016$ & $0.426\pm0.012$ & $3.419\pm0.070$ & $0$ \\
      0.40--0.80 & $1.068\pm0.029$ & $0.957\pm0.025$ & $0.883\pm0.012$ & $0.414\pm0.009$ & $3.434\pm0.081$ & $0$ \\
      0.80--1.60 & $1.043\pm0.036$ & $0.914\pm0.028$ & $0.825\pm0.013$ & $0.402\pm0.009$ & $3.625\pm0.110$ & $0$ \\
      1.60--3.20 & $1.174\pm0.055$ & $0.883\pm0.040$ & $0.809\pm0.018$ & $0.412\pm0.013$ & $3.767\pm0.154$ & $0$ \\
      3.20--6.40 & $1.103\pm0.100$ & $0.942\pm0.072$ & $0.908\pm0.033$ & $0.454\pm0.024$ & $3.873\pm0.230$ & $0$ \\
      6.40--12.8 & $1.361\pm0.189$ & $0.964\pm0.136$ & $0.816\pm0.063$ & $0.417\pm0.045$ & $4.150\pm0.352$ & $0$ \\
      12.8--25.6 & $1.417\pm0.381$ & $1.011\pm0.271$ & $0.850\pm0.121$ & $0.370\pm0.087$ & $4.344\pm0.565$ & $0$ \\
      \hline
    \end{tabular}}\label{tab:first}
\begin{tabnote}
\end{tabnote}
\end{table}

\clearpage

\subsection{Orion Clouds}
The Orion A and B clouds are the archetypes of local giant molecular clouds (cloud complex with a molecular mass
of ${\sim}10^{5}~M_{\odot}$) where interstellar gas condenses and stars are formed.
The clouds have been studied in various wavebands (e.g., \cite{Wilson2005}, \cite{Nishimura2015})
to study star-forming activities in the solar neighborhood. Thanks to their proximity (${\sim}$400~pc;
see \cite{FermiOrion} for the summary of distance measurements) and
large mass, they have been studied in $\gamma$-rays since the COS-B era \citep{Bloemen1984}. 
The past study by Fermi-LAT can be found in \citet{FermiOrion}.

We defined an ROI as
$195^{\circ} \le l \le 225^{\circ}$ and $-10^{\circ} \le b \le -35^{\circ}$, similar to that adopted in \citet{FermiOrion}.
A radiance-based residual gas template gives a much worse fit to $\gamma$-rays, likely because the strong star-forming activity affects 
the dust radiance and the effect was not removed even with our masking infrared sources (Appendix~4).
Like other clouds studied here,
we found that narrow {\HI} gives larger emissivity than broad {\HI}, with a correction factor of 1.41 for the revised $\tau_{353}$ map.
We adopted a Std-SA0 IC model that best fits $\gamma$-ray data.
The obtained results with the final model configuration are summarized in Figure~11 and Table~5.
The {\HI} emissivity above 100~MeV and 400~MeV are $(1.464 \pm 0.010) \times 10^{-26}$ and $(0.471 \pm 0.003) \times 10^{-26}$ in units of 
${\rm ph~s^{-1}~sr^{-1}}$ per H atom, respectively.
$X_\mathrm{CO}$ is $(1.296 \pm 0.010) \times 10^{20}~\mathrm{cm^{-2}~(K~km~s^{-1})^{-1}}.$

\begin{figure}[htbp]
\begin{tabular}{cc}
\begin{minipage}{0.5\textwidth}
\centering
\begin{overpic}[width=\textwidth]{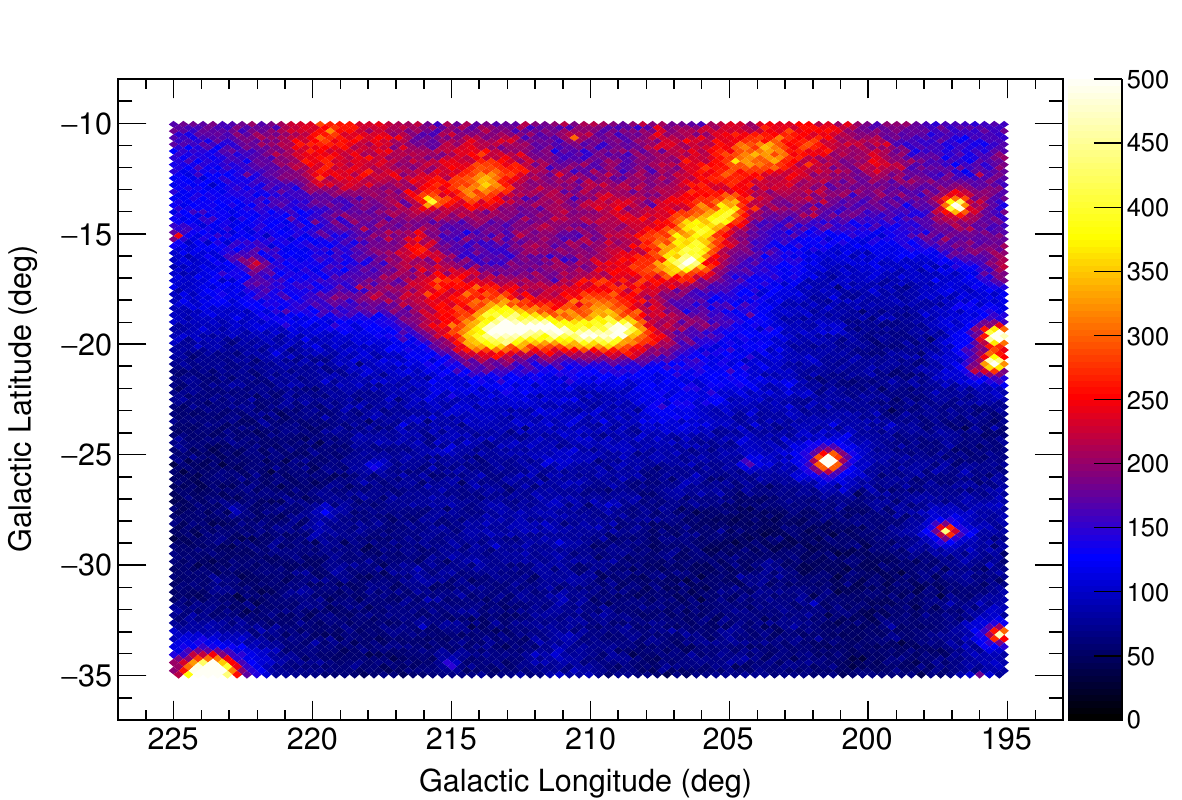}
\put(15,63){(a)}
\end{overpic}
\end{minipage}
\begin{minipage}{0.5\textwidth}
\centering
\begin{overpic}[width=\textwidth]{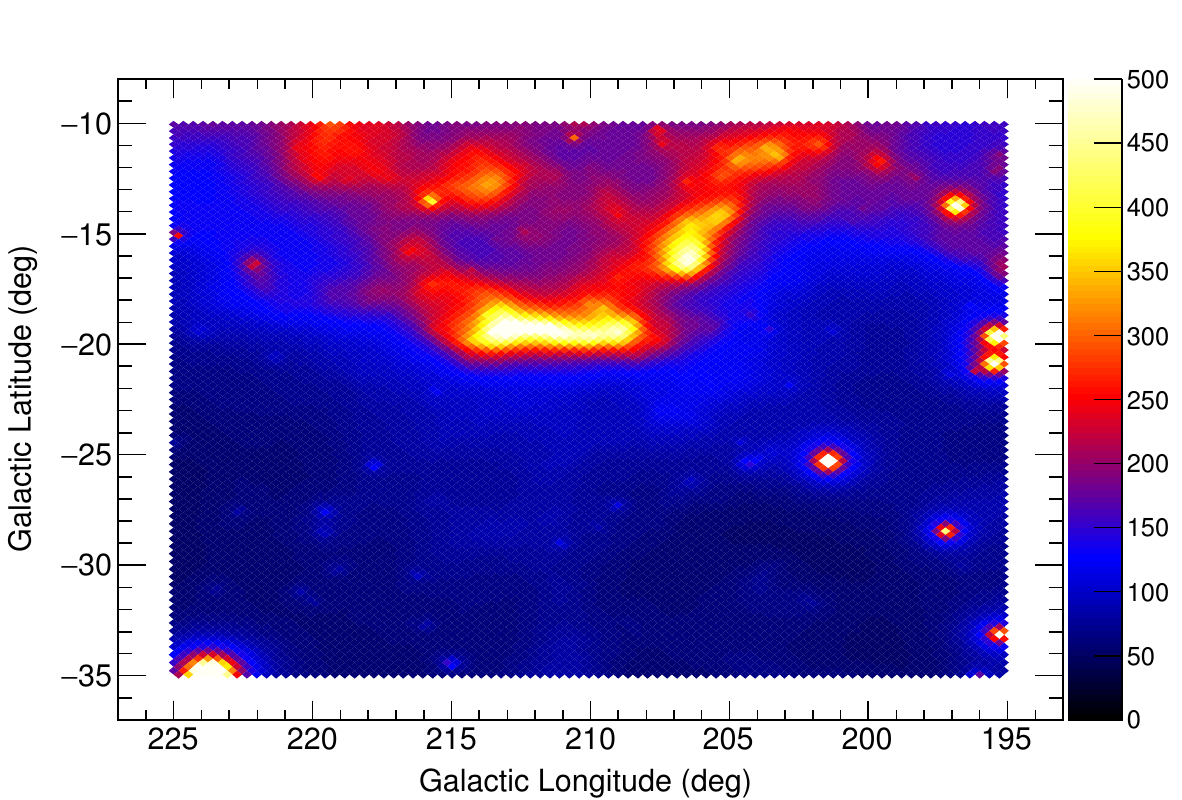}
\put(15,63){(b)}
\end{overpic}
\end{minipage}\\
\\
\begin{minipage}{0.5\textwidth}
\centering
\begin{overpic}[width=\textwidth]{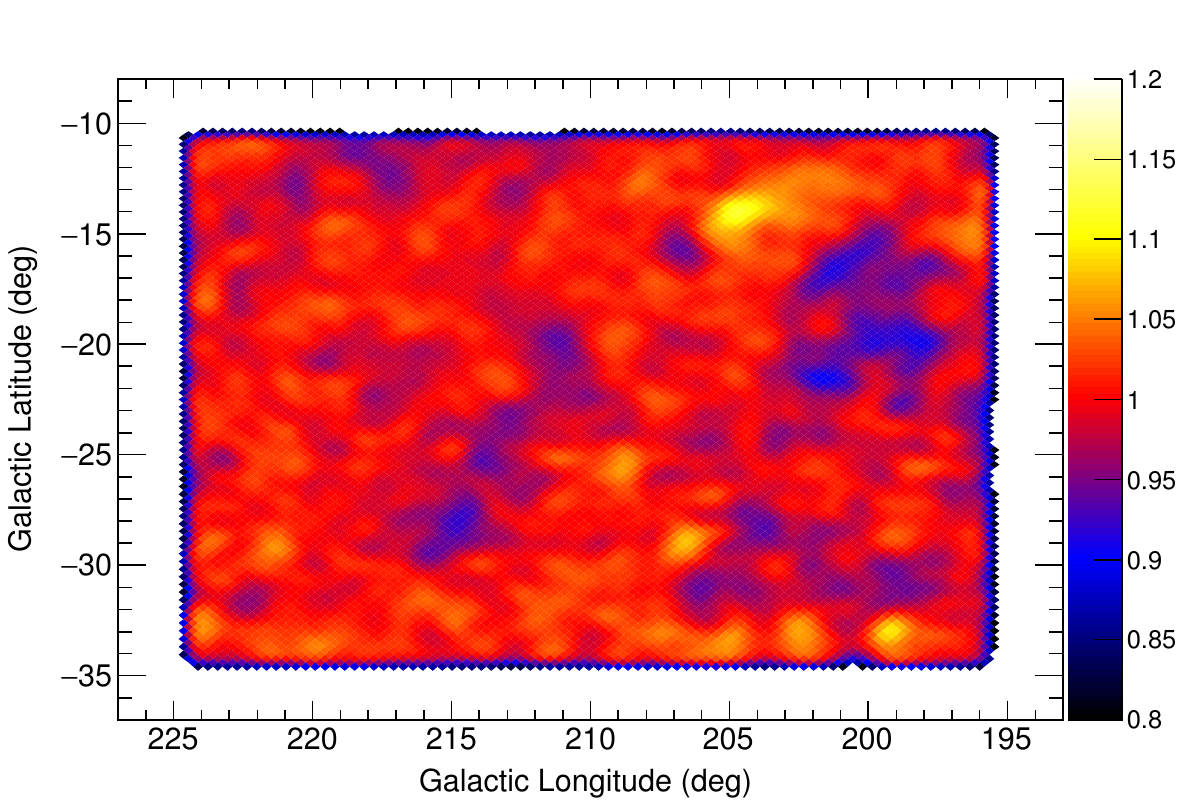}
\put(15,63){(c)}
\end{overpic}
\end{minipage}
\begin{minipage}{0.5\textwidth}
\centering
\begin{overpic}[width=\textwidth]{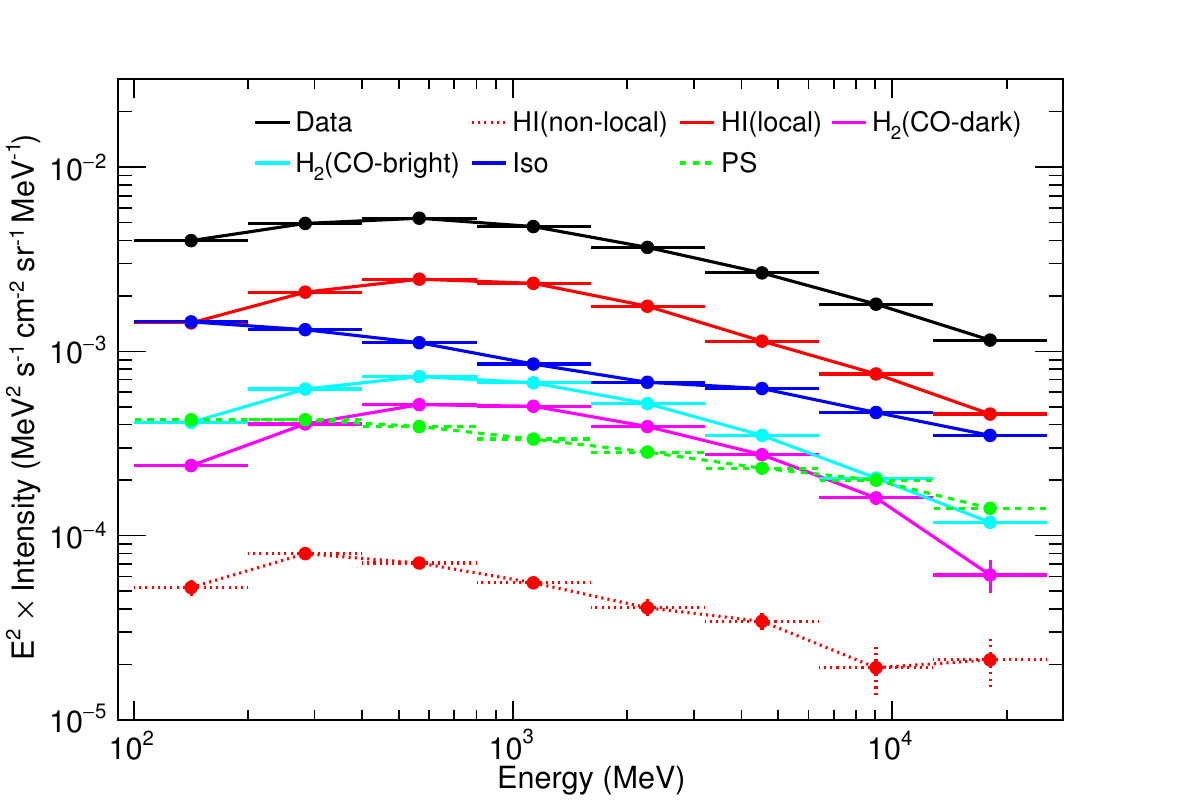}
\put(15,63){(d)}
\end{overpic}
\end{minipage}\\
\end{tabular}
\caption{The same as Figure~7, but for the Orion region instead of the MBM/Pegasus region. 
{Alt text: Three maps and one plot showing the spectrum of the Orion region.}}\label{..}
\end{figure}

\begin{table}[htbp]
  \tbl{Best-fit parameters, with 1$\sigma$ statistical uncertainties, obtained by the final modeling for the Orion region.
Parameters for IC are fixed to 0 since best-fit values are very small.}{%
  \begin{tabular}{ccccccc}
      \hline
      Energy & $\CHIzro$   & $\CHIonetwo$ & $C_\mathrm{CO}$ & $C_\mathrm{dust}$ & $C_\mathrm{IC}$ & $C_\mathrm{iso}$ \\ 
      (GeV)  & (non-local) & (local)      &                 &                   &                 &                  \\
      \hline
      0.10--0.20 & $1.059\pm0.103$ & $1.098\pm0.016$ & $1.452\pm0.035$ & $0.373\pm0.019$ & $0$ & $1.598\pm0.015$ \\
      0.20--0.40 & $1.212\pm0.075$ & $1.058\pm0.011$ & $1.412\pm0.019$ & $0.407\pm0.011$ & $0$ & $1.625\pm0.019$ \\
      0.40--0.80 & $0.890\pm0.051$ & $1.020\pm0.008$ & $1.339\pm0.013$ & $0.421\pm0.008$ & $0$ & $1.592\pm0.019$ \\
      0.80--1.60 & $0.764\pm0.056$ & $1.009\pm0.009$ & $1.283\pm0.013$ & $0.430\pm0.008$ & $0$ & $1.561\pm0.026$ \\
      1.60--3.20 & $0.753\pm0.081$ & $0.996\pm0.012$ & $1.300\pm0.018$ & $0.437\pm0.011$ & $0$ & $1.636\pm0.039$ \\
      3.20--6.40 & $1.018\pm0.151$ & $1.025\pm0.023$ & $1.383\pm0.032$ & $0.487\pm0.020$ & $0$ & $1.730\pm0.053$ \\
      6.40--12.8 & $0.965\pm0.289$ & $1.143\pm0.046$ & $1.347\pm0.061$ & $0.476\pm0.037$ & $0$ & $1.419\pm0.071$ \\
      12.8--25.6 & $1.814\pm0.582$ & $1.186\pm0.089$ & $1.347\pm0.111$ & $0.311\pm0.064$ & $0$ & $1.320\pm0.102$ \\
      \hline
    \end{tabular}}\label{tab:first}
\begin{tabnote}
\end{tabnote}
\end{table}

\clearpage
\section{Discussion}

\subsection{CR Spectrum in Solar Neighborhood}
Most previous works used {\HI} templates with uniform $T_\mathrm{s}$ and reported $\gamma$-ray emissivity 
larger than expected from directly-measured CR spectra (e.g., discussion in \cite{Mizuno2022}). 
The summary of {\HI} emissivity above 400~MeV 
observed through this study is given in Figure~12 as a function of the position of clouds.
Past results of the same clouds summarized in \citet{Planck2015} are also plotted for comparison. 
We assume the distance between the Sun and the Galactic center to be 8.5~kpc. We also
assume the distances of 150~pc for the R CrA and Chamaeleon clouds, and 450~pc for the Cep/Pol and Orion clouds,
giving the Galactocentric radius very close to the values in \citet{Planck2015}
with $(l, b) \sim$ (\timeform{0D}, \timeform{-20D}) for R CrA, $\sim$ (\timeform{300D}, \timeform{-16D}) for Chamaeleon,
$\sim$ (\timeform{110D}, \timeform{17D}) for Cep/Pol, and $\sim$ (\timeform{210D}, \timeform{-17D}) for Orion.
We adopt the distance to be 150~pc for MBM/Pegasus, giving 
the Galactocentric radius of 8.6~kpc and the height of -105~pc
with $(l, b) \sim (\timeform{90D}, \timeform{-35D})$.
Since we required the fit coefficient ratio of the narrow {\HI} to broad {\HI} to be 1.0, 
statistical errors of the latter (1--3\% depending on regions) contribute to the systematic uncertainty of the emissivity.
By adding the systematics due to the parameters for constructing the residual gas template (${\sim}$3\%; Section~3.1) in quadrature, 
our emissivities have ${\sim}$4\% systematic uncertainty.
Therefore, the measured emissivities are significantly smaller (by 5--30\%) than previous results and agree better with the 
adopted model ($0.464 \times 10^{-27}~{\rm ph~s^{-1}~sr^{-1}}$ per H atom) based on directly-measured CR spectra.

In Figure~12, we observe a higher emissivity (CR intensity) in clouds closer to the inner Galaxy and/or Galactic plane.
As a gauge of the emissivity at ${\sim}$1~GeV, we evaluated the CR proton intensity at 10~GeV. Specifically, we referred to our GALPROP model of Std-SA0
(assuming the axisymmetric CR source distribution by \citet{CRdist} and CR halo height of 6~kpc) and plot the intensity distributions
on the Galactic plane 
(for the Galactocentric radius dependence) and along the line passing through the Sun (for the height dependence) in the figure. 
There, the intensity distributions are scaled to the model emissivity at the position of the Sun. 
While the predicted height dependence is too small to explain the data, a 10\% decrease as a function of Galactocentric radius
is predicted for the distance scale studied here, and may be compatible with the measured negative dependence.
The marked dip for the Cep/Pol region may be due to the additional systematic uncertainty caused by a considerable (${\sim}$20\%) contribution
from non-local {\HI}. Alternatively, non-axisymmetric CR source distributions (SA50 and SA100) predict
even larger (at the 20--30\% level) position dependence of the CR proton intensity within ${\sim}$500~pc and may agree better with our results.
Comparison with such a model, however, is beyond the scope of this study.

\begin{figure}[htbp]
\begin{tabular}{cc}
\begin{minipage}{0.5\textwidth}
\centering
\begin{overpic}[width=\textwidth]{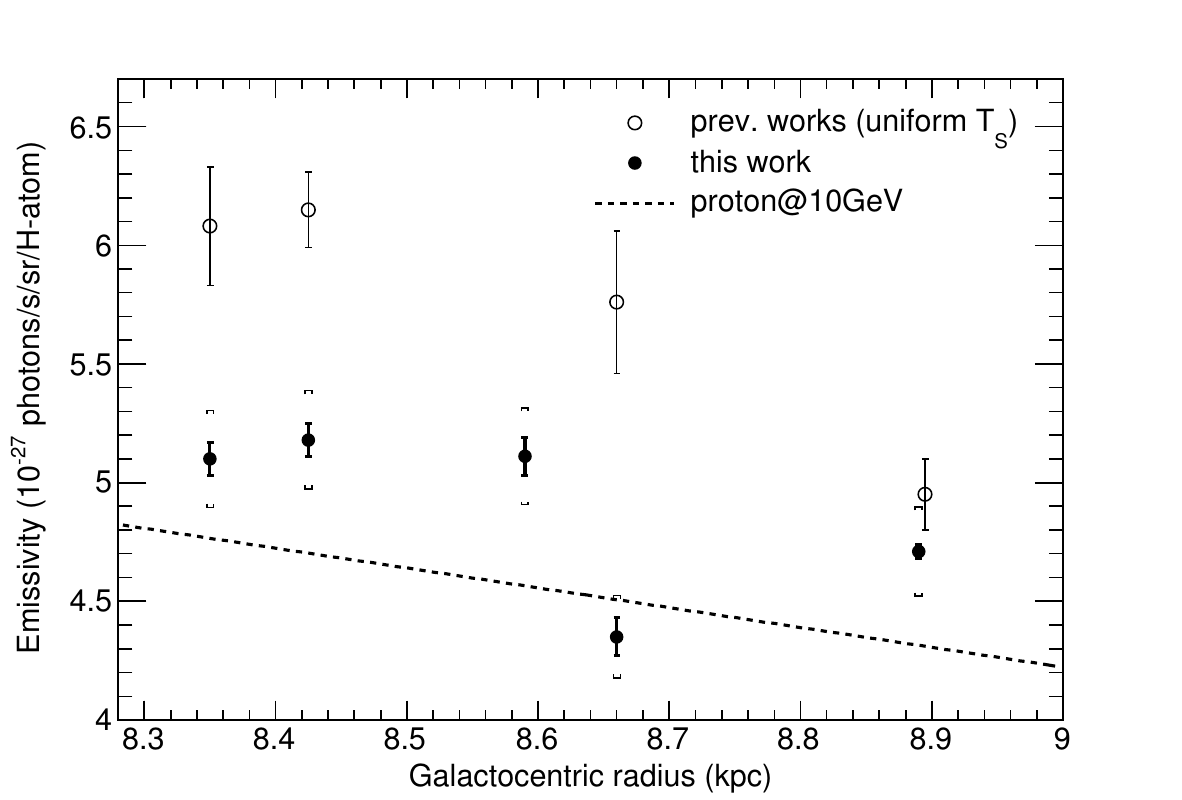}
\put(15,63){(a)}
\end{overpic}
\end{minipage}
\begin{minipage}{0.5\textwidth}
\centering
\begin{overpic}[width=\textwidth]{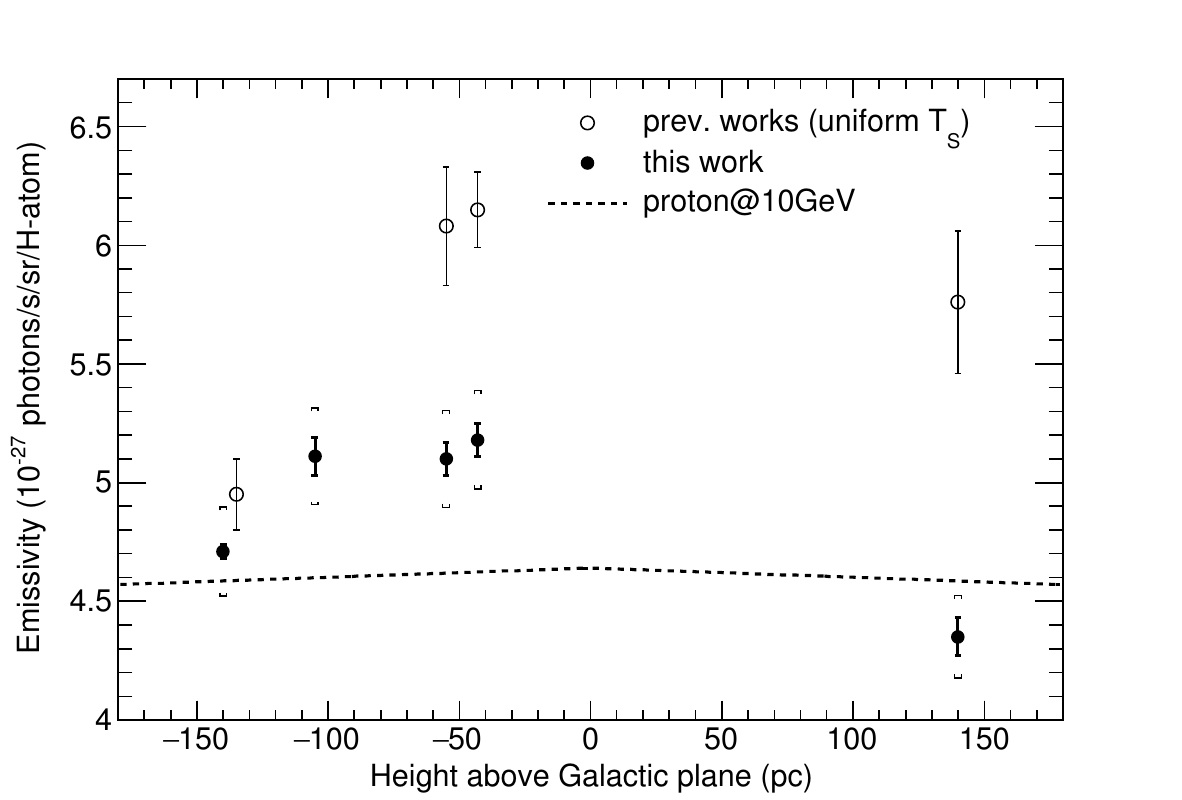}
\put(15,63){(b)}
\end{overpic}
\end{minipage}
\end{tabular}
\caption{Summary of {\HI} emissivities above 400~MeV obtained through this study,
(a) as a function of the Galactocentric radius and (b) height from the Galactic plane.
Emissivities by previous works summarized in \citet{Planck2015} are also plotted for comparison,
with the points for the Orion region shifted horizontally for clarity.
Square brackets represent the systematic errors.
In the left panel, data points (from left to right) correspond to the R CrA, Chamaeleon, MBM/Pegasus, Cep/Pol, and Orion regions.
In the right panel, data points (from left to right) correspond to the Orion, MBM/Pegasus, R CrA, Chamaeleon, and Cep/Pol regions.
Position dependence of the CR proton intensity at 10 GeV, scaled to
the model emissivity at the position of the Sun, are shown by dashed lines. See the text for details.
{Alt text: Two plots showing emissivities of regions studied.}
}\label{..}
\end{figure}

\clearpage

\subsection{ISM Gas Properties in Solar Neighborhood}
Assuming a uniform CR intensity of each region, we can evaluate the column density of each gas phase, and
we tabulated $\NH$ integrated over each ROI and each gas phase in Table~6. There, 
$\int \NH \,d\Omega$ for broad {\HI} and non-local {\HI} were calculated assuming the optically thin case, and 
$\int \NH \,d\Omega$ for other phases were calculated from the average of fit coefficients ratio (in 0.1-25.6~GeV) about broad {\HI}.
The correction over the optically-thin case is shown separately for narrow {\HI}.
The integral for the $D_\mathrm{em, res}$ map shows 
the "dark gas" contribution that {\HI} cannot trace even with linewidth taken into account, 
and we interpret this to be CO-dark {\Htwo}.
The integral of CO-bright {\Htwo} is equal to $2 \WCO \int X_\mathrm{CO} \,d\Omega$, where $X_\mathrm{CO}$ is summarized in Appendix~7.

While the previous work \citep{Mizuno2022} gives a similar amount of ISM gas in {\HI} optical depth 
correction and CO-dark {\Htwo} (residual gas)
for the MBM/Pegasus region, our new results show that the latter is ${\ge}$1.5 times larger. The difference is that, 
while \citet{Mizuno2022} used radiance to construct the residual gas template 
(since it gives a better fit to $\gamma$-ray data as a single ISM gas template), we used the revised $\tau_{353}$ map that gives the best fit
when $\ln{L}$ is compared in a final fit configuration (i.e., dust is taken into account as a residual gas template).
We interpret this because $\tau_{353}$ and radiance are better gas tracers in high-density and low-density areas, respectively.
Although we believe our new results are more accurate,
neither $\tau_{353}$ nor radiance is a perfect tracer of the ISM gas. 
Suppose the contribution of CO-dark {\Htwo} to the integral of the dark gas column density is
equal to the contribution of the optically thick {\HI} correction. In that case, the integral will be 7.0 instead of 9.0, 
and we adopt the difference (${\sim}$25\%) as a systematic uncertainty of each dark-gas phase.

From the table we can see several common properties of the ISM gas. Firstly,
the ratio of CO-dark {\Htwo} to the optical depth correction is 2--8, indicating that
dark gas is mainly CO-dark {\Htwo}. The conclusion is valid even if we take the 25\% systematic uncertainty into account.
Secondly, the ratio of CO-dark {\Htwo} to CO-bright {\Htwo} anticorrelates with the molecular cloud mass traced by $\WCO$ ($\MHtwoCO$). 
To visualize this, we plot the ratio as a function of $\MHtwoCO$
in Figure~13. There, the mass is calculated as
\begin{equation}
M = \mu m_\mathrm{H} d^{2} \int \NH \,d\Omega~,
\end{equation}
where $d$ is the distance to the cloud, $m_\mathrm{H}$ is the mass of the hydrogen atom, and $\mu=1.41$ is the mean atomic mass per H atom.
For convenience, 
$\int \NH \,d\Omega = 10^{22}~\mathrm{cm^{-2}~deg^{2}}$ corresponds to ${\sim}$740~$M_{\odot}$ for $d=150~\mathrm{pc}$.
When discussing the CO-dark {\Htwo} to CO-bright {\Htwo} ratio for the R CrA region, care must be taken.
As described in Section~3.2 we adopted 
a large ROI to separate the ISM gas components and the FB, and the ROI may include CO-dark {\Htwo}
that is not physically related to CO-bright {\Htwo} seen only toward $l \sim \timeform{0D}$.
We therefore reevaluated the integral within $|l| \le \timeform{20D}$ (also shown in Table~6) and used the ratio in Figure~13.
We observe that, while massive systems 
($\MHtwoCO \sim 10^{5}~M_{\odot}$) have similar amount of
CO-dark {\Htwo} and CO-bright {\Htwo}, small systems ($\MHtwoCO \sim 10^{3}~M_{\odot}$) are
CO-dark {\Htwo} rich with a ratio of 5--10.
A similar trend was reported in anticenter clouds by \citet{Remy2018}, therefore the anti-correlation between the
CO-dark {\Htwo} to CO-bright {\Htwo} ratio and $\MHtwoCO$ mass is commonly seen within a single system and among systems.
A possible explanation for the anti-correlation is less shielding for CO photodissociation in small systems.
Theoretical works predicted CO-dark {\Htwo} fraction anticorrelates with visual extinction \citep{Wolfire2010}
and the molecular gas column density \citep{Smith2014}.
Small-mass systems have a smaller column density of gas/dust and hence smaller shielding for UV radiation,
resulting in a larger fraction of CO-dark {\Htwo}.

Our result on ISM gas is also relevant to the study of Galactic CRs. Molecular clouds traced by $\WCO$
in the vicinity of potential CR accelerators have often been used to evaluate the CR energy density
in GeV/TeV ${\gamma}$-ray observations. For example, \citet{Amenomori2021} studied a potential PeVatron supernova remnant
G106.3+2.7 using CO data by \citet{Kothes2006} that gives the cloud mass of only ${\sim}100~M_{\odot}$.
With the limited spatial resolution of $\gamma$-rays, one may easily miss the contribution of
CO-dark {\Htwo} and overestimate the total energy of CRs in studying such CR accelerators.
CO-dark {\Htwo} is seen in the peripherals of CO-bright one; hence their contribution also affects
the morphology of $\gamma$-rays and the discussion of CR transport in the vicinity of each object.

\begin{table}[htbp]
  \tbl{Integral for $\NH$ ($\int \NH \,d\Omega$ in units of $10^{22}~\mathrm{cm^{-2}~deg^{2}}$) of each gas phase of each region}{%
  \begin{tabular}{cccccc}
      \hline
phase & MBM/Pegasus & R CrA$^{\ddag}$ & Chamaeleon & Cep/Pol & Orion \\ 
\hline
broad {\HI} & 39.9 & 59.2 & 37.3 & 19.1 & 57,2 \\
narrow {\HI}$^{*}$ & 18.0+5.0 & 18.5+2.2 & 16.0+3.4 & 7.8+4.7 & 19.9+7.7 \\
non-local {\HI}$^{\dag}$ & 2.8 & & 0.7 & 4.2 & 2.5 \\
residual gas & 9.0 & 17.3 (10.3) & 10.5 & 10.5 & 21.4 \\
CO-bright {\Htwo} & 1.1 & 2.5 (2.2) & 7.7 & 10.8 & 25.4 \\
      \hline
    \end{tabular}}\label{tab:first}
\begin{tabnote}
\footnotemark[$*$] optically thin case plus correction \\ 
\footnotemark[$\dag$] optically thin assumed \\
\footnotemark[$\ddag$]  values in $|l|\le\timeform{20D}$ are given in bracket \\ 
\end{tabnote}
\end{table}

\begin{figure}[htbp]
\begin{tabular}{cc}
\begin{minipage}{0.5\textwidth}
\centering
\includegraphics[width=\textwidth]{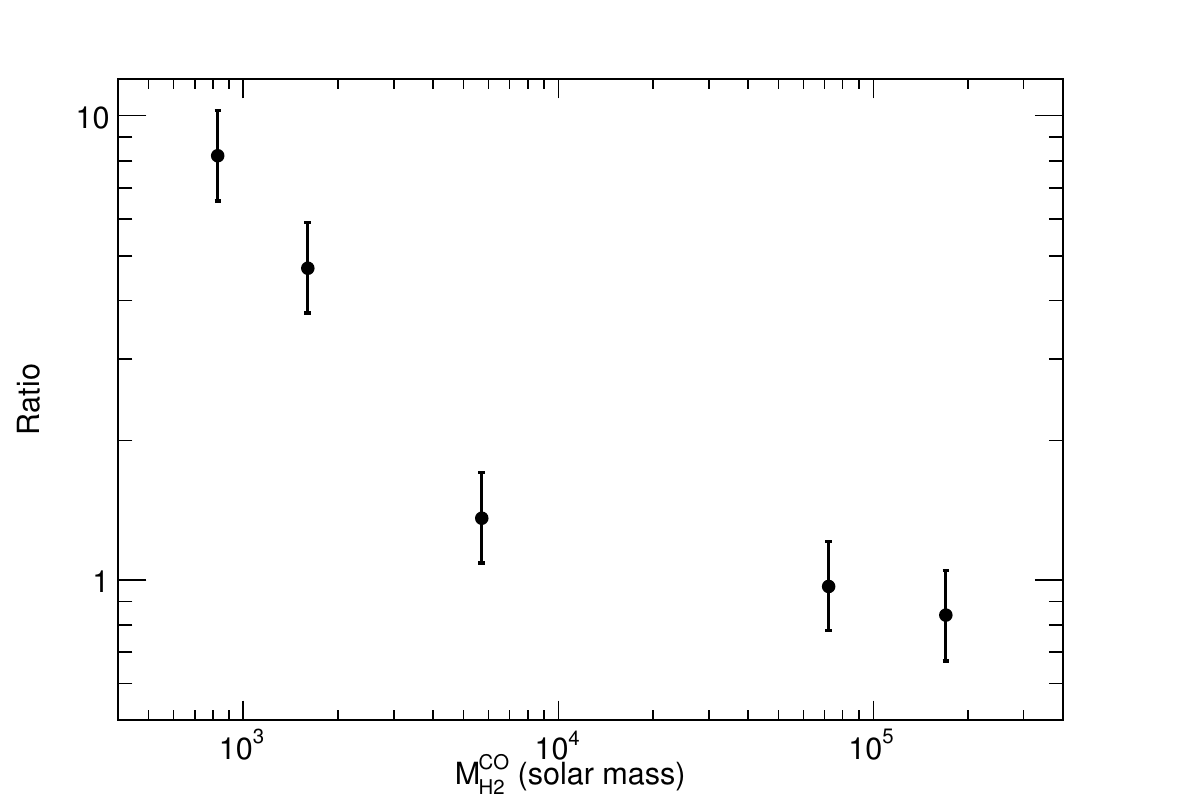}
\end{minipage}
\end{tabular}
\caption{Ratio of the CO-dark {\Htwo} to CO-bright {\Htwo} as a function of {\Htwo} mass traced by $\WCO$.
Error bars represent 25\% systematic uncertainties.
{Alt text: A plot showing the ratio of the CO-dark H2 to CO-bright H2.}}
\label{..}
\end{figure}

\clearpage
\section{Summary and Prospects}

We carried out a systematic study of the Fermi-LAT ${\gamma}$-ray data in the 0.1--25.6~GeV
energy range for five nearby molecular cloud regions, to investigate the properties of the ISM gas and Galactic CRs in the solar neighborhood.
We improved the ISM gas modeling over most previous works by using a component decomposition of the 21 cm {\HI} emission line,
allowing for identifying optically thin {\HI}. 
We also employed a correlation among the Planck dust emission model, narrow-line {\HI}, and broad-line {\HI}, to trace CO-dark {\Htwo}.
Through ${\gamma}$-ray data analysis, we confirmed that
narrow {\HI} is optically thick, and succeeded in distinguishing non-local {\HI}, optically-thin {\HI},
optically-thick {\HI}, CO-dark {\Htwo}, and CO-bright {\Htwo} for all five regions.
On CRs, 
we found that obtained emissivities above 0.4~GeV (proportional to the CR intensity above a few GeV) 
are systematically smaller than previous results and agree better with a model based on directly-measured CRs.
We also found a ${\sim}$10\% decrease in emissivity as a function of Galactocentric radius;
the level of decline is compatible with predictions by a canonical model of Galactic CR propagation.
On the ISM gas, we found that CO-dark {\Htwo} dominates the dark gas.
We also found that the CO-dark {\Htwo} to CO-bright {\Htwo} ratio
anti-correlates with the cloud mass traced by $\WCO$;
the ratio reaches 5--10 for small systems of $\sim 1000~M_{\odot}$. 
The latter finding indicates that CO-dark {\Htwo} should be considered
in studying CRs in the vicinity of individual accelerators.

\begin{ack}
We thank Dmitry Malyshev for providing the FB template file and valuable comments.

The \textit{Fermi} LAT Collaboration acknowledges generous ongoing support
from a number of agencies and institutes that have supported both the
development and the operation of the LAT as well as scientific data analysis.
These include the National Aeronautics and Space Administration and the
Department of Energy in the United States, the Commissariat \`a l'Energie Atomique
and the Centre National de la Recherche Scientifique / Institut National de Physique
Nucl\'eaire et de Physique des Particules in France, the Agenzia Spaziale Italiana
and the Istituto Nazionale di Fisica Nucleare in Italy, the Ministry of Education,
Culture, Sports, Science and Technology (MEXT), High Energy Accelerator Research
Organization (KEK) and Japan Aerospace Exploration Agency (JAXA) in Japan, and
the K.~A.~Wallenberg Foundation, the Swedish Research Council and the
Swedish National Space Board in Sweden.
 
Additional support for science analysis during the operations phase is gratefully
acknowledged from the Istituto Nazionale di Astrofisica in Italy and the Centre
National d'\'Etudes Spatiales in France. This work performed in part under DOE
Contract DE-AC02-76SF00515.

\end{ack}

\section*{Funding}
Part of this work was supported by JSPS KAKENHI Grant Numbers 23K25882 and 23H04895 (T.M.).
Partial support from NASA grant No. 80NSSC22K0495 is also acknowledged.


\clearpage

\appendix 

\section{Regions for the Analysis}
In Figure 14, we show the all-sky revised $\tau_{353}$ map with ROIs overlaid.

\begin{figure}[htbp]
\begin{tabular}{cc}
\centering
\includegraphics[width=\textwidth]{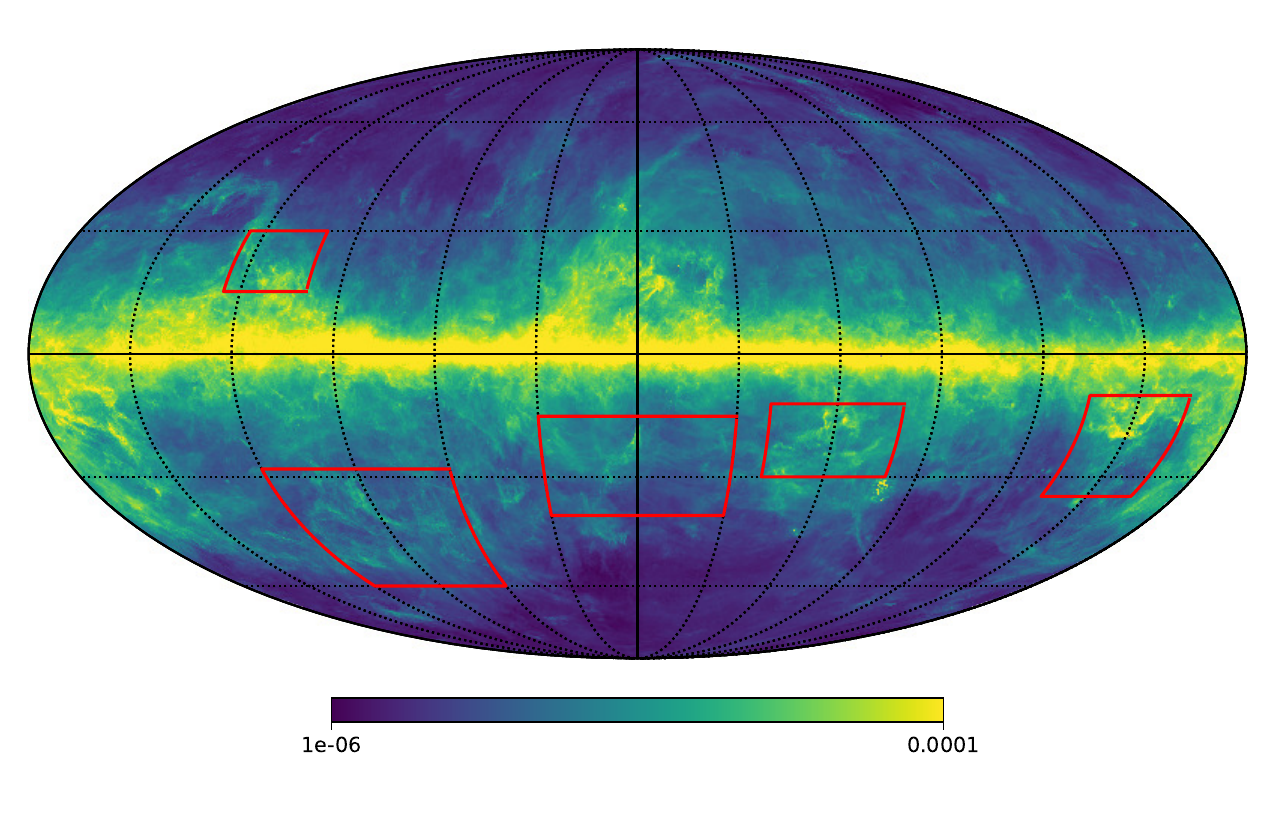}
\end{tabular}
\caption{Five ROIs overlaid on the revised $\tau_{353}$ map (in logarithmic scale). The ROI in positive latitude corresponds to the Cep/Pol region,
and ROIs in negative latitude (from left to right) correspond to the MBM/Pegasus,
R CrA, Chamaeleon, and Orion regions.
{Alt text: A map showing five ROIs.}}
\label{..}
\end{figure}

\section{Threshold to Separate Narrow {\HI} and Broad {\HI}}
{\HI} gas is in thermal equilibrium in most cases, and pure absorption and stimulated emission contribute to the net absorption.
Then,
as discussed by \citet{Hayashi2019}, the {\HI} optical depth $\tau_{\nu}$ at frequency $\nu$ is given by
\begin{equation}
\tau_{\nu} = \frac{3 c^{2} h}{32 \pi \nu k} \cdot \frac{A}{T_\mathrm{s}} \cdot \phi(\nu)
\end{equation}
where $c$, $h$, and $k$ are the speed of light, the Planck constant, and the Boltzmann constant, respectively,
$A$ and $\phi(\nu)$ are the Einstein coefficient for spontaneous emission and the line shape function of
{\HI} 21~cm line, respectively. 
The ratio of the number of atoms in upper and lower energy states deviates slightly from the ratio of the number of states (3:1),
making $\tau_{\nu}$ inversely proportional to $T_\mathrm{s}$.
If the kinetic temperature ($T_\mathrm{k}$) of the {\HI} gas is large, the gas will be optically thin since
$\tau_{\nu} \propto T_\mathrm{s}^{-1} \sim T_\mathrm{k}^{-1}$ and $\tau_{\nu} \propto \phi({\nu}) \propto T_\mathrm{D}^{-0.5} \sim T_\mathrm{k}^{-0.5}$.
Specifically, if we approximate the line profile (of the FWHM $\Delta \VHI$) with a top-hat function with a width of $\Delta \VHI$, $\tau_{\nu}$ can be evaluated as
\begin{equation}
\tau_{\nu} \sim \frac{1}{1.82 \times 10^{18}} \left( \frac{T_\mathrm{s}}{\mathrm{K}} \right)^{-1} \cdot \left( \frac{\NHI}{\mathrm{cm^{-2}}} \right)
\cdot \left( \frac{\Delta \VHI}{\mathrm{km~s^{-1}}} \right)^{-1}~.
\end{equation}
Therefore, with $T_\mathrm{D} \sim T_\mathrm{k} \sim T_\mathrm{s}=1000~\mathrm{K}$, $\Delta \VHI \sim 6.8~\mathrm{km~s^{-1}}$ and 
$\tau_{\nu} \sim \frac{\NHI}{10^{22}~\mathrm{cm^{-2}}}$, and we can assume that the ISM {\HI} gas is optically thin 
up to a column density of
$(2\mbox{--}3) \times 10^{21}~\mathrm{cm^{-2}}$.

\section{Non-local $\WHI$ Maps}
In Figure~15 we summarize non-local $\WHI$ maps prepared in Section~2.2.1.

\begin{figure}[htbp]
\begin{tabular}{cc}
\begin{minipage}{0.5\textwidth}
\centering
\begin{overpic}[width=\textwidth]{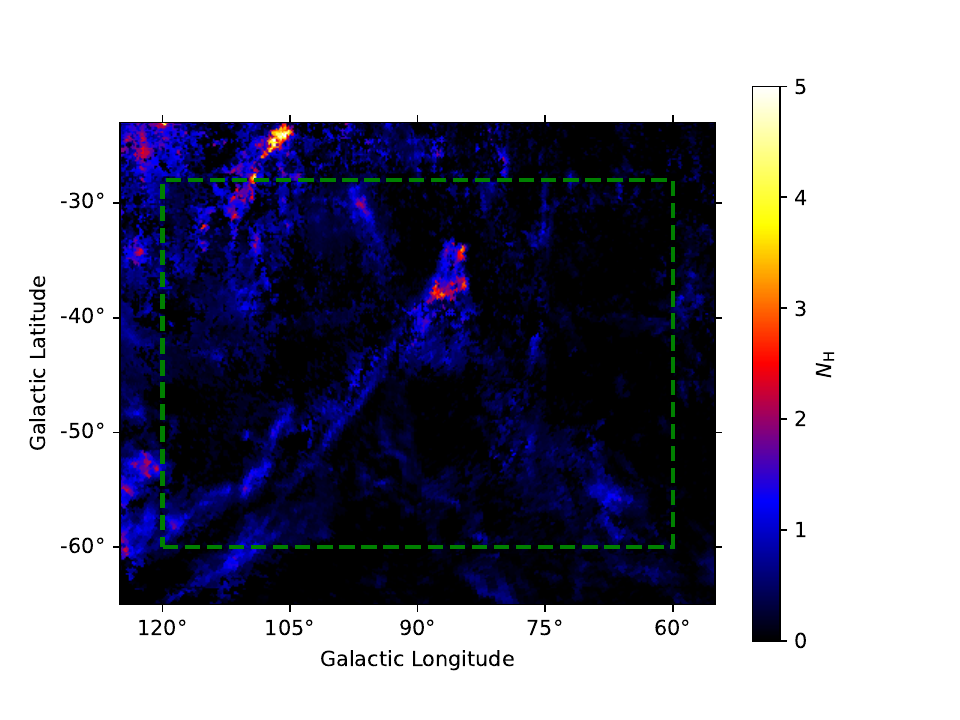}
\put(15,65){(a)}
\end{overpic}
\end{minipage}
\begin{minipage}{0.5\textwidth}
\centering
\begin{overpic}[width=\textwidth]{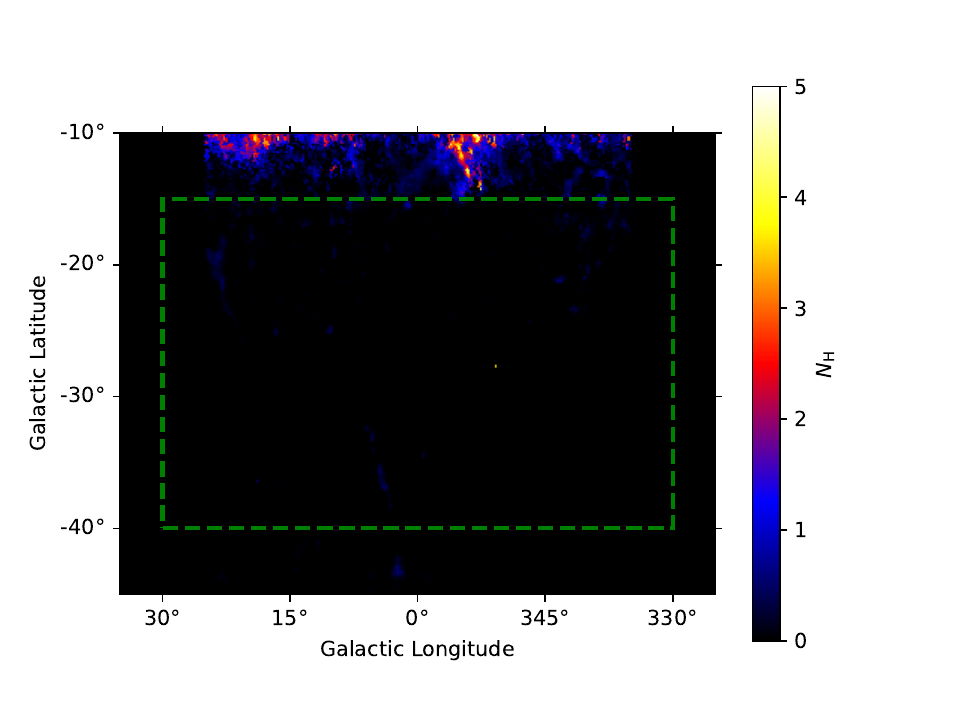}
\put(15,65){(b)}
\end{overpic}
\end{minipage} \\
\\
\begin{minipage}{0.5\textwidth}
\centering
\begin{overpic}[width=\textwidth]{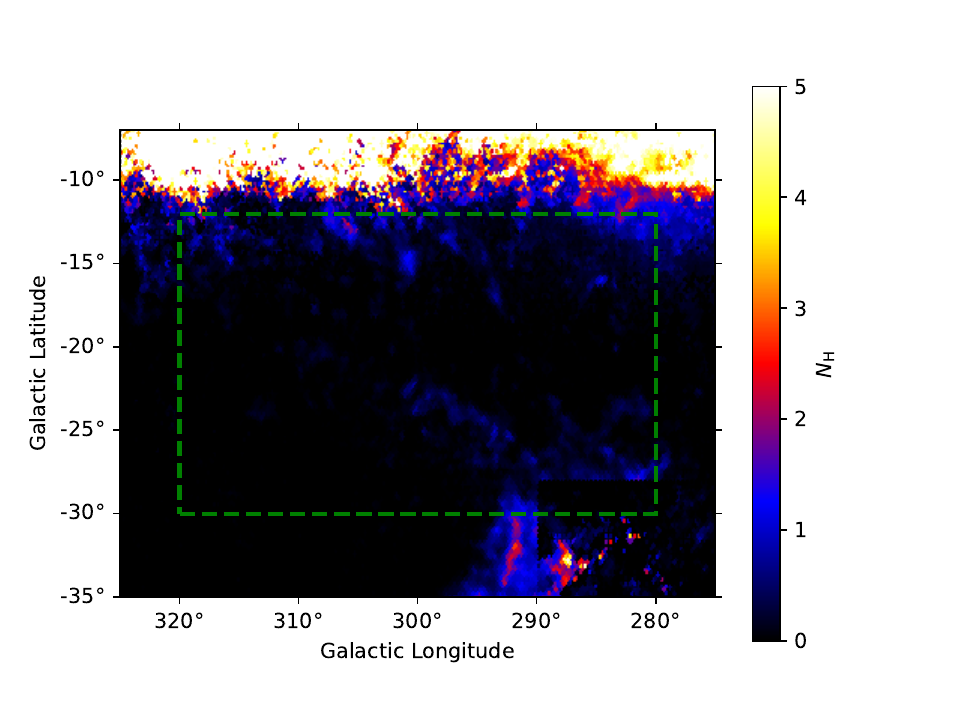}
\put(15,65){(c)}
\end{overpic}
\end{minipage}
\begin{minipage}{0.5\textwidth}
\centering
\begin{overpic}[width=\textwidth]{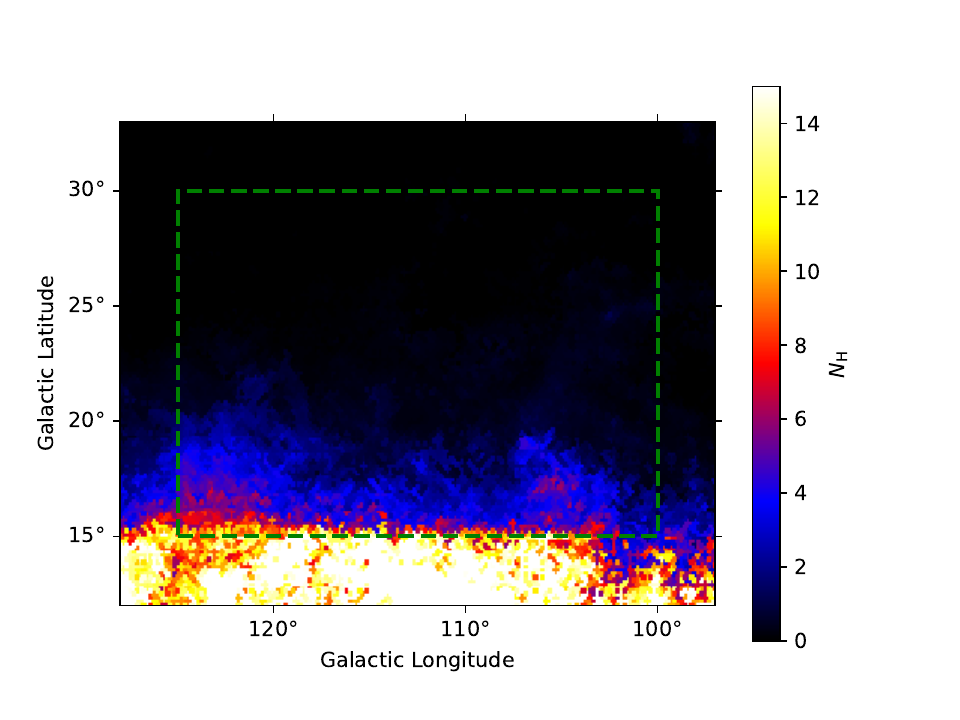}
\put(15,65){(d)}
\end{overpic}
\end{minipage} \\
\\
\begin{minipage}{0.5\textwidth}
\centering
\begin{overpic}[width=\textwidth]{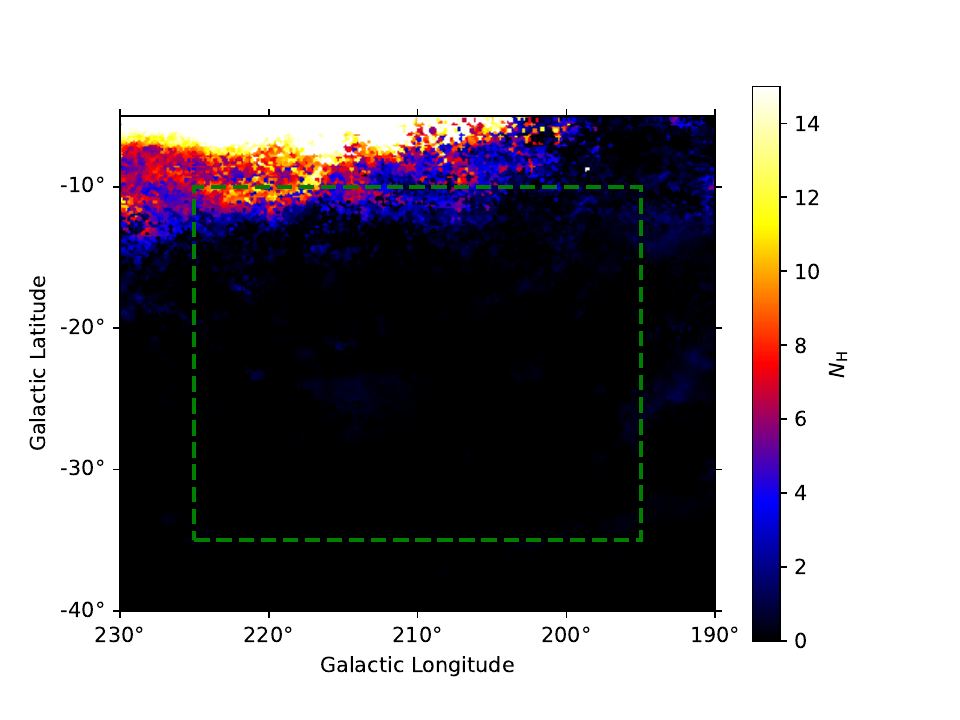}
\put(15,65){(e)}
\end{overpic}
\end{minipage}
\end{tabular}
\caption{$\NHI$ maps of non-local {\HI}, for the MBM/Pegasus region (top left), R CrA region (top right),
Chamaeleon region (middle left), Cep/Pol region (middle right), and Orion region (bottom). 
All maps are converted from $\WHI$ assuming an optically thin case and in $10^{20}~\mathrm{cm^{-2}}$.
{Alt text: Five maps showing non-local HI templates.}}\label{..}
\end{figure}

\clearpage

\section{Treatment of the Infrared Sources in the Orion Region}
In the Planck dust-model maps from Data Release 2, we identified several regions with high $T_{\rm d}$,
$\tau_{353}$, and radiance in the Orion region, indicating localized heating by stars.
We refilled these areas in the $T_{\rm d}$ , $\tau_{353}$, and radiance maps
with the average of the peripheral pixels:
values in a circular region of radius $r_{1}$ are filled with the average
of the pixels in an annulus with inner radius $r_{1}$ and outer radius $r_{2}$.
The central positions in ($l, b$), $r_{1}$, and $r_{2}$ are summarized in Table~7. 
The first two sources are removed from the $\tau_{353}$ map,
and the latter five are removed from the radiance map. All seven sources are removed from the $T_\mathrm{d}$ map.

\begin{table}[htbp]
\tbl{Infrared sources excised and interpolated across
in the \textit{Planck} dust maps}{
\begin{tabular}{cccc}
\hline
$l$(deg) & $b$(deg) & $r_{1}$(deg) & $r_{2}$(deg) \\
\hline \hline
205.1 & $-14.05$ & 0.25 & 0.35 \\
206.9 & $-16.60$ & 0.25 & 0.35 \\
206.5 & $-16.20$ & 0.70 & 0.80 \\
209.0 & $-19.20$ & 0.80 & 0.90 \\
205.3 & $-14.40$ & 0.25 & 0.35 \\
213.7 & $-12.60$ & 0.25 & 0.35 \\
219.2 & $-08.90$ & 0.25 & 0.35 \\
\hline
\end{tabular}}
\end{table}

\section{Spin-Temperature Correction to narrow {\HI}}

In addition to using a single scaling factor (Section~3), we also applied a spin-temperature correction to narrow {\HI} map
based on the procedure by \citet{Mizuno2022}.
For simplicity, we assumed that the peak brightness temperature 
is representative of the
brightness temperature of each {\HI} line along the line of sight. Then we evaluated the optical depth and calculated the gas column density
for each of {\HI} lines of the MBM/Pegasus region. 
We constructed corrected $\NHI$ maps of narrow {\HI} assuming $T_\mathrm{s}=100~\mathrm{K}$ to 40K in 10~K steps.
We used those $\NHI$ maps instead of the original map in the $\gamma$-ray data analysis and found that $T_\mathrm{s} = 50~\mathrm{K}$ gives the
best fit in terms of $\ln{L}$. The fit coefficient ratios of narrow {\HI} and broad {\HI} differ by 2~\% and we take this ratio into account
to construct a single $\NHI$ map. The final model configuration gives the {\HI} emissivity ($\epsilon_{\mathrm m}$) above 400~MeV to be 
$(0.513 \pm 0.008) \times 10^{-26}$ in units of ${\rm ph~s^{-1}~sr^{-1}}$ per H atom that is very close to the one
obtained with a single scaling factor. We applied the same procedure for other regions and
found that the effect is ${\le}$5~\% with $T_\mathrm{s}$ giving the best fit to be 50-90~K. The summary of $T_\mathrm{s}$ and {\HI} emissivity are give in 
Table~8 and Figure~16.

Several previous works evaluated $T_\mathrm{s}$ using $\gamma$-ray data under the assuption of uniform temperature and obtained larger values.
For example, \citet{Planck2015} obtained ${\ge}$300~K for Chamaeleon region and \citet{FermiHI2} obtained 140~K for 
$10^{\circ} \le |b| \le 70^{\circ}$. We understand that the difference mainly comes from that those pionnering works assumed
uniform $T_\mathrm{s}$ to obtain the average value of the temperature across their ROIs, and that narrow {\HI} is
rather optically thick ($T_\mathrm{s} \le 100~\mathrm{K}$). We also remind that there are several caveats in our procedure
(see discussion in \cite{Mizuno2022}) and the values of $T_\mathrm{s}$ presented here should not be taken at face value.

\begin{table}[htbp]
  \tbl{Summary of $T_\mathrm{s}$ and $\epsilon_{\mathrm m}$ above 400~MeV in unit of $10^{>-26}~\mathrm{ph~s^{-1}~sr^{-1}}$}{%
  \begin{tabular}{ccccccc}
      \hline
& MBM/Pegasus & R CrA & Chamaeleon & Cep/Pol & Orion \\ 
\hline
$T_\mathrm{s}(\mathrm{K})$ & 50 & 50 & 60 & 70 & 90 \\
$\epsilon_{\mathrm m}$ & $0.513 \pm 0.008$ & $0.532 \pm 0.007$ & $0.538 \pm 0.008$ & $0.458 \pm 0.008$ & $0.476 \pm 0.003$ \\
      \hline
    \end{tabular}}\label{tab:first}
\end{table}

\begin{figure}[htbp]
\begin{tabular}{cc}
\begin{minipage}{0.5\textwidth}
\centering
\begin{overpic}[width=\textwidth]{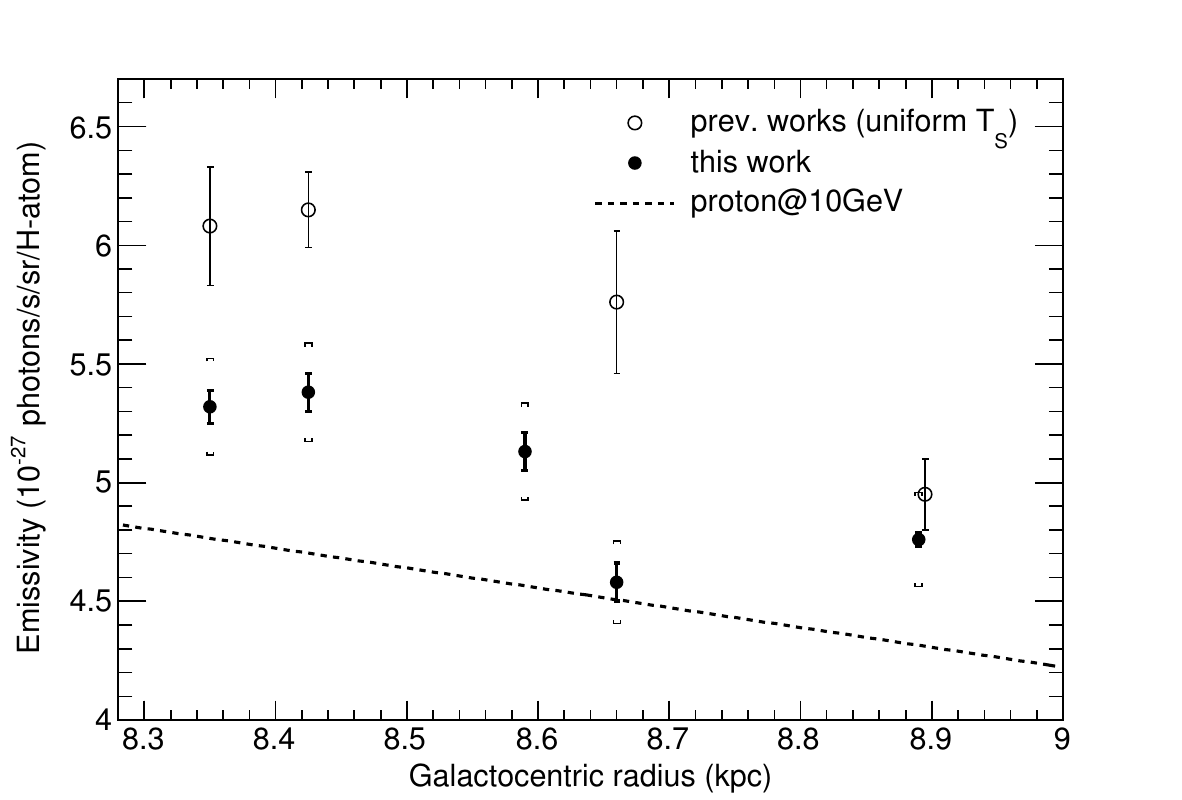}
\put(15,65){(a)}
\end{overpic}
\end{minipage}
\begin{minipage}{0.5\textwidth}
\centering
\begin{overpic}[width=\textwidth]{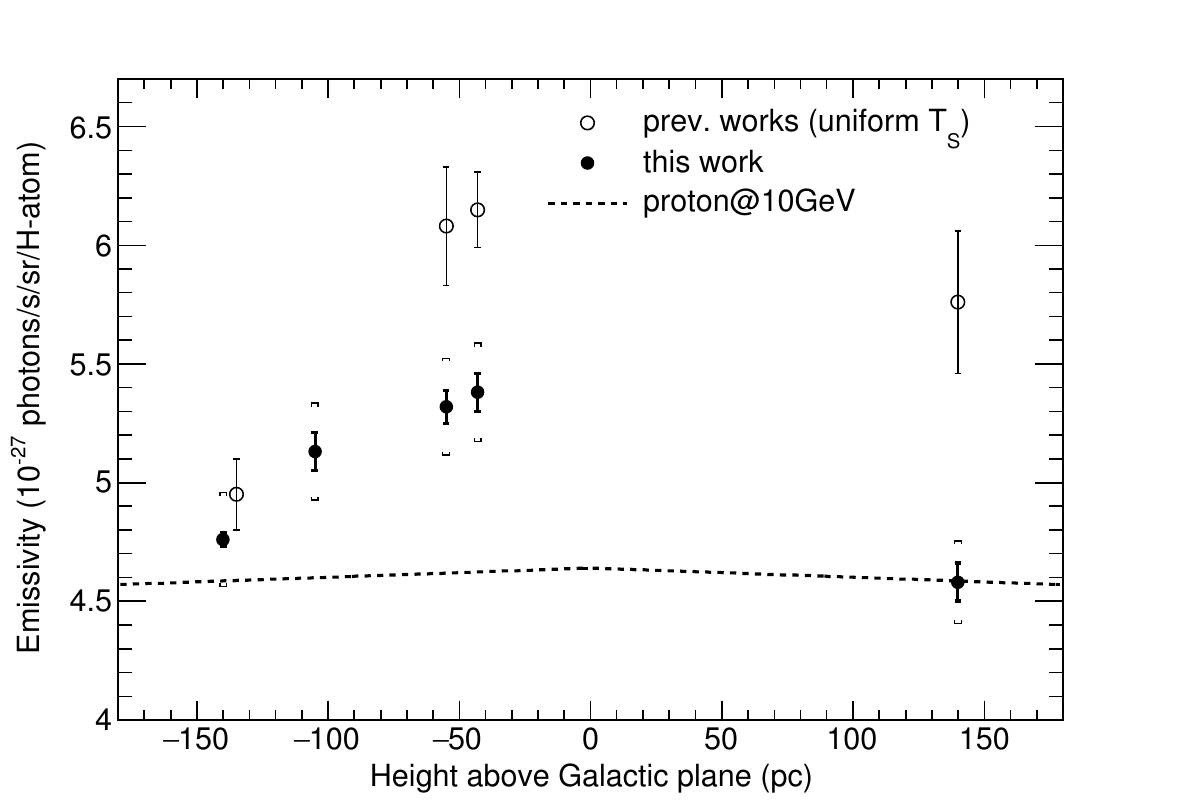}
\put(15,65){(b)}
\end{overpic}
\end{minipage}
\end{tabular}
\caption{The same as Figure~12 but with $T_\mathrm{s}$ correction for narrow {\HI} is applied instead of 
a single scaling factor.
{Alt text: Two plots showing emissivities of regions studied.}
}\label{..}
\end{figure}

\clearpage 

\section{Fermi bubble template}
In analyzing the R CrA region, we employed a template of FB developed by \citet{Ackermann2017} (Figure~8 of the reference). Whatever dust emission model we used,
we observed coherent positive residuals in $(l, b) \sim (\timeform{0D}, \timeform{-30D})$ and $\sim (\timeform{-10D}, \timeform{-32D})$, 
and coherent negative residuals in $l=\timeform{330D} \mbox{--} \timeform{334D}$.
We found that positive residuals correspond to holes in the template map that positionally coincide with brobs in 
the soft component of $\gamma$-rays toward the Galactic center (Figure~7 in \cite{Ackermann2017}), 
and negative residuals are at the peripherals of the
template. Accordingly we filled holes in the template, and removed the peripherals.
The new FB template, shown in Figure~16,
gives a much-improved fit to $\gamma$-rays (with $\Delta \ln{L}$ more than 600).

\begin{figure}[htbp]
\begin{tabular}{cc}
\begin{minipage}{0.5\textwidth}
\centering
\includegraphics[width=\textwidth]{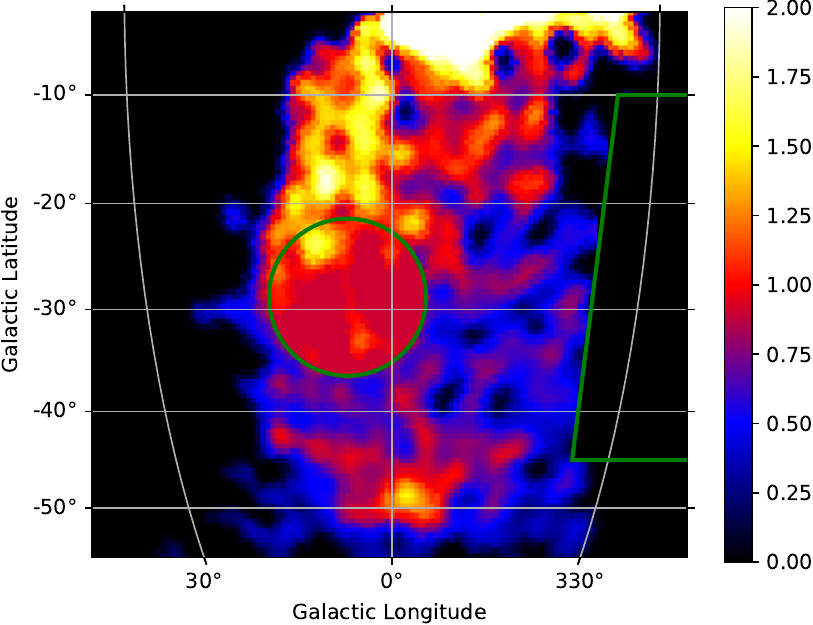}
\end{minipage}
\end{tabular}
\caption{FB template map (in an arbitrary unit) used in this study. 
Original map in \citet{Ackermann2017} was modified in a circle and polygon presented.
{Alt text: A map showing the FB template.}
}
\label{..}
\end{figure}

\section{Summary of $X_\mathrm{CO}$}
For convenience we summarize $X_\mathrm{CO}$ for each region. We remind that errors are statistical only.

\begin{table}[htbp]
  \tbl{Summary of $X_\mathrm{CO}$ in unit of $10^{20}~\mathrm{cm^{-2}~(K~km~s^{-1})^{-1}}$}{%
  \begin{tabular}{cccccc}
      \hline
MBM/Pegasus & R CrA & Chamaeleon & Cep/Pol & Orion \\ 
\hline
$0.559 \pm 0.038$ & $1.668 \pm 0.034$ & $0.932 \pm 0.016$ & $0.912 \pm 0.017$ & $1.296 \pm 0.010$ \\
      \hline
    \end{tabular}}\label{tab:first}
\end{table}


\clearpage

\begin{thebibliography}{}
\bibitem[Abdo et al.(2009)]{Abdo2009} Abdo, A.~A., Ackermann, M., Ajello, M., et al.\ 2009, Astropart. Phys., 32, 193
\bibitem[Abdollahi et al.(2022)]{Fermi4FGLDR3} Abdollahi, S., Acero, F., Baldini, L., et al.\ 2022, \apjs, 260, 53
\bibitem[Ackermann et al.(2012a)]{FermiPaper2} Ackermann, M., Ajello, M., Atwood, W.~B., et al.\ 2012a, \apj, 750, 3
\bibitem[Ackermann et al.(2012b)]{Ackermann2012} Ackermann, M., et al.\ 2012b, \apj, 755, 22
\bibitem[Ackermann et al.(2012c)]{FermiOrion} Ackermann, M., et al.\ 2012c, \apj, 756, 4
\bibitem[Ackermann et al.(2017)]{Ackermann2017} Ackermann, M. et al.\ 2017, \apj, 840, 43
\bibitem[Amenomori et al.(2021)]{Amenomori2021} Amenomori, M. et al.\ 2021, Nature Astronomy, 5, 460
\bibitem[Atwood et al.(2009)]{Atwood2009} Atwood, W.~B., Abdo, A.~A., Ackermann, M., et al.\ 2009, \apj, 697, 1071
\bibitem[Atwood et al.(2013)]{Atwood2013} Atwood, W.~B., Albert, L., Baldini, L., et al.\ 2013, arXiv:1303.3514
\bibitem[Bloemen et al.(1984)]{Bloemen1984} Bloemen, J.~B.~G.~M.\ et al.\ 1984, \aap, 139, 37
\bibitem[Bruel et al.(2018)]{Bruel2018} Bruel, P., Burnett, T.~H., Digel, S.~W., et al.\ 2018, arXiv:1810.11394
\bibitem[Casandjian(2015)]{FermiHI2} Casandjian, J.-M.\ 2015, \apj, 806, 240
\bibitem[Casandjian et al.(2022)]{Casandjian2022} Casandjian, J.-M. et al.\ 2022, \apj, 940, 116
\bibitem[Dame et al.(1987)]{Dame1987} Dame, T.~M., et al.\ 1987, \apj, 332, 706
\bibitem[Dame et al.(2001)]{Dame2001} Dame, T.~M., Hartmann, D., \& Thaddeus, P.\ 2001, \apj, 547, 792
\bibitem[Dame(2011)]{Dame2011} Dame, T.~M. \ 2011, arXiv:1101.1499
\bibitem[Dickey \& Lockman(1990)]{Dickey1990} Dickey, J.~M., \& Lockman, F.~J.\ 1990, \araa, 28, 215
\bibitem[Ferriere(2001)]{Ferriere2001} Ferriere, K.~M.\ 2001, Rev. Mod. Phys., 73, 1031
\bibitem[Fukui et al.(2015)]{Fukui2015} Fukui, Y., Torii, K., Onishi, T., et al.\ 2015, \apj, 798, 6
\bibitem[Fukui et al.(2014)]{Fukui2014} Fukui, Y., Okamoto, R., Kaji, et al.\ 2014, \apj, 796, 59
\bibitem[Galli et al.(2020)]{Gali2020} Galli P. A. B., Bouy H., Olivares J. et al. 2020, \aap, 634 98
\bibitem[G\'{o}rski et al.(2005)]{Gorski2005} G\'{o}rski, K.~M., Hivon, E., Banday, A.~J., et al.\ 2005, \apj, 622, 759
\bibitem[Grenier et al.(2005)]{Grenier2005} Grenier, I.~A., Casandjian, J.-M., \& Terrier, R.\ 2005, Science, 307, 1292
\bibitem[Grenier et al.(2015)]{Grenier2015} Grenier, I.~A., Black, J.~H., \& Strong, A.~W.\ 2015, \araa, 53, 199
\bibitem[Hayashi et al.(2019)]{Hayashi2019} Hayashi, K., Mizuno, T., Fukui, Y., et al.\ 2019, \apj, 884, 130
\bibitem[HI4PI Collaboration(2016)]{HI4PI} HI4PI Collaboration 2016, \aap, 594, 116
\bibitem[Kachelriess et al.(2019)]{AAfrag} Kachelriess, M., Moskalenko, I.~V., \& Ostapchenko, S.\ 2019, Computer Physics Communications, 245, 106846
\bibitem[Kalberla \& Kerp(2009)]{Kalberla2009} Kalberla, P.M.~W., \& Kerp, J. \ 2009, \araa, 47, 27
\bibitem[Kalberla \& Haud(2018)]{Kalberla2018} Kalberla, P.M.~W., \& Haud, U. \ 2018, \aap, 619, 58
\bibitem[Kalberla et al.(2020)]{Kalberla2020} Kalberla, P.M.~W., Kerp, J., \& Haud, U. \ 2020, \aap, 639, 26
\bibitem[Kothes et al.(2006)]{Kothes2006} Kothes, R., et al.\ 2006, \apj, 638, 225
\bibitem[Luhman et al.(2008)]{Luhman2008} Luhman, K.~L., et al. \ 2008, in Handbook of Star Forming Regions, Volume ~II: THe Southern Sky, Vol. 5, ed. B. Reipurth
(San Francisco, SA: ASP), 169
\bibitem[Mattox et al.(1996)]{Mattox1996} Mattox, J.~R., Bertsch, D.~L., Chiang, J., et al.\ 1996, \apj, 461, 396
\bibitem[Maurin et al.(2014)]{Maurin2014} Maurin, D., Melot, F., \& Taillet, R.\ 2014, \aap, 569, 32
\bibitem[Mizuno et al.(2001)]{Mizuno2001} Mizuno, A., et al.\ 2001, \pasj, 53, 1071
\bibitem[Mizuno et al.(2016)]{Mizuno2016} Mizuno, T., Abdollahi, S., Fukui, Y., et al.\ 2016, \apj, 833, 278
\bibitem[Mizuno et al.(2022)]{Mizuno2022} Mizuno, T., et al.\ 2022, \apj, 935, 97
\bibitem[Moskalenko et al.(2006)]{Moskalenko2006} Moskalenko, I.~V., Porter, T., Strong W.\ 2006, \apjl, 640, 155
\bibitem[Murray et al.(2018)]{Murray2018} Murray, C.~M., Peek, J.E.~G., Lee, M.-Y., et al.\ 2018, \apj, 862, 131
\bibitem[Nishimura et al.(2015)]{Nishimura2015} Nishimura, A., et al.\ 2015, \apjs, 216, 18
\bibitem[Orlando(2018)]{Orlando2018} Orlando, E.\ 2018, \mnras, 475, 2724
\bibitem[Panopoulou et al.(2016)]{Panopoulou2016} Panooulou, G.~V., et al.\ 2016, \mnras, 462, 1517
\bibitem[Perrot et al.(2003)]{Perrot2003} Perrot, C.~A., et al.\ 2003, \aap, 404, 519
\bibitem[Planck Collaboration XIX(2011)]{Planck2011} Planck Collaboration XIX 2011, \aap, 536, 19
\bibitem[Planck Collaboration XI(2014)]{Planck2014} Planck Collaboration XI 2014, \aap, 571, 11
\bibitem[Planck Collaboration XI(2015)]{Planck2015} Planck Collaboration XXVIII 2015, \aap, 582, 31
\bibitem[Porter et al.(2017)]{Porter2017} Porter, T.~A., J\'{o}hannesson, G., Moskalenko, I.~V. \ 2017, \apj, 846, 23
\bibitem[Reach et al.(1994)]{Reach1994} Reach, W.~T., Bon-Chul, K., \& Carl, H.\ 1994, \apj, 429, 672
\bibitem[Remy et al.(2017)]{Remy2017} Remy, Q., et al.\ 2017, \aap, 601, 78
\bibitem[Remy et al.(2018)]{Remy2018} Remy, Q., et al.\ 2018, \aap, 611, 51
\bibitem[Schlegel et al.(1998)]{Schlegel1998} Schlegel, D.~J., Finkbeiner, D.~P., \& Davis, M. \ 1998, \apj, 500, 525
\bibitem[Smith et al.(2014)]{Smith2014} Smith, R.~J., et al.\ 2014, \mnras, 441, 1628
\bibitem[Strong \& Moskalenko(1998)]{Galprop1} Strong, A.~W., \& Moskalenko, I.\ 1998, \apj, 509, 212
\bibitem[Strong et al.(2007)]{Galprop2} Strong, A.~W., Moskalenko, I.V., \& Ptuskin, V.~S.\ 2007, \araa, 57, 285
\bibitem[Su et al.(2010)]{Su2010} Su, M., et al.\ 2010, \apj 724, 1044
\bibitem[Tachihara et al.(2024)]{Tachihara2024} Tachihara, K. et al.\ 2024, \apj, 968, 131
\bibitem[Welty et al.(1989)]{Welty1989} Welty, D.~E., Hobbs, L.~M., \& Penprase, B.~E.\ 1989, \apj, 346, 232
\bibitem[Wolfire et al.(2010)]{Wolfire2010} Wolfire, M.~G., Hollenbach, D., \& McKee, C.~F. \ 2010, \apj, 716, 1191
\bibitem[Yamamoto et al.(2006)]{Yamamoto2006} Yamamoto, H., Kawamura, A., Tachihara, K., et al.\ 2006, \apj, 642, 307
\bibitem[Yang et al.(2014)]{Yang2014} Yang, R.-z., et al.\ 2014, \aap, 566, 142
\bibitem[Yusifov \& K\"{u}c\"{u}k(2004)] {CRdist}Yusifov, I., \& K\"{u}c\"{u}k, I. \ 2004, \aap, 422, 553
\bibitem[Wakker(2001)]{Wakker2001} Wakker, B.~P.\ 2001, \apjs, 136, 463
\bibitem[Welty et al.(1989)]{Welty1989} Welty, D.~E., Hobbs, L.~M., \& Penprase, B.~E.\ 1989, \apj, 346, 232
\bibitem[Wilson et al.(2005)]{Wilson2005} Wilson, B.~A., et al.\ 2005, \aap, 430, 523
\end{thebibliography}
\end{document}